\documentclass[11pt, twoside]{uwthesis}
\usepackage{bm}
\usepackage{amsmath}
\usepackage{graphicx}
\usepackage{amsmath}
\usepackage{amsfonts}
\usepackage{feynmp}
\usepackage{amssymb}
\usepackage{latexcad}

\begin{document}
\ifx\href\undefined\else\hypersetup{linktocpage=true}\fi 

\prelimpages

\Title{Quarks and Antiquarks in Nuclei} \Author{Jason Smith}
\Year{2005} \Program{Physics}

\titlepage

\Chair{Gerald Miller}{Professor}{Department of Physics}

\Signature{Gerald Miller}

\Signature{Aurel Bulgac}

\Signature{Stephen Ellis}

\signaturepage

\doctoralquoteslip

\setcounter{page}{-1}

\abstract{The Chiral Quark-Soliton model provides the quark and
antiquark substructure of the nucleon, which is embedded in
nuclear matter. This provides a new way to asses the effects of
the nuclear medium on the nucleon. We calculate nuclear binding
and saturation, describe the European Muon Collaboration effect
consistently with Drell-Yan experiments, and predict modifications
to the polarized parton distributions. We also calculate the
electromagnetic form factors of a bound proton, and find
significant modifications of the electromagnetic form factors in
the nuclear environment. In every case, the properties of the sea
of quark-antiquark pairs in the nucleon are very important, and
serve to mitigate the valence quark effect. The changes in the sea
quarks when the nucleon is immersed in the nuclear medium are the
primary mode by which consistency is maintained with experimental
constraints (Drell-Yan data, magnetic moment), while still
maintaining a significant effect needed to explain the deep
inelastic scattering and polarization transfer data.}

\tableofcontents

\listoffigures

\listoftables

\chapter*{Glossary}
\addcontentsline{toc}{chapter}{Glossary} \thispagestyle{plain}

\begin{glossary}
\item[DGLAP] Dokshitzer-Gribov-Lipatov-Altarelli-Parisi
\item[DIS] Deep Inelastic Scattering
\item[EMC] European Muon Collaboration (effect)
\item[FNAL] Fermi National Accelerator Laboratory
\item[GPD] Generalized Parton Distribution
\item[HERMES] HERA fixed target experiment to explore spin
\item[IA] Impulse Approximation
\item[KR] Kahana-Ripka (basis)
\item[MFT] Mean Field Theory:
\item[MIT] Massachusetts Institute of Technology (bag model)
\item[NJL] Nambu-Jona-Lasinio (model)
\item[NM] (Infinite) Nuclear Matter
\item[OPE] Operator Product Expansion
\item[PV] Pauli-Villars (subtraction)
\item[QCD] Quantum Chromodynamics
\item[QMC] Quark-Meson Coupling (model)
\item[RDWIA] Relativistic Distorted-Wave Impulse Approximation
\item[RMFT] Relativistic Mean Field Theory
\item[RMSGA] Relativistic Multiple-Scattering Glauber Approximation
\item[RPWIA] Relativistic Plane-Wave Impulse Approximation
\item[SLAC] Stanford Linear Accelerator Center
\item[TJNAF] Thomas Jefferson National Accelerator Facility
\item[$\chi$PT] Chiral Perturbation Theory
\item[$\chi$QS] Chiral Quark-Soliton (model)
\end{glossary}


\textpages

\chapter{Introduction}
\label{ch:intro}

The past century has seen the birth and incredible progress of the
physics of hadrons and their strong interactions from the
discovery of the nucleus \cite{Rutherford:1911} to the recognition
of Quantum Chromodynamics (QCD) as the fundamental theory of the
strong interactions with the 2004 Nobel prize in physics
\cite{Gross:1973id,Politzer:1973fx}. Despite being the fundamental
theory, describing a proton, much less an entire nucleus, in the
context of the quark and gluon degrees of freedom of QCD is still
out of reach. The fundamental theory is non-perturbative at the
low energy, long wavelength scales of nuclear physics, but it does
tell us some very useful information about the physics of hadrons.
It gives us a scale $\Lambda_{QCD}$ that separates perturbative
and non-perturbative physics, tells us how to translate our
knowledge at one scale to another through the evolution equations,
gives us an approximate chiral symmetry implying its spontaneous
breaking at low scales, and hints at the non-perturbative degrees
of freedom with the instanton solutions to the QCD field equations
providing a possible mechanism for that symmetry breaking
\cite{'tHooft:1976up}. The enormous effect of chiral symmetry has
been encapsulated in the very successful low energy effective
theory, Chiral Perturbation Theory ($\chi$PT)
\cite{Gasser:1983yg}.

These high and low energy theories have yet to meet, and a
frontier of hadronic physics lies in the intermediate range of
length scales available to present experiments where neither QCD
nor $\chi$PT have useful perturbative expansions. Neither
fundamental quarks and gluons nor point-like hadrons provide a
complete description, so including the non-perturbative
information that hadrons are bound states of valence quarks in a
polarized vacuum is necessary. One way to probe these intermediate
length scales and this non-perturbative physics is to examine the
short distance structure of a large object. The prime example is
the European Muon Collaboration (EMC) effect \cite{Aubert:1983xm}
measured in lepton-nucleus Deep Inelastic Scattering (DIS) where
the short distance ($\sim 5$ GeV, or $\sim 10^{-2}\text{ fm}$)
structure of nuclei differs from that of a collection of free
nucleons. This measurement showed that the quarks and antiquarks
in bound nucleons carry less momentum than in free ones, and
implied that the medium modification of nucleon structure could be
significant for many nuclear observables \cite{Geesaman:1995yd}.
Explanations of the EMC effect have remained a diverse and
controversial theoretical subject. Until a measurement of dimuon
production in proton-nucleus scattering through the Drell-Yan
process \cite{Alde:im} demonstrated separately that the antiquarks
in nucleons are virtually unchanged in a nuclear system, there was
virtually no other handle on medium modifications in DIS. Many
explanations either fail to describe the Drell-Yan data, fail to
address antiquarks at the non-perturbative scale of nuclei, or
fail to address antiquarks altogether.

There is a hint to the solution from evolution in QCD; dynamical
rescaling \cite{Close:1984zn} suggests that the EMC effect is due
to a lowering of the starting scale for the renormalization group
equations. While successful at describing the phenomenology of
both the EMC effect and Drell-Yan experiments, like the situation
for QCD and chiral symmetry breaking, it is silent on the actual
mechanism and scale.

The first evidence that the protons and neutrons had structure
came from the measurement of their electromagnetic form factors in
elastic lepton-nucleon scattering \cite{Mcallister:1956ng}. Recent
polarization transfer experiments at TJNAF (Thomas Jefferson
National Accelerator Facility, also referred to as JLab)
\cite{Strauch:2002wu} observed a difference in the electromagnetic
form factors of a proton bound in a Helium nucleus compared to a
free one. This seems to add to the evidence in favor of the
modification of hadrons in the nuclear medium.

We would like to address the subject of medium modifications with
a theory that is consistent with QCD and $\chi$PT. Any description
of the EMC effect must be consistent with the constraints set by
both DIS and Drell-Yan data. Thus a successful model must include
antiquarks as well as quarks, and show how the medium modifies
both the valence and sea quark distributions. Our purpose is to
provide a mechanism for that modification within the Chiral
Quark-Soliton ($\chi$QS) model
\cite{Kahana:dx,Birse:1983gm,Diakonov:2000pa,Christov:1995vm}. It
is essentially a relativistic mean field approximation approach to
baryons motivated by the Instanton Liquid Model (ILM) of the QCD
vacuum and the large $N_{C}$ expansion (discussed in Chapter
\ref{ch:cqsm}). This phenomenological model has many desirable
qualities: the ability to describe a wide class of hadron
observables with surprising accuracy given its simplicity, the
inclusion of antiquarks, the positivity of quark distributions
\cite{Diakonov:1997vc}, polynomiality of Generalized Parton
Distributions (GPD) \cite{Schweitzer:2002nm}, and a basis in QCD
\cite{Diakonov:2000pa}. The model also predicted the mass and
narrow width \cite{Diakonov:1997mm} of the $\theta^{+}$ pentaquark
for which there was encouraging early evidence
\cite{Barmin:2003vv,Nakano:2003qx}, but remains a controversial
subject with many experiments showing null results
\cite{Dzierba:2004db}. One of the drawbacks of the model is that,
in the form presented here, it lacks explicit confinement. This is
the property of QCD that quarks and gluons are never seen as free
particles. The lack of confinement in the $\chi$QS model has not
prevented it from describing a wide variety of free nucleon
observables \cite{Christov:1995vm}. Here we show how the model
describes nuclear saturation properties, reproduces the EMC
effect, satisfies the bounds on nuclear antiquark enhancement
provided by Drell-Yan experiments, and predicts large effects in
spin polarized structure functions and electromagnetic form
factors.

We will continue this introduction to the subject with a more
detailed description of deep inelastic scattering and the EMC
effect in Section \ref{sec:dis} and dimuon production in Section
\ref{sec:dy}. We will then present the $\chi$QS model in Chapter
\ref{ch:cqsm}, and use it to build a nuclei out of quarks and
antiquarks in Chapter \ref{ch:np}. The medium modifications of the
quark distributions and electromagnetic form factors are
calculated in Chapters \ref{ch:dist} and \ref{ch:emff},
respectively. In the Appendices, we elaborate on the connection of
the $\chi$QS model to $\chi$PT and QCD via lattice calculations of
the non-perturbative quark propagator, provide background on the
small binding effect predicted by relativistic mean field theory
for finite nuclei and nuclear matter, and present some details for
the numerical work. This work is built on results that have been
published previously elsewhere
\cite{Miller:2001tg,Smith:2002ci,Smith:2003hu,Smith:2004dn}.

\section{Deep Inelastic Scattering and the EMC Effect}
\label{sec:dis}

Deep inelastic lepton-nucleon scattering occurs when a virtual
photon, mediating the interaction, of high energy and momentum $q$
($-q^{2} \gg \Lambda_{QCD}^{2}$) fragments the target nucleon $N$
of mass $M_{N}$ into a host of particles labelled $X$. The
electromagnetic interaction couples to all of the electrically
charged constituents in the target nucleon. Deep inelastic
scattering experiments see the valence quarks as well as the
virtual quarks and antiquarks in the polarized QCD vacuum
simultaneously.

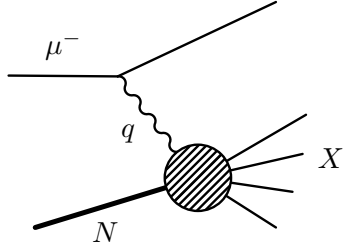
\begin{figure}\centering
\begin{fmffile}{basicdis}
\begin{fmfgraph*}(40,30)
    \fmfleft{nucleon1,y2,muon1,y1}
    \fmf{plain,label=$\mu^{-}$,label.side=left,tension=2}{muon1,v1}
    \fmf{plain,label=$N$,width=2,tension=2}{nucleon1,blob1}
    \fmf{photon,label=$q$,tension=0.75}{v1,blob1}
    \fmf{plain}{v1,muon2}
    \fmfblob{25}{blob1}
    \fmf{plain}{blob1,x1}
    \fmf{plain}{blob1,x2}
    \fmf{plain}{blob1,x3}
    \fmf{plain}{blob1,x4}
    \fmfright{x1,x2,x3,x4,x5,x6,muon2}
    \fmfv{label=$X$}{x3}
\end{fmfgraph*}
\end{fmffile} \caption{Basic diagram of the deep inelastic scattering
process $\mu^{-}+N\rightarrow\mu^{-}+X$ with momentum transfer
$q$.}\label{fig:basicdis}
\end{figure}

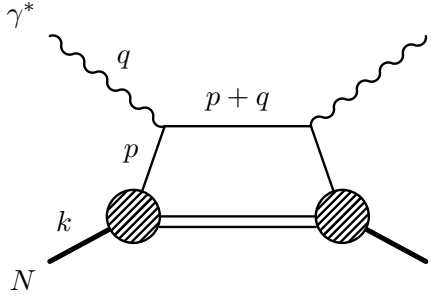
\begin{figure}\centering
\begin{fmffile}{parton}
\begin{fmfgraph*}(50,30)
    \fmfstraight
    \fmfpen{thin}
    \fmfleft{nin,gin,gin2,gin3}
    \fmfright{nout,gout,gout2,gout3}
    \fmftop{p1,p2}
    \fmf{phantom,tension=4}{nin,b1}
    \fmf{phantom,tension=2}{b1,b2}
    \fmf{phantom,tension=4}{b2,nout}
    \fmf{phantom}{p1,b1}
    \fmf{phantom}{p2,b2}
    \fmffreeze
    \fmfv{label=$N$}{nin}
    \fmfv{label=$\gamma^{*}$}{p1}
    \fmfblob{20}{b1,b2}
    \fmf{photon,label=$q$,label.side=left}{p1,v1}
    \fmf{photon}{v2,p2}
    \fmf{plain,label=$p$}{b1,v1}
    \fmf{plain,label=$p+q$,label.side=left}{v1,v2}
    \fmf{plain}{v2,b2}
    \fmf{plain,width=2,label=$k$,label.side=left}{nin,b1}
    \fmf{plain,width=2}{b2,nout}
    \fmf{plain}{b1,b2}
    \fmfi{plain}{vpath (__b1,__b2) shifted (thick* (0,-2))}
\end{fmfgraph*}
\end{fmffile} \caption{The handbag diagram in the parton model.
The virtual photon $\gamma^{*}$ transfers momentum $q$ to the
nucleon. The nucleon has momentum $k$, and the struck quark has
momentum $p$.} \label{fig:parton}
\end{figure}

The case where the lepton is a muon is illustrated in the Feynman
diagram in Fig.~\ref{fig:basicdis}.
The use of the optical theorem allows us to package our ignorance
of the photon-nucleon vertex and fragmentation function in
Fig.~\ref{fig:basicdis}. In inclusive inelastic scattering (where
one measures the cross section with any final state $X$), we can
write the sum on the final states $X$ of the diagram in
Fig.~\ref{fig:basicdis} as
\[
\begin{fmffile}{optical}
\sum_{X} \left|
\parbox{27mm}{
\begin{fmfgraph*}(20,10)
    \fmfleft{n1,p1}
    \fmf{plain,width=2,tension=2}{n1,b1}
    \fmf{photon,tension=2}{p1,b1}
    \fmfblob{20}{b1}
    \fmf{plain}{b1,x1}
    \fmf{plain}{b1,x2}
    \fmf{plain}{b1,x3}
    \fmf{plain}{b1,x4}
    \fmfright{xa,xb,x1,x2,x3,x4,xc,xd}
    \fmfv{label=$X$}{x3}
\end{fmfgraph*}}\right|^{2} = 2 \text{ Im } \left(
\parbox{20mm}{
\begin{fmfgraph*}(20,10)
    \fmfleft{n1,p1}
    \fmf{plain,width=2,tension=2}{n1,b1}
    \fmf{photon,tension=2}{p1,b1}
    \fmfblob{20}{b1}
    \fmf{plain,width=2,tension=2}{b1,n2}
    \fmf{photon,tension=2}{b1,p2}
    \fmfright{n2,p2}
\end{fmfgraph*}} \right) \equiv 2 \text{ Im } W^{\mu\nu}(k,q)
\end{fmffile}
\]
We can use Lorentz covariance and current conservation to write
down the form of $W^{\mu\nu}$
\begin{equation}
W^{\mu\nu}(k,q)=\left(g^{\mu\nu}-\frac{q^{\mu}q^{\nu}}{q^{2}}\right)W_{1}(k,q)
+\frac{1}{M_{N}^{2}}\left(k^{\mu}-q^{\mu} \frac{k\cdot q
}{q^{2}}\right)\left(k^{\nu}-q^{\nu} \frac{k\cdot q
}{q^{2}}\right)W_{2}(k,q),
\end{equation}
but we need a model to calculate $W^{\mu\nu}$, specifically the
functions $W_{1}$ and $W_{2}$, explicitly. For example, we can use
the parton model of Bjorken and Feynman as in
Fig.~\ref{fig:parton}, where the nucleon is composed of massless
(or nearly massless) point-like, non-interacting, mostly spin-1/2
constituents. This model is justified by the asymptotic freedom
\cite{Politzer:1973fx,Gross:1973id} of QCD, where at high energy
the strong interaction coupling $\alpha_{s}$ between the quarks
and gluons goes to zero.

It is conventional to use the momenta in Fig.~\ref{fig:parton} to
define the variables
\begin{eqnarray}
Q^{2} & \equiv &  -q^{2}\\
\nu  & \equiv &  \frac{k\cdot q}{M_{N}}\\
x & \equiv &  \frac{-q^{2}}{2 k\cdot q} = \frac{Q^{2}}{2 M_{N}
\nu} \simeq \frac{p^{+}}{k^{+}} \label{eq:xbj}
\end{eqnarray}
where $x$ is called the Bjorken variable. It has the
interpretation in the parton model of the fraction of the nucleon
momentum carried by the struck parton. The parton model is defined
by the constituents with momentum $p$ in Fig.~\ref{fig:parton}
satisfying
\begin{eqnarray}
(p + q )^{2} & = & p^{2}\\
\Rightarrow 2 p\cdot q - Q^{2} & = & 0
\end{eqnarray}
so that if $p=x k$, then $x$ is given by Eq.~(\ref{eq:xbj}). We
have also given $x$ in terms of light cone variables, $A^{\pm} =
A^{0} \pm A^{3}$, $\bm{A}_{\perp} = (A^{1},A^{2})$ that are
closely connected to the infinite momentum frame where the parton
model is defined. The approximation in Eq.~(\ref{eq:xbj}) is that
at large $Q^{2}$ and $\nu$, the struck quark has effectively zero
transverse momentum after the interaction with the photon.

The Bjorken limit \cite{Bjorken:1968dy} is defined as
$Q^{2},\nu\rightarrow\infty$ with $Q^{2}/\nu$ finite, whereupon
the functions $W_{i}$ exhibit Bjorken-Feynman scaling, and we
define the structure function of the nucleon $F_{2}^{(N)}$
\begin{equation} \frac{\nu}{M_{N}}
W_{2}(\nu,Q^{2}) \rightarrow \sum_{i}e^{2}_{i} \frac{Q^{2}}{2
M_{N} \nu} q_{i}(Q^{2}/2 M_{N} \nu) \equiv F_{2}^{(N)}(x)
\label{eq:F2def}
\end{equation}
where $i$ runs over the quark flavors. This scaling breaks down
experimentally as the function $F_{2}^{(N)}$ acquires additional
$Q^{2}$ dependence. The Dokshitzer-Gribov-Lipatov-Altarelli-Parisi
(DGLAP) evolution equations
\cite{Dokshitzer:1977sg,Gribov:1972ri,Altarelli:1977zs} (see
Appendix \ref{ch:evolve}) derived from QCD describe the scaling
violation by allowing one to calculate how the distribution of
momentum $q_{i}(x)$ among the quarks (and gluons) changes with the
momentum scale $Q^{2}$. It is beyond the parton model, and is one
of the most successful results of QCD. The structure function
Eq.~(\ref{eq:F2def}) is defined in terms of the quark distribution
functions $q_{i}(x)$, which, in the parton model in
Fig.~\ref{fig:parton}, is given by
\begin{equation}
q_{i}(x) =  k^{+} \int \frac{dz}{2\pi}^{-} e^{i x k^{+}z^{-}}
\langle k | \overline{\psi}(0) \gamma^{+} \psi (z^{-}) | k \rangle
\label{eq:pdf1}
\end{equation}
The $\psi$ are quark field operators, and $\overline{\psi} \equiv
\psi^{\dag} \gamma^{0}$ (we will use $\Psi$ for nucleon field
operators). There are two interpretations of quark distributions:
\begin{itemize}
\item Probability distributions of partons carrying fraction $x$ of the nucleon
momentum in the infinite momentum frame, and
\item QCD matrix elements of quark bilinears with light-like
separation and the Operator Product Expansion (OPE) for the
moments of the distributions.
\end{itemize}
We have so far used the first one, but these two definitions are
equivalent in the $\chi$QS model because Feynman's hypothesis that
the transverse momenta of partons do not grow with $Q^{2}$ is
satisfied \cite{Diakonov:1997vc}.

The second approach is advantageous for the discussion of
dynamical rescaling \cite{Close:1984zn}. We can define the moments
of the structure function of a nucleus $F_{2}^{(A)}$ with momentum
$P$ in Wilson's operator product expansion
\begin{eqnarray}
M_{n}^{(A)}(Q^{2}) & = &  \frac{1}{A} \int_{0}^{A} dx\, x^{n-2} F_{2}^{(A)}(x,Q^{2})\nonumber\\
& = & \frac{A^{n-2}}{Q^{2}} \sum_{n,i} c_{n,i}(Q^{2},\mu_{A}^{2})
\bar{\mathcal{O}}_{n,i}^{(A)}(\mu_{A}^{2}) \label{eq:ope}
\end{eqnarray}
where the sum runs on all twist-2 operators contributing to
$F_{2}$ (twist of an operator is given by $t = d_{i} - n$, where
$d_{i}$ is the mass dimension of the operator), and
$\bar{\mathcal{O}}$ is the matrix element
\begin{equation}
\langle P |
\mathcal{O}_{i,\mu_{1}\ldots\mu_{n}}^{(A)}(\mu_{A}^{2}) | P\rangle
= \bar{\mathcal{O}}_{n,i}^{(A)}(\mu_{A}^{2}) P_{\mu_{1}}\ldots
P_{\mu_{n}}
\end{equation}
The essence of the dynamical rescaling description of the EMC
effect is that the renormalization scale for nuclei is less than
that for a free nucleon
\[
\mu_{A}^{2} < \mu_{1}^{2} \ll Q^{2}
\]
so that the nuclear operators $\mathcal{O}^{(A)}$ undergo `extra
evolution' going up to the DIS scale $Q^{2}$, which accounts for
the EMC effect. QCD does not tell us what $\mu_{A}^{2}$ is, nor
its relationship to $\mu_{1}^{A}$, but numerical studies
\cite{Close:1984zn} show that this process can describe both the
EMC effect and Drell-Yan experiments \cite{Alde:im}.

The Wilson coefficients $c$ in Eq.~(\ref{eq:ope}) in the OPE are
dependent on the starting renormalization scale, and the essence
of the dynamical rescaling picture of the EMC effect is that the
scale is different in nuclei than for a free nucleon. Nuclear
interactions are effective QCD interactions which renormalize a
free nucleon into a bound one, but as we cannot yet in QCD
calculate the parton distributions of a free nucleon, much less
one in a nuclear medium. We need a model to probe the mechanism of
that `renormalization'.

In the handbag diagram Fig.~\ref{fig:parton}, the struck quark in
the intermediate state is renormalized by QCD from the same
starting scale as the other quarks in the nucleon, and hence all
feel the effects of nuclear interactions at the low scale. In
Chapter \ref{ch:dist} we will calculate the effect of the presence
of other nucleons on the quark distributions. One can imagine that
we are calculating the effect of different starting scales for
free and bound nucleons by showing what a bound nucleon parton
distribution looks like at the scale of a free nucleon.

The use of a manifestly covariant formulation to derive the
expression for the nuclear structure function appearing in
Eq.~(\ref{eq:ope}) leads to a convolution formula
\cite{Jung:1988jw}
\begin{equation}
\frac{F_{2}^{(A)}(x)}{A} = \int_{x}^{A} dy f(y) F_{2}^{(N)}(x/y)
\label{eq:conv-intro}
\end{equation}
The function $f(y)$ is the nucleon momentum distribution in the
nucleus, and is given by
\begin{equation}
f(y)=\int \frac{d^{4}k}{(2\pi)^4} \delta\left(y-\frac{k^{+}}
{M_{N}}\right) \text{ Tr } \left[ \gamma^{+}
\chi^{(A)}(k,P)\right] \label{eq:fn}
\end{equation}
where the function $\chi^{(A)}(k,P)$ is the Fourier transform of
the connected part of the nucleon Green's function. The
calculation of Eq.~(\ref{eq:fn}) is detailed in Appendix
\ref{ch:finite}.

The properly normalized nuclear structure function
$F_{2}^{(A)}(x)/A$ is smaller than the free nucleon structure
function $F_{2}^{(N)}(x)$ for values of $0.3\lesssim x \lesssim
0.7$. This is the EMC effect \cite{Aubert:1983xm}, and it has been
known for more than twenty years. The underlying mechanism
responsible for the change of the distribution of momenta among
the constituents of the nucleus remains beyond our ability to
calculate in QCD. We must resort to models, and there is a clear
interpretation within the parton model: a valence quark in a bound
nucleon carries less momentum than a valence quark in a free one.
One popular mechanism for this reduction in momentum involves
ordinary nuclear binding which, in its simplest form, is
represented by using the nucleon momentum distribution
Eq.~(\ref{eq:fn}) to evaluate the free nucleon structure function
at a value of $x$ increased by a factor of the average separation
energy divided by the nucleon mass $\bar{\epsilon}/M_{N} \simeq
0.04$. This picture encapsulates any nuclear effects in the
nuclear structure $F_{2}^{(A)}/A$ in the function $f(y)$.

The validity of the binding effect has been questioned in the
reviews
\cite{Arneodo:1992wf,Geesaman:1995yd,Piller:1999wx,Frankfurt:nt},
and we have shown \cite{Miller:2001tg,Smith:2002ci} that
conventional relativistic nuclear mean field theory fails to
describe the phenomenology of the EMC effect in terms of $f(y)$ in
Eq.~(\ref{eq:conv-intro}). These results are summarized in
Appendix \ref{ch:finite}. We believe that this failure to describe
the EMC effect provides additional motivation to study quarks and
antiquarks in nuclei beyond trying to connect QCD to nuclear
physics, and to look for medium modifications of the bound nucleon
in the only other part of Eq.~(\ref{eq:conv-intro}) available: the
structure function $F_{2}^{(N)}$. We must also remain consistent
with the Drell-Yan experiments, which we describe next.

\section{The Drell-Yan Process and Dimuon Production}
\label{sec:dy}

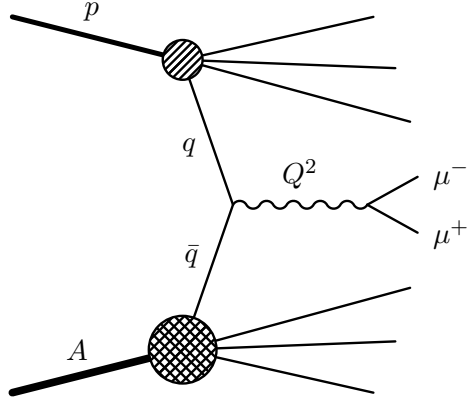
\begin{figure}
\centering
\begin{fmffile}{basicdy}
\begin{fmfgraph*}(60,50)
    \fmfleft{nucleus1,proton1}
    \fmf{plain,label=$p$,label.side=left,tension=4,width=2}{proton1,blob1}
    \fmf{plain,label=$A$,width=3,tension=4}{nucleus1,blob2}
    \fmf{plain,label=$q$,label.side=right}{blob1,v1}
    \fmf{plain,label=$\bar{q}$,label.side=left}{blob2,v1}
    \fmf{photon,label=$Q^{2}$,tension=0.75}{v1,v2}
    \fmf{plain}{v2,mu1}
    \fmf{plain}{v2,mu2}
    \fmfblob{15}{blob1}
    \fmfv{decor.shape=circle,decor.filled=hatched,decor.size=.15w}{blob2}
    \fmf{plain}{blob1,x1}
    \fmf{plain}{blob1,x2}
    \fmf{plain}{blob1,x3}
    \fmf{plain}{blob2,x4}
    \fmf{plain}{blob2,x5}
    \fmf{plain}{blob2,x6}
    \fmfright{x4,x5,x6,mu1,mu2,x1,x2,x3}
    \fmfv{label=$\mu^{+}$}{mu1}
    \fmfv{label=$\mu^{-}$}{mu2}
\end{fmfgraph*}
\end{fmffile}
\caption{Drell-Yan process $p+A\rightarrow \mu^{+}\mu^{-}+X$. A
quark $q$ from the proton, and an antiquark $\bar{q}$ from the
nucleus annihilate to form a muon pair.} \label{fig:basicdy}
\end{figure}

The proton-nucleus Drell-Yan cross-section can tell us about the
antiquark content of the nucleus, and hence serves as another
handle on nuclear parton distributions. The experimental data are
consistent with no modification of the antiquark distribution
\cite{Alde:im}. Many models fail to describe, or even address,
dimuon production in the Drell-Yan process depicted in
Fig.~\ref{fig:basicdy}. Dynamical rescaling matches the trend of
the data \cite{Alde:im}, but, as in its description of DIS data,
it does not give any insight as to the mechanism.
If $x_{p}$ is the momentum fraction carried by the (anti)quark in
the proton, and $x_{A}$ is the same for the nucleus, then one
finds the cross section from Fig.~\ref{fig:basicdy}
\cite{Bickerstaff:1984ax}
\begin{equation}
\frac{d^{2}\sigma}{dx_{p}dx_{A}} = \frac{4\pi \alpha^{2} K}{9 s}
\frac{1}{x_{p} x_{A}} \sum_{i} e^{2}_{i} \left[
q_{i}^{(p)}(x_{p},Q^{2})\bar{q}_{i}^{(A)}(x_{A},Q^{2})+\bar{q}_{i}^{(p)}(x_{p},Q^{2})q_{i}^{(A)}(x_{A},Q^{2})\right]
\end{equation}
where $\alpha \simeq 1/137$ is the electromagnetic coupling, and
$\sqrt{s}$ is the energy in the center-of-mass frame of the proton
and nucleus. The factor $K$ is an enhancement factor accounting
for other dilepton production modes independent of $x_{p}$ and
$x_{A}$ and the target and projectile, see
\cite{Bickerstaff:1984ax}, and references therein. If we look
experimentally at $x_{p} \gtrsim x_{A} + 0.2$, then at high
$Q^{2}$ the first term dominates (the antiquark distribution for
the proton is effectively zero for $x_{p}\gtrsim 0.2$ and $Q^{2}$
large). This can be achieved by controlling the kinematics of the
experiment, so we can use the Drell-Yan process to examine the
antiquark content of the nucleus. In order describe the
experimental data, we must have a model that includes both quarks
and antiquarks at the model scale\footnote{The evolution equations
in Appendix \ref{ch:evolve} generate antiquark, as well as gluon,
distributions as the scale changes. While different starting
distributions will generate different antiquark distributions, a
lack of antiquarks in the model ignores possible nuclear effects
at the starting scale. A common reason for failing to describe the
Drell-Yan data is that models produce too large an enhancement in
the antiquark content, so addressing antiquarks with a model at
the model scale is essential.}. We introduce such a model in
Chapter \ref{ch:cqsm}.

\chapter{The Chiral Quark-Soliton Model}
\label{ch:cqsm}

We will begin our study in earnest by providing motivation for the
$\chi$QS model, following the reviews
\cite{Diakonov:2000pa,Diakonov:2002fq}. First, we will discuss
instantons, and how they break chiral symmetry. Constructing the
model itself, and deriving some of its basic properties in the
meson and baryon sectors is next. The meson sector connects us to
$\chi$PT, and the baryon sector connects us to our description of
nuclear elastic, and deep inelastic scattering experiments.

\section{Results From The Instanton Vacuum}
\label{sec:instanton}

Quantum Chromodynamics is a Yang-Mills gauge theory with gauge
symmetry $SU(N_{C})$, where the number of colors is $N_{C} = 3$.
In Euclidean space\footnote{Euclidean spacetime related to
ordinary Minkowski spacetime with time $t$ by taking $t = i \tau$
with $\tau$ being the Euclidean time.}, the QCD field equation for
the gluons contains solutions called instantons. These solutions
never appear at any order in the perturbation theory expansion in
the strong coupling $\alpha_{s}$, and hence may provide insight
into non-perturbative QCD. Instantons can exist concentrated at a
single point in spacetime and are sometimes called pseudoparticles
as they do not follow spacetime worldlines like ordinary
particles.
Lattice simulations provide evidence that the QCD vacuum is
populated by instantons with a size $\bar{\rho} \sim 0.3 \text{
fm}$, which is approximately the scale of chiral symmetry
breaking, and a separation $\bar{R} = (N/V)^{1/4} \sim 1 \text{
fm}$ where $N/V$ is the instanton number density. The instanton
packing fraction, equal to the four dimensional instanton volume
$\pi^{2}\bar{\rho}^{4}$ times the instanton number density, is
then
\[
\frac{\pi^{2}\bar{\rho}^{4}}{\bar{R}^{4}} \simeq \frac{1}{8},
\]
which gives a small parameter aside from $\alpha_{s}$ in
perturbative QCD to use in expansions. This is the basis for the
Instanton Liquid Model (ILM) picture of the QCD vacuum. Though
dilute based on the packing fraction, the instantons are still
strongly interacting, therefore a liquid, not a gas, picture is
more accurate.

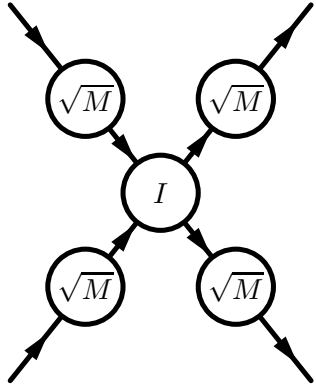
\begin{figure}
\centering
\begin{fmffile}{thooft}
\begin{fmfgraph*}(50,50)
\fmfpen{thick} \fmfleft{in1,in2} \fmfright{out1,out2}
\fmf{fermion}{in1,f1,i1,f2,out1} \fmf{fermion}{in2,f3,i1,f4,out2}
\fmfv{decor.shape=circle,decor.filled=empty,decor.size=.2w,label=$I$,label.dist=0}{i1}
\fmfv{decor.shape=circle,decor.filled=empty,decor.size=.2w,label=$\sqrt{M}$,label.dist=0}{f1}
\fmfv{decor.shape=circle,decor.filled=empty,decor.size=.2w,label=$\sqrt{M}$,label.dist=0}{f2}
\fmfv{decor.shape=circle,decor.filled=empty,decor.size=.2w,label=$\sqrt{M}$,label.dist=0}{f3}
\fmfv{decor.shape=circle,decor.filled=empty,decor.size=.2w,label=$\sqrt{M}$,label.dist=0}{f4}
\end{fmfgraph*}
\end{fmffile}
\caption{A 't Hooft vertex \cite{'tHooft:1976up} for $N_{f} = 2$.
The instanton is labelled by $I$, and the quark zero mode wave
functions are represented by form factors $\sqrt{M}$. The arrows
indicate the helicity of the quarks.}. \label{fig:thooft}
\end{figure}

Because of their topological properties, the instantons are
accompanied by $2N_{f}$ zero modes of chiral fermions ($N_{f}$
each of left and right handed fermions), where $N_{f}$ is the
number of flavors of quarks. Instantons are localized near a
single point in spacetime and have $2 N_{f}$ fermions meeting at
that point, so they look like interaction vertices appearing in
Feynman rules. In the case of $N_{f}=2$, we have effective
four-quark interactions. These are the so-called 't Hooft vertices
\cite{'tHooft:1976up}; a Feynman diagram for one with $N_{f}=2$
(which we will use throughout) is in Fig.~\ref{fig:thooft}. A
random ensemble of instantons \cite{Diakonov:1983hh} provides a
mechanism to break chiral symmetry via these zero modes.
Effectively one creates a helicity-flipping term in the Lagrangian
that behaves like a mass term via scattering of chiral quarks by
the 't Hooft vertices. They create a dynamical quark mass
\begin{eqnarray}
M(p) & = & \lambda \sqrt{\frac{N \pi^{2}\bar{\rho}^{2}}{V N_{C}}} \phi (p\bar{\rho}/2)^{2}\\
\phi (z) & \equiv & 2 z \left[
I_{0}(z)K_{1}(z)-I_{1}(z)K_{0}(z)-\frac{1}{z} I_{1}(z)K_{1}(z)
\right] \label{eq:ilmmass}
\end{eqnarray}
where $\phi (z)$ is the Fourier transform of the quark zero mode
wave function bound to each instanton, $N/V$ is the instanton
density, and $N_{C}$ is the number of colors in the gauge theory.
The constant $\lambda$ is obtained from a detailed numerical
calculation \cite{Diakonov:2002fq}, which results in a dynamical
constituent quark mass with $M(0)\simeq 350\text{ MeV}$. The
dynamical quark mass falls to zero for $p \gg 1/\bar{\rho} \simeq
650 \text{ MeV}$ which acts as an effective momentum cutoff for
the theory.

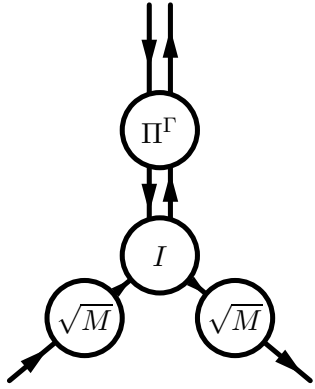
\begin{figure}
\centering
\begin{fmffile}{meson}
\begin{fmfgraph*}(50,50)
\fmfpen{thick} \fmfleft{in1,in2} \fmfright{out1,out2} \fmftop{m1}
\fmf{fermion}{in1,f1,i1,f2,out1} \fmf{phantom}{i1,f3,m1}
\fmf{phantom,tension=0}{m1,f3,i1}
\fmfv{decor.shape=circle,decor.filled=empty,decor.size=.2w,label=$I$,label.dist=0}{i1}
\fmfv{decor.shape=circle,decor.filled=empty,decor.size=.2w,label=$\sqrt{M}$,label.dist=0}{f1}
\fmfv{decor.shape=circle,decor.filled=empty,decor.size=.2w,label=$\sqrt{M}$,label.dist=0}{f2}
\fmfv{decor.shape=circle,decor.filled=empty,decor.size=.2w,label=$\Pi^{\Gamma}$,label.dist=0}{f3}
\fmffreeze \fmfi{fermion}{vpath (__i1,__f3) shifted (thick*
(2,0))} \fmfi{fermion}{vpath (__f3,__i1) shifted (thick* (-2,0))}
\fmfi{fermion}{vpath (__f3,__m1) shifted (thick* (2,0))}
\fmfi{fermion}{vpath (__m1,__f3) shifted (thick* (-2,0))}
\end{fmfgraph*}
\end{fmffile}
\caption{Quark-meson vertex formed from a 't Hooft vertex
(Fig.~\ref{fig:thooft}), where $\Gamma = 1, \gamma_{5},
\gamma^{\mu}, \gamma_{5}\gamma^{\mu}, \sigma^{\mu\nu}$ represent
the channels onto which the quark zero modes are projected.}
\label{fig:meson}
\end{figure}

If one combines a quark and an antiquark line from a 't Hooft
vertex Fig.~\ref{fig:thooft}, one obtains a quark-meson vertex
pictured in Fig.~\ref{fig:meson} where the type of meson depends
on the quantum numbers. If one considers the correlation function
of two quarks
\begin{eqnarray}
\Pi^{\Gamma}(p) & = & -\int d^{4}x e^{-ip\cdot x}
\text{ tr } \overline{S(x,0) \Gamma S(0,x) \Gamma}\\
\Gamma & = &
1,\gamma_{5},\gamma^{\mu},\gamma_{5}\gamma^{\mu},\sigma_{\mu\nu}
\end{eqnarray}
where $S(x,y)$ is the quark Green's function and the bar indicates
an average over the instanton ensemble in the ILM, one finds a
pole in the pseudoscalar channel at $p^{2}=0$ corresponding to our
massless pions resulting from the dynamical breaking of chiral
symmetry, and in the scalar channel at $p^{2}\simeq 4 M^{2}(0)$
which is taken as the mass a scalar meson. There are no poles in
the vector or tensor channels \cite{Diakonov:1983hh}. We will
denote scalar, pseudoscalar, and vector with subscripts $s$, $ps$,
and $v$ in the following.

\section{The Model Lagrangian}
\label{sec:chqsm}

\begin{figure}
\centering
\includegraphics[scale=0.7]{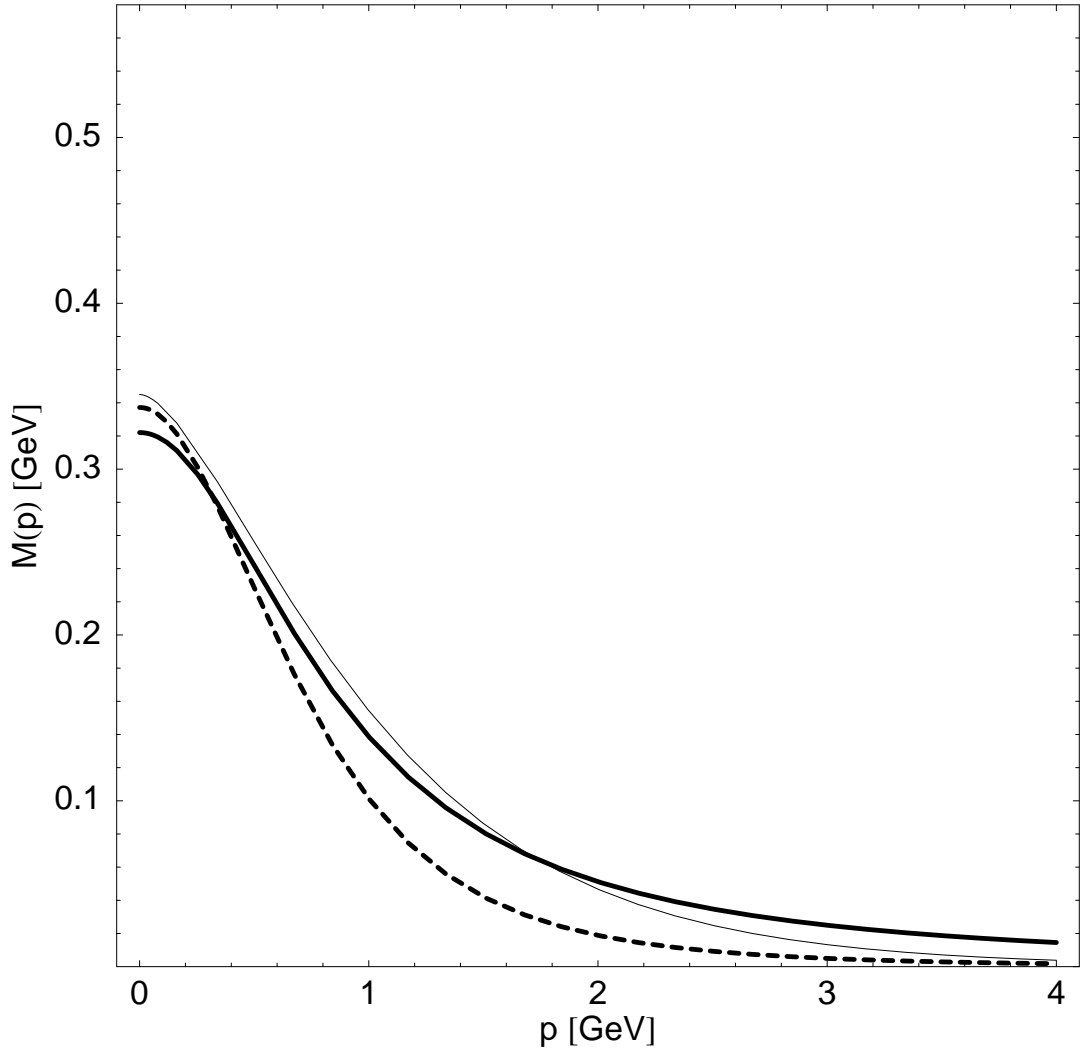}
\caption{The mass function $M(p)$ using one Pauli-Villars
regulator fit to the pion decay constant (heavy line), the
instanton model (light line) \cite{Diakonov:1983hh} and the
extrapolation of fits to quenched lattice data
\cite{RuizArriola:2004en} to the chiral limit using three
Pauli-Villars propagators (heavy dashed line), see Appendix
\ref{ch:npqp}.}\label{fig:mp}
\end{figure}

Using the quark meson vertex in Fig.~\ref{fig:meson}, we can write
down a Lagrangian \cite{Diakonov:2000pa} with constituent
(anti)quarks $\overline{\psi},\psi$ and pions $\bm{\pi}(x)$
\begin{eqnarray}
\mathcal{L}[\overline{\psi},\psi,\bm{\pi}] & = &  \overline{\psi}
i \partial \!\!\!\!\!\:/\, \psi - \int d^{4}p_{1}d^{4}p_{2} e^{i
(p_{1}-p_{2})\cdot x} \overline{\psi}(p_{1}) \sqrt{M(p_{1})}
U^{\gamma_{5}}(x) \sqrt{M(p_{2})} \psi(p_{2})
\nonumber\\
U^{\gamma_{5}}(x) & \equiv & \exp \left( i \gamma_{5}
\frac{\bm{\tau}\cdot\bm{\pi}(x)}{f_{\pi}} \right)
\label{eq:nonlocal}
\end{eqnarray}
where $f_{\pi} \simeq 93 \text{ MeV}$ is the pion decay constant
and $\tau_{i} = \sigma_{i}/2$ are the isospin operators
($\sigma_{i}$ are the Pauli spin matrices). This theory exhibits
asymptotic freedom like QCD; the interaction disappears at large
momenta with the function $M(p)$ falling to zero leaving a theory
of free massless quarks. We would like to replace this complicated
nonlocal theory with a local one; this is done by replacing the
momentum dependent mass $M(p)$ with a constant $M$. The
ultraviolet cutoff $M(p\rightarrow\infty)\rightarrow 0 $ is
replaced by a Pauli-Villars regularization fixed by the physical
value of \vspace{4mm}
\begin{eqnarray}
\parbox[30mm]{20mm}{
\begin{fmffile}{fpi}
\begin{fmfgraph*}(20,15)
    \fmftop{a1}
    \fmf{plain,left,tension=1}{a1,b1}
    \fmf{plain,left,tension=1}{b1,a1}
    \fmfdot{a1}\fmfdot{b1}
    \fmfbottom{b1}
    \fmfv{label=$i\tau^{a} \gamma^{5}$}{a1}
    \fmfv{label=$i\tau^{b} \gamma^{5}$}{b1}
\end{fmfgraph*}
\end{fmffile}} = f_{\pi}^{2} & = & 4 N_{C} \int \frac{d^{4}k}{(2
\pi)^{4}} \frac{M^{2}}{(k^{2}-M^{2})^{2}} -
\frac{M^{2}}{M_{PV}^{2}}\frac{M_{PV}^{2}}{(k^{2}-M_{PV}^{2})^{2}}
\nonumber\\
& = & \frac{N_{C}M^{2}}{4 \pi^{2}} \log \frac{M_{PV}^{2}}{M^{2}}
\label{eq:fpi}
\end{eqnarray}
Using $M = 420\text{ MeV}$, the resulting value for $M_{PV} \simeq
580\text{ MeV}$ is comparable to the instanton size cutoff, and
the resulting mass function is close not only to the instanton
model Eq.~(\ref{eq:ilmmass}), but to an extrapolation of quenched
lattice data to the chiral limit (see Appendix \ref{ch:npqp}). The
three functions are plotted in Fig.~\ref{fig:mp}. This value for
the constituent quark mass has also been shown to reproduce many
of the properties of free nucleons \cite{Christov:1995vm}, which
will be discussed in Chapters \ref{ch:dist} and \ref{ch:emff}.
We introduce the notation
\[
A - PV \equiv A(M) - \frac{M^{2}}{M_{PV}^{2}} A(M_{PV})
\]
as a shorthand to represent the Pauli-Villars subtraction, where
$A$ is any logarithmically divergent quantity in the $\chi$QS
model. We use this method to regulate the divergences as it
preserves the completeness of the quark states, which is important
in the calculation of the quark distributions
\cite{Diakonov:1997vc}.

In the case of the nucleon state the pion field is taken to be the
`hedgehog' \textit{ansatz\"{e}} with profile function $\Theta(r)$
and unit winding number.
\begin{eqnarray}
U(x) & = & \exp \left( i \bm{n}\cdot\bm{\tau} \Theta(r) \right)\\
\Theta(0) & = & -\pi \label{eq:unit}\\
\Theta(r\rightarrow \infty) & = & 0
\end{eqnarray}
Unit winding number is guaranteed by the condition
Eq.~(\ref{eq:unit}). The local version of the theory
(\ref{eq:nonlocal}) with constant mass, Pauli-Villars regulator
and hedgehog profile function results in the Lagrangian for the
Chiral Quark-Soliton ($\chi$QS) model
\begin{equation}
\mathcal{L} =  \overline{\psi} ( i \partial \!\!\!\!\!\:/\, - M
e^{  i \gamma_{5}\bm{n}\cdot\bm{\tau} \Theta(r) } ) \psi
\label{eq:lagrangian}
\end{equation}
The pion field configuration corresponds corresponds to the
soliton (Skyrmion) in the Skyrme model, hence its appearance in
the name of the model (quarks, however, do not appear as degrees
of freedom in the Skyrme model). This is a non-renormalizable
effective field theory, so that the ultraviolet cutoff, the
Pauli-Villars mass, has a physical meaning (observables can depend
on it), namely the the instanton size. The constituent quark mass
$M$ could be taken to be a free parameter when we tackle medium
modifications in Chapters \ref{ch:dist} and \ref{ch:emff}, but we
have chosen to fix it at $420\text{ MeV}$. This value best
reproduces a wide variety of nucleon observables
\cite{Christov:1995hr,Christov:1995vm}, and produces a mass
function similar to Eq.~(\ref{eq:ilmmass}) in Fig.~\ref{fig:mp}.
It is worth noting that we could have just written down the
Lagrangian (\ref{eq:lagrangian}) without motivation from the
instanton picture of the QCD vacuum as the simplest field theory
of constituent quarks consistent with chiral symmetry (a chiral
transformation $\psi \rightarrow e^{i \gamma_{5} \alpha (x)}\psi$
can be absorbed into a redefinition of the pion field $\bm{\pi}$).

Quarks are described by single particle wave functions $\psi_{n}$
that are solutions to the Dirac equation
\begin{eqnarray}
( i \partial \!\!\!\!\!\:/\, - M e^{  i \gamma_{5}\bm{n}\cdot\bm{\tau} \Theta(r) } )
\psi_{n}(\bm{r}) e^{-i E_{n} t} & = & 0\\
e^{i E_{n} t} \overline{\psi}_{n}(\bm{r}) ( i \partial
\!\!\!\!\!\:/\, - M e^{ i \gamma_{5}\bm{n}\cdot\bm{\tau} \Theta(r)
} ) & = & 0
\end{eqnarray}
with energy eigenvalues $E_{n}$. They are represented as a
discrete set labelled by $n$ for simplicity, but the spectrum of
the Dirac operator contains a single bound state (which we will
refer to as the valence level, and label with a superscript $v$),
as well as a positive and negative energy continua for energies
$|E|>M$. At most one bound state is guaranteed the the unit
winding number Eq.~(\ref{eq:unit}) of the hedgehog field (there
may be no bound state solution for certain values of $M$). Unit
winding number is a necessary, not sufficient, condition to obtain
unit baryon number given by the number of quarks in the valence
level; the winding number is not the baryon number as it is in the
Skyrme model.

The states in the negative continuum are filled in both the vacuum
sector with $U=1$ and the nucleon sector with profile function
$\Theta (r)$. For any observable, one must subtract the
contribution from the vacuum ($U=1$) to obtain a physical result.
These observables can still be logarithmically divergent after the
vacuum subtraction, requiring Pauli-Villars regularization. It is
the filled negative energy continuum that allows one to describe
the nucleon quark sea: the virtual quarks and antiquarks that
appear out of the polarized QCD vacuum in the presence of a
nucleon. Positive energy antiquarks appear as a hole: a lower
density of states in the negative energy continuum. Conversely,
negative energy quarks appear as a higher density of states. The
pions in the theory are lighter than $2 M$ because they are
density fluctuations in the negative continuum consisting of a
positive energy antiquark and a negative energy quark so that
their mass is given by $m_{\pi} \simeq M - M = 0$
\cite{Diakonov:2005eq}.

There is an equivalence in calculating observables where one can
use sums over the occupied states (the valence level and negative
energy continuum) or the unoccupied states (the positive
continuum) \cite{Diakonov:1997vc}. This equivalence is deeply
rooted in the fact that the model is a relativistic quantum field
theory, and is needed to preserve general conditions such as the
positivity of the antiquark distribution and the baryon and
momentum sum rules discussed in Chapter \ref{ch:dist}. It can be
spoiled by the regulator, such as Schwinger's proper-time
regulator or sharp cutoffs, but Pauli-Villars regularization
preserves the completeness of the quark states and maintains the
equivalence \cite{Diakonov:1997vc}.

The field equation for the profile function follows directly from
the Lagrangian (\ref{eq:lagrangian}) and is given by
\begin{equation}
\rho_{s}(r)\sin \Theta(r)-\rho_{ps}(r)\cos \Theta(r)=0
\label{eq:thetafe1}
\end{equation}
where the quark scalar and pseudoscalar densities are
\begin{eqnarray}
\rho_{s,ps}(r) & = & \sum_{n} \overline{\psi}_{n}(r)
\Gamma \psi_{n}(r)\\
\Gamma & = & 1, i\gamma_{5}\bm{n}\cdot\bm{\tau}\nonumber
\end{eqnarray}
For a single nucleon this sum is over all occupied states which
includes the negative energy continuum and the single valence
level. In the nuclear medium, this becomes more complicated, and
the general case would be exceedingly difficult. In symmetric
infinite nuclear matter, with equal constant densities of protons
and neutrons filling all space, we will be able to make some
simplifications. This is an often used approximation to describe
heavy nuclei \cite{Walecka:qa}.

\section{The Nucleon and the Large $N_{C}$ Limit}
\label{sec:nmass}

Aside from the gauge coupling used in perturbative QCD, there is
another `small' dimensionless parameter that we can expand in: the
inverse number of colors $N_{C}^{-1}$. It is believed that many of
the important properties of QCD follow into the limit where
$N_{C}$ is large; we will leave the basics of this expansion
regarding baryons to Ref.~\cite{Witten:1979}. We will use the
limit here to justify the $\chi$QS model as a relativistic mean
field theory of baryons \cite{Diakonov:2005eq}, simplifying the
calculation of the nucleon mass, form factors and quark
distributions. We will follow Ref.~\cite{Diakonov:1987ty} to
obtain an expression for the nucleon mass.

In Euclidean space (in which we will work in this section), the
partition function for the $\chi$QS model is
\begin{equation}
Z = \int D\pi D\overline{\psi} D\psi \exp - \int d^{4}x
\overline{\psi}(i
\partial\!\!\!/+ i MU^{\gamma_{5}})\psi
\end{equation}
The integration over the quark fields can be done, giving the
effective action for the pion fields (including the vacuum
subtraction)
\begin{equation}
S_{\text{eff}}[\pi] = -N_{C}\log \det \left(
\frac{i\partial\!\!\!/ + i MU^{\gamma_{5}}}{i\partial\!\!\!/ + iM}
\right) \label{eq:Seff}
\end{equation}
One can expand this in derivatives
\begin{eqnarray}
\text{Re }S_{\text{eff}}^{(2)}[\pi] & = & \frac{N_{C}}{4}\int
d^{4}x\int \frac{d^{4}k}{(2\pi)^{4}}\text{ tr }\left(
\frac{M\partial\!\!\!/U^{\gamma_{5}}}{k^{2}+M^{2}} \right)^{2}\nonumber\\
& = & \frac{1}{4} f_{\pi}^{2} \int d^{4}x \text{ tr }
\partial_{\mu}U\partial^{\mu}U^{\dag}\label{eq:GLp2}
\end{eqnarray}
and see that one obtains the $\mathcal{O}(p^2)$ term in the
momentum expansion in $\chi$PT (note this also defines
Eq.~(\ref{eq:fpi}) for $f_{\pi}$). If one includes a current quark
mass, one obtains a Lagrangian consistent with the
$\mathcal{O}(p^{4})$ Gasser-Leutwyler Lagrangian and large
$N_{C}$. This is elaborated on in Appendix \ref{ch:npqp}, but we
have already made the connection between $\chi$PT and QCD through
the $\chi$QS model.

From the effective action, we can also determine the energy in the
background pion field in the limit of long Euclidean times $T$. We
write Eq.~(\ref{eq:Seff}) in terms the Hamiltonian $H$, and
$H^{(0)}$ with $U = 1$
\begin{eqnarray}
S_{\text{eff}}[\pi] & = & - N_{C} \text{ Tr } \left[ \log \left( i
\partial_{4} + i H \right) - \log \left( i
\partial_{4} + i H^{(0)} \right) \right] \\
& = & - N_{C} T \int \frac{dE}{2 \pi} \text{ Tr } \left[ \log
\left( E + i H \right) - \log \left( E + i H^{(0)} \right) \right] \\
& = & - N_{C} T \int \frac{dE}{2 \pi} \text{ Tr } \left(
\frac{ E }{ E + i H } - \frac{E}{ E + i H^{(0)} } \right) \\
& = & - N_{C} T \left(  \sum_{E_{n}<0} E_{n} -
\sum_{E_{n}^{(0)}<0} E_{n}^{(0)} \right) \\
& \equiv & - N_{C} T E_{\text{field}}[\pi]
\end{eqnarray}

One expects the limit of the correlation function of two nucleon
currents to give the nucleon mass at long times $T$
\begin{equation}
\lim_{T\rightarrow\infty}\langle J(0,T)J^{\dag}(0,0) \rangle \sim
e^{-M_{N}T}
\end{equation}
where the nucleon current is given by
\begin{equation}
J(x)=\frac{1}{N_{C}!} \varepsilon^{\alpha_{1}\cdots\alpha_{N_{C}}}
\Gamma_{JJ_{3}TT_{3}}^{f_{1}\cdots f_{N_{C}}}
\psi_{\alpha_{1}f_{1}}(x) \cdots \psi_{\alpha_{N_{C}}f_{N_{C}}}(x)
\end{equation}
where $\alpha_{i}$ are color indices, $f_{i}$ are flavor indices,
$J$ and $J_{3}$ are angular momentum quantum numbers, and $T$ and
$T_{3}$ are the isospin. We subsume the description of the nucleon
quantum numbers in terms of its constituent quarks into the matrix
$\Gamma$. The correlation function is defined in terms of the
partition function, and the integration over the quark fields can
be done. This results in the effective action and the quark
Green's function in the background pion field
\begin{eqnarray}
\langle J(0,T)J^{\dag}(0,0) \rangle & = & \frac{1}{Z}\int D\pi
D\psi D\psi^{\dag} J(0,T)J^{\dag}(0,0) e^{-S[\pi,\psi,\psi^{\dag}]}\nonumber\\
& = & \Gamma^{\{f\}}_{\{JT\}}\Gamma^{\{g\}}_{\{JT\}} \int D\pi
\prod_{i}^{N_{C}}
G_{f_{i}g_{i}}(T,0|\pi)e^{-S_{\text{eff}}[\pi]}\nonumber
\end{eqnarray}
The long time limit of the quark Green's function in the
background pion field gives us the valence energy level
\begin{equation}
\lim_{T\rightarrow\infty}G_{f_{i}g_{i}}(T,0|\pi)\sim
e^{-TE^{v}[\pi]}
\end{equation}
Putting this together we obtain
\begin{eqnarray}
\lim_{T\rightarrow\infty}\langle J(0,T)J^{\dag}(0,0) \rangle &
\sim & \int D\pi e^{-T N_{C}(E^{v}[\pi] +
E_{\text{field}}[\pi])} \nonumber\\
& \sim & e^{-M_{N}T} \nonumber
\end{eqnarray}
which gives us, in the saddle point approximation (leading order
in $N_{C}$), the nucleon mass
\begin{equation}
M_{N}[\pi(x)] = \min_{\{\pi(x)\}} N_{C} \left( E^{v}[\pi] +
E_{\text{field}}[\pi] \right) \label{eq:mn-gen}
\end{equation}
We have neglected quark loop corrections at large $N_{C}$ in the
saddle point approximation (and therefore quantum fluctuations of
the pion field, which propagate through quark loops). In this
sense, the large $N_{C}$ limit effectively represents a
relativistic mean field approximation. The abstract functional
Eq.~(\ref{eq:mn-gen}) is minimized by solving the system
self-consistently with the pion field replaced by the profile
function $\Theta (r)$. This is the subject of Chapter \ref{ch:np}.

\chapter{Quarks and Antiquarks in Nuclei}
\label{ch:np}

In order to describe nuclei in terms of the quarks and antiquarks
in the $\chi$QS model, there needs to be a way to connect the
nuclear physics and the quark physics. This has been done in the
context of the MIT bag model \cite{Saito:1994ki} to create a
so-called Quark-Meson Coupling (QMC) model. We will use some of
the ideas of that model, but also include some phenomenology of
medium modifications and physics of the QCD vacuum. We begin with
the role of the vacuum quark scalar condensate, and then construct
a two-tier self-consistent calculation. We are able to obtain
nuclear binding and saturation, and reproduce standard properties
of nuclear matter. We therefore create a model of nuclei that
includes the physics of valence and sea quarks.

\section{The Effect of the Medium}
\label{sec:nuclear}

We will begin with some motivation for our procedure to couple the
quark substructure of the nucleon to the nuclear medium. Through
the use of QCD sum rules, Ioffe \cite{Ioffe:kw} derived a
relationship between the vacuum scalar condensate, $\langle
\overline{\psi}\psi\rangle_{0}$, and the nucleon mass. One can
re-derive this estimate in a constituent quark field theory such
as we are using here. We begin with the scalar condensate
\begin{eqnarray}
\parbox{15mm}{
\begin{fmffile}{qqbar}
\begin{fmfgraph*}(15,10)
    \fmftop{a1}
    \fmf{plain,left,tension=1}{a1,b1}
    \fmf{plain,left,tension=1}{b1,a1}
    \fmfdot{b1}
    \fmfbottom{b1}
    \fmfv{label=$1$}{b1}
\end{fmfgraph*}
\end{fmffile}} & = & \langle \overline{\psi}\psi\rangle_{0} = - \text{tr}
\int^{\Lambda} \frac{d^{4}p}{(2 \pi)^{4}} \frac{1}{p\!\!\!\!\;/\,
- M}
\nonumber\\
& \sim & -\frac{N_{C}M \Lambda^{2}}{4 \pi^{2}}, \label{eq:qqvac}
\end{eqnarray}
where the divergent loop integral is regulated by a momentum
cutoff (playing the role of the Borel mass in the QCD sum rule
approach). Using the fact that constituent quarks are essentially
defined as having a mass $\sim M_{N}/N_{C}$, we can rewrite
Eq.~(\ref{eq:qqvac}) as
\begin{eqnarray}
M_{N} & \sim & -\frac{4 \pi^{2}}{\Lambda^{2}} \langle
\overline{\psi}\psi\rangle_{0}. \label{eq:ioffe}
\end{eqnarray}
Although Eq.~(\ref{eq:ioffe}) is not a very accurate estimate, it
does highlight the role of the condensate. It will be modified in
the presence of other nucleons.

The condensate at finite density, proportional to $k_{F}^{3}$
where $k_{F}$ is the Fermi momentum, can be written in terms of
the nuclear scalar density $\rho_{s}^{N}$ and the nucleon sigma
term $\sigma_{N}$ \cite{Cohen:1991nk} as
\begin{eqnarray}
\langle \overline{\psi}\psi\rangle_{k_{F}} & = &
\langle\overline{\psi}\psi\rangle_{0} -
\langle\overline{\psi}\psi\rangle_{0}\frac{\sigma_{N}}{m_{\pi}^{2}f_{\pi}^{2}}
\rho_{s}^{N}\label{eq:qqmed}.
\end{eqnarray}
We can then substitute Eq.~(\ref{eq:qqmed}) into
Eq.~(\ref{eq:ioffe}) to obtain a schematic picture of the effect
of the nuclear medium on the nucleon mass
\begin{eqnarray}
M_{N}(k_{F}) & \sim & -\frac{4 \pi^{2}}{\Lambda^{2}}
\langle\overline{\psi}\psi\rangle_{0} \left[ 1 -
\frac{\sigma_{N}}{m_{\pi}^{2}f_{\pi}^{2}} \rho_{s}^{N} \right],
\label{eq:ioffemed}
\end{eqnarray}
We will effectively replace the combination of the vacuum
condensate, pion mass, decay constant and the the sigma term in
Eq.~(\ref{eq:ioffemed}) with a dimensionless scalar coupling
constant $g_{s}$, which we take as a free parameter. The value of
$g_{s}$ is determined from a fit to nuclear density and binding
energy.

\begin{figure}
\centering
\includegraphics[scale=0.7]{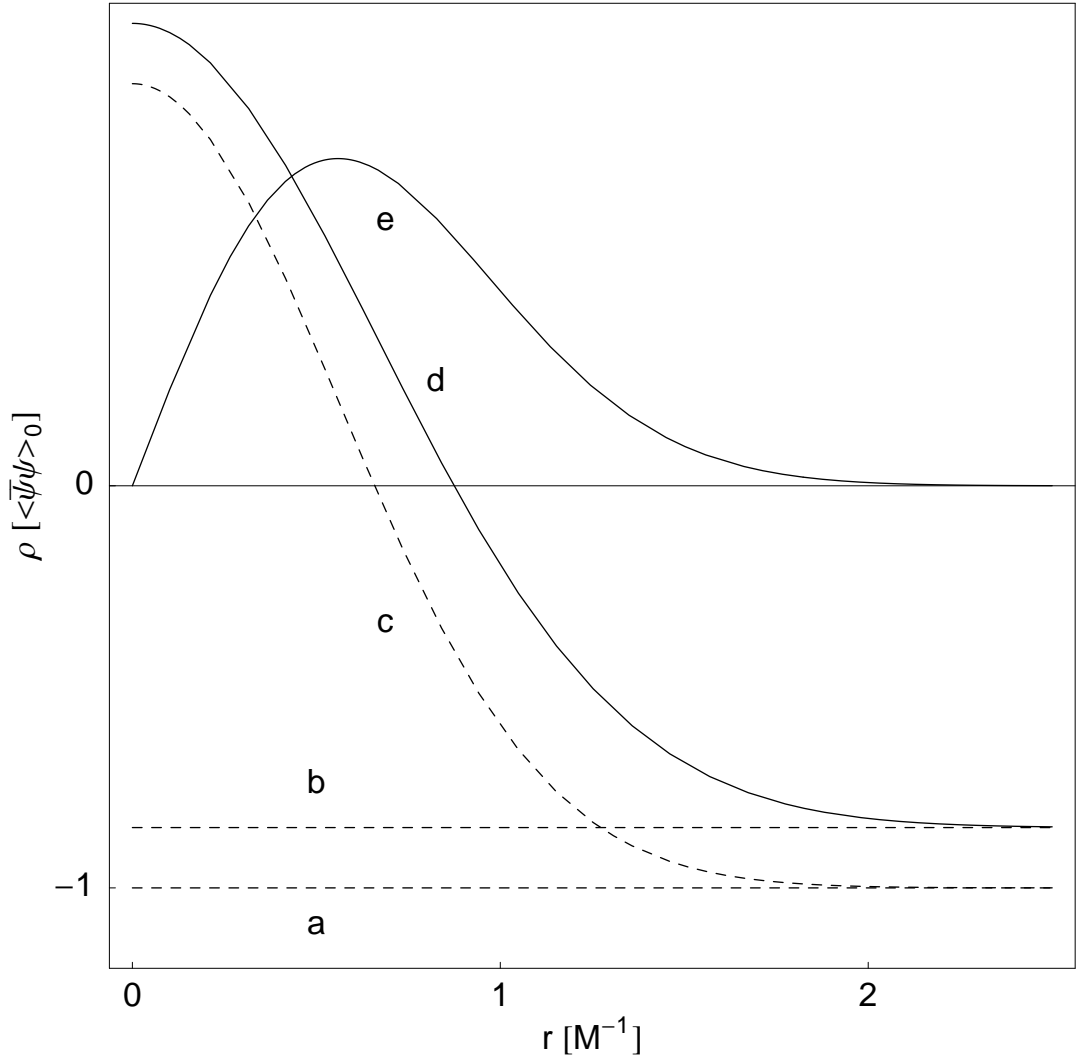}
\caption{Cartoon of the scalar and pseudoscalar densities as a
function of radius from the center of the nucleon. The inverse
constituent quark mass sets the typical size scale. Pictured are
(a) the vacuum condensate, (b) the effective condensate
Eq.~(\ref{eq:effcond}), (c) the free nucleon scalar density, (d)
the bound nucleon scalar density, and (e) the pseudoscalar
density.}\label{fig:scalar}
\end{figure}

Using this dependence of the nucleon mass on the nuclear medium as
a guide, we incorporate the medium dependence in the model by
simply letting the quark scalar density in the field equation
(\ref{eq:thetafe1}) contain a (constant) contribution arising from
other nucleons present in symmetric nuclear matter. This models a
scalar interaction via the exchange of multiple pairs of pions
between nucleons. We take the scalar density to consist of three
terms: 1) the constant condensate value $\langle
\overline{\psi}\psi\rangle_{0}$ (in the vacuum or at large
distances from a free nucleon), 2) the valence contribution
$\rho_{s}^{v}$ and 3) the contribution from the medium which takes
the form of the convolution of the nucleon $\rho_{s}^{N}$ and
valence quark scalar densities as in the QMC model
\cite{Lu:1998tn}
\begin{subequations}
\label{eq:rhos}
\begin{eqnarray}
\rho_{s}^{q}(\bm{r}) & \simeq & \langle
\overline{\psi}\psi\rangle_{0} + \rho_{s}^{v}(\bm{r}) + g_{s} \int
d\bm{r}'\rho_{s}^{N}(\bm{r}-\bm{r}')\rho_{s}^{v}(\bm{r}')\\
& = & \langle \overline{\psi}\psi\rangle_{0} + \rho_{s}^{v}(r) +
g_{s} P_{s}^{N}(k_{F})
\end{eqnarray}
\end{subequations}
We define an effective condensate in the presence of nuclear
matter by
\begin{equation}
\langle \overline{\psi}\psi\rangle_{0} + g_{s} P_{s}^{N}(k_{F}).
\label{eq:effcond}
\end{equation}
We take the pseudoscalar density to have only the valence term
$\rho_{ps}^{q} \simeq \rho_{ps}^{v}$; the two other contributions
analogous to the first and third terms of Eq.~(\ref{eq:rhos})
vanish due to symmetries of the QCD vacuum and nuclear matter.
These approximations to the densities neglect the precise form of
the negative continuum wave functions. We will see that the
resulting free nucleon profile function has little discernible
difference from a fully self-consistent treatment (see
Fig.~\ref{fig:profapprox} in Section \ref{sec:numresults}),
demonstrating the excellence of this approximation. We provide a
schematic picture of the scalar density in a free and bound
nucleon in Fig.~\ref{fig:scalar}.
Incorporating this dependence on nuclear matter in the field
equation Eq.~(\ref{eq:thetafe1}), the in-medium field equation for
the profile function becomes
\begin{equation}
\Theta(r,k_{F}) = \arctan
\frac{\rho_{ps}^{q}(r)}{\rho_{s}^{q}(r)+g_{s} P_{s}^{N}(k_{F})}
\label{eq:thetafe},
\end{equation}
which is similar to the way the pion mass would enter, except with
the opposite sign (\textit{i.e.~}medium effects serve to bring us
closer to chiral symmetry restoration).

The (constant) nucleon scalar density in Eq.~(\ref{eq:rhos}) is
determined by solving the nuclear self-consistency equation
\cite{Walecka:qa}
\begin{equation}
\rho_{s}^{N} = 4 \int^{k_{F}} \frac{d^{3}k}{(2\pi)^{3}}
\frac{M_{N}(\rho_{s}^{N})}{\sqrt{k^{2}+M_{N}(\rho_{s}^{N})^{2}}}.\label{eq:nsc}
\end{equation}
The dependence of the nucleon mass, and any other properties
calculable in the model, on the Fermi momentum $k_{F}$ enters
through Eq.~(\ref{eq:nsc}). Thus there are two coupled
self-consistency equations: one for the profile,
Eq.~(\ref{eq:thetafe}), and one for the density,
Eq.~(\ref{eq:nsc}). These are iterated until the change in the
nucleon mass Eq.~(\ref{eq:mn}) is as small as desired (in our
case, $\Delta M_{N}\leq 0.1 \text{ MeV}$) for each value of the
Fermi momentum.


We introduce a phenomenological vector meson field $V^{\mu}$ (with
mass fixed at $m_{v}=0.77\text{ GeV}$ and coupling $g_{v}$)
\cite{Walecka:qa} exchanged between nucleons, but not quarks in
the same nucleon (\textit{i.e.}~we ignore the spatial dependence
of the vector field in the vicinity of a nucleon, treating only
the nuclear mean field). In the mean field approximation (MFA)
$V^{i = 1,2,3} = 0$, and the vector potential has a simple field
equation \cite{Walecka:qa}
\begin{equation}
V^{0} = \frac{g_v}{m_{v}^{2}} P_{v}^{N}(k_{F})
\end{equation}
where the vector meson couples to the vector density
\begin{equation}
P_{v}^{N}(k_{F}) = \int d^{3}r' \rho_{v}^{N}(r')\rho_{v}^{q}(r-r')
= \frac{2 k_{F}^{3}}{3\pi^{2}}.
\end{equation}
It is a proxy for uncalculated soliton-soliton interactions used
to obtain the necessary short distance repulsion which stabilizes
the nucleus. The resulting energy per nucleon is
\begin{equation}
\frac{E}{A} = \frac{4}{\rho_{B}(k_{F})} \int^{k_{F}}
\frac{d^{3}k}{(2\pi)^{3}} \sqrt{k^{2} + M_{N}(k_{F})^{2}}
+\frac{1}{2}\frac{g_{v}^{2}}{m_{v}^{2}}P_{v}^{N}(k_{F})
\label{eq:epn}.
\end{equation}
The single-particle Dirac Hamiltonian for the quarks in the
hedgehog pion field, after including the vector interaction, is
\begin{eqnarray}
H[\Theta ] \psi_{n} (\bm{r}) & \equiv & [-i
\gamma^{0}\gamma^{k}(\partial_{k} + i g_{v} V_{k}) + g_{v}V_{0} +
M \gamma^{0} (\cos \Theta + i\gamma_{5}\bm{\tau}\cdot\bm{n}\sin
\Theta)]
\psi_{n} (\bm{r})\nonumber \\
& = & E_{n} \psi_{n} (\bm{r})\label{eq:hamiltonian}
\end{eqnarray}
where the effect of the nuclear medium, the dependence on $k_{F}$,
enters through the vector potential, as well as the profile
function $\Theta = \Theta (r, k_{F})$.

\section{The Numerical Procedure}
\label{sec:numproc}

The Dirac Hamiltonian is diagonalized in the discrete Kahana-Ripka
basis \cite{Kahana:1984be} (see Appendix \ref{ch:KR}). The basis
functions $\phi_{i}$ are defined by the solutions to the free
single-particle Dirac Hamiltonian
\begin{equation}
H^{(0)} \phi_{i} (\bm{r}) =  [-i \gamma^{0}\gamma^{k}\partial_{k}
+ M \gamma^{0}] \phi_{i} (\bm{r}) = E_{i}^{(0)} \phi_{i}
(\bm{r})\label{eq:H0}
\end{equation}
The full Hamiltonian is then solved by diagonalizing the matrix
equation
\begin{equation}
\sum_{j} H_{ij}[\Theta] c_{nj} = E_{n} c_{ni}
\end{equation}
where
\[
H_{ij}[\Theta] = \int d\bm{r} \phi^{\dag}_{i}(\bm{r}) H[\Theta]
\phi_{j}(\bm{r})
\]
and
\[
\psi_{n} (\bm{r}) = \sum_{i} c_{ni} \phi_{i}(\bm{r})
\]

The use of a finite basis turns the positive and negative energy
continua into a discrete set of states. Our interpretation of
these continua as quarks and antiquarks in the nucleon sea remains
the same. Antiquarks (holes) appear as lower densities of discrete
energy levels in the negative energy states. In the finite basis,
we can write the nucleon mass Eq.~(\ref{eq:mn-gen}) as
\begin{equation}
M_{N}[\Theta (r, k_{F})] = N_{C} \left[ E^{v} + \sum_{E_{n}<0}
E_{n} - \sum_{E_{n}^{(0)}<0} E_{n}^{(0)} \right] - PV.
\label{eq:mn}
\end{equation}

Starting at zero Fermi momentum and an initial guess for the
profile function as the first step in the self consistent
calculation, the valence level wave function is used to calculate
the quark scalar and pseudoscalar densities (there is no medium
contribution at $k_{F}=0$). These are used in
Eq.~(\ref{eq:thetafe}) to calculate a new profile function.

This new profile function then gives us new quark densities after
diagonalizing the Hamiltonian Eq.~(\ref{eq:hamiltonian}). This
procedure is iterated until self consistency is reached when the
change in the valence energy level $\Delta E^{v} \leq 0.01 \text{
 MeV}$ (the first self-consistent loop). Next, the eigenvalues are
used to calculate $M_{N}$. At $k_{F}=0$, the procedure is done
here and $k_{F}$ is incremented a step $\Delta k_{F} = 0.01 \text{
fm}$.

\begin{figure}
\centering
\input{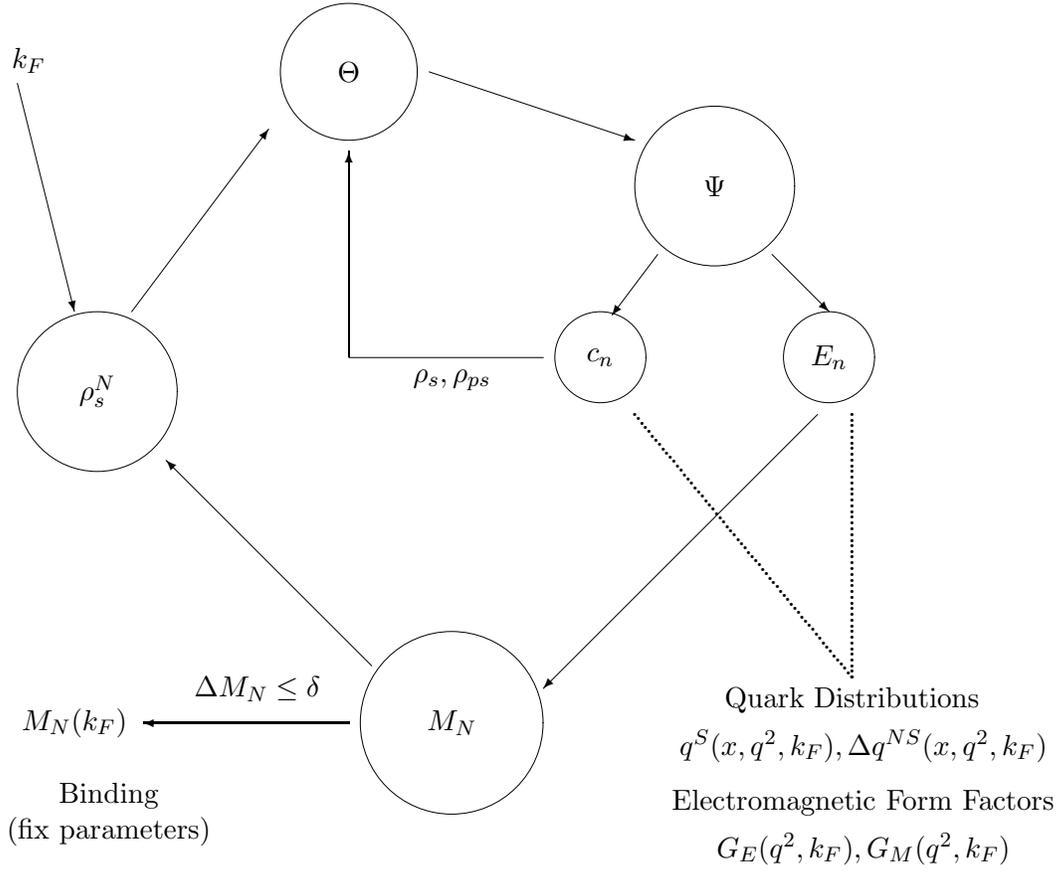}
\caption{Diagram of the two-tier self-consistent loop. One inputs
a value for $k_{F}$ and a guess for the profile function. $\Theta$
represents the field equation for the profile. $\Psi$ represents
the Dirac equation solved in the finite basis, which takes the
profile function as an input, and outputs a set of energy
eigenvalues $E_{n}$ and wave functions. The $E_{n}$ are then used
to calculate a the nucleon mass $M_{N}$ which is then used to
calculate a new nucleon scalar density.} \label{fig:scloop}
\end{figure}

With $k_{F}\neq 0$, there is now a contribution to the scalar
density from the medium in the profile field equation
(\ref{eq:thetafe}), as well as a non-zero vector potential. We
calculate the nucleon scalar density via Eq.~(\ref{eq:nsc}), using
the nucleon mass from the previous step, $M_{N}(k_{F}-\Delta
k_{F})$, as the initial guess. The Hamiltonian is again
diagonalized and the profile is calculated self consistently using
the valence level and the medium contribution. The nucleon mass is
then calculated, and the nucleon scalar density is recalculated
using the new nucleon mass. This is repeated at each $k_{F}$ until
self consistency is reached when the change in the nucleon mass is
$\Delta M_{n} \leq 0.1 \text{ MeV}$ (the second self-consistent
loop). Then, $k_{F}$ is incremented another step, and the process
continues until one reaches any value of $k_{F}$ desired. A
diagram of this process is in Fig.~\ref{fig:scloop}, which
clarifies the double self-consistent loop structure.
\setlength{\unitlength}{0.76mm}
\setlength{\unitlength}{1mm}

\section{The Numerical Results}
\label{sec:numresults}

\begin{figure}
\centering
\includegraphics[scale=0.7]{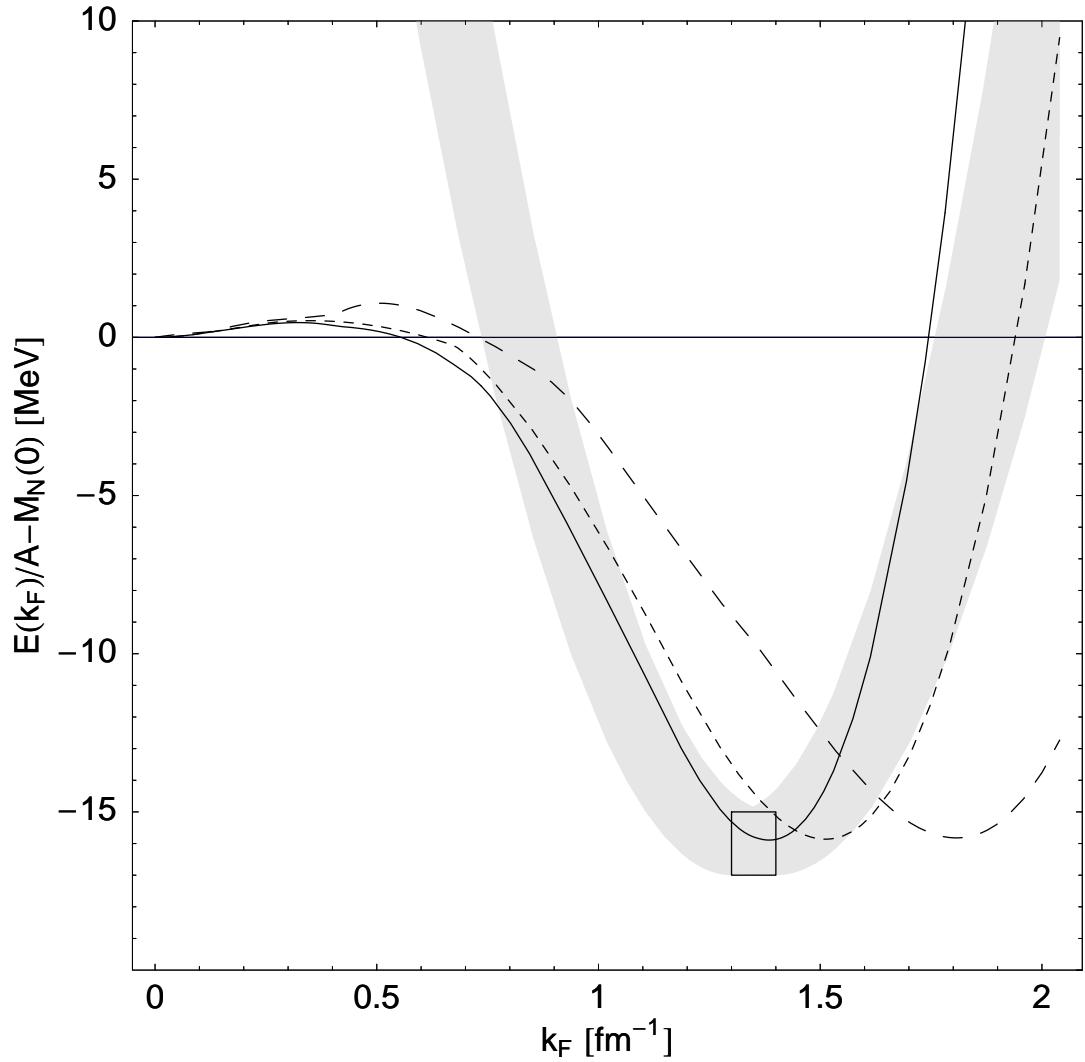}
\caption{Binding energy per nucleon $B = E/A - M_{N}$ as a
function of Fermi momentum fit to $B=-15.75\text{ MeV}$ at the
minimum for $g_{s} = 0.89$ (long dashed), $1.10$ (short dashed)
and $1.27$ (solid). The box and shaded region are the experimental
uncertainty \cite{Blaizot:tw} in the binding energy, density and
compressibility of nuclear matter.} \label{fig:bepn}
\end{figure}

\begin{figure}
\centering
\includegraphics[scale=0.7]{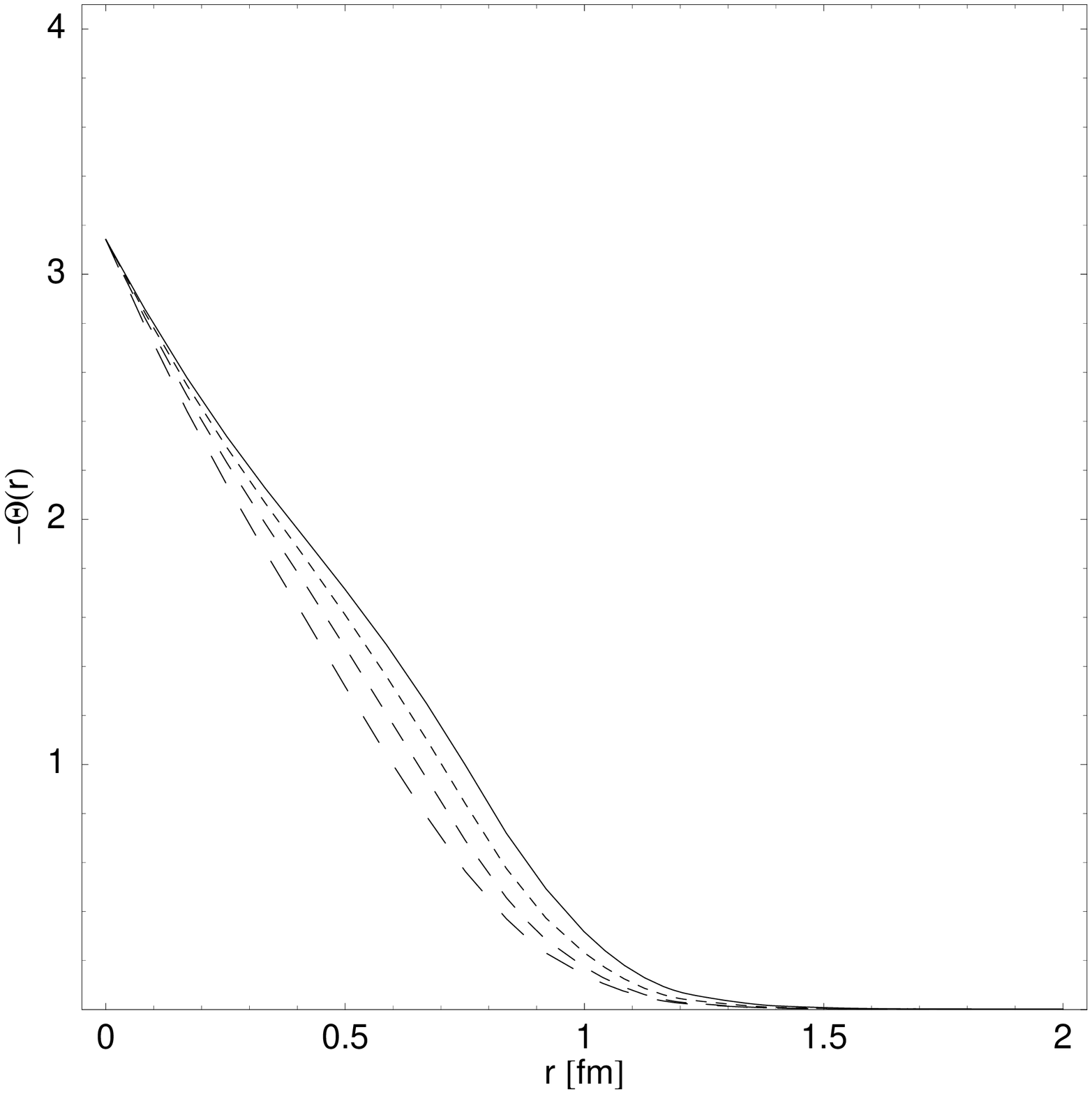}
\caption{Profile functions in nuclear matter. The solid line is
the profile function for $1.5\rho_{0}$; the curves with
progressively longer dashes correspond to $1.0\rho_{0}$,
$0.5\rho_{0}$ and zero density respectively.} \label{fig:profile}
\end{figure}

The mass of a free nucleon is computed to be $M_{N}(k_{F}=0) =
1209\text{ MeV}$. The $\sim 30\%$ difference is as expected in the
model at leading order in $N_{C}$. We evaluate the nucleon mass
Eq.~(\ref{eq:mn}) and energy per nucleon Eq.~(\ref{eq:epn}) as a
function of $k_F$. We choose our free parameters to fit $E/A -
M_{N}(0) \equiv B = -15.75 \text{ MeV}$ at the minimum. The
parameters are summarized in Table \ref{tab:params}. We use the
value $g_{s} = 1.27$ (corresponding to $\sigma_{N} = 41.4\text{
MeV}$), and vector coupling $g_{v}^{2}/4\pi = 10.55$, which gives
a Fermi momentum of $k_{F} = 1.38\text{ fm}^{-1}$ in nuclear
matter consistent with the known value $k_{F} = 1.35 \pm 0.05
\text{ fm}^{-1}$ \cite{Blaizot:tw}. This value of $g_{s}$ is used
for the calculation of all of the parton distributions and
electromagnetic form factors.
\begin{table}
\centering \caption{Model parameters to fit binding energy
$B=-15.75$ MeV, the location of the minima, compressiblilty, and
the resulting density of nuclear matter in units of
$\rho_{0}=0.178\text{ fm}^{-3}$ consistent with the experimental
value of $0.167\pm 0.018\text{ fm}^{-3}$ \cite{Blaizot:tw}.}
\label{tab:params}
\begin{tabular}{ccccc}
$ g_{s}$ & $ g_{v}^{2}/4\pi$ & $ k_{F}$ & $K$ & $\rho/\rho_{0}$ \\
$[\text{MeV}]$ & & $[\text{fm}^{-1}]$ & $[\text{MeV}]$ &  \\
\hline
0.89 & 7.22 & 1.81 & 291.7 & 2.26 \\
1.10 & 8.96 & 1.51 & 312.5 & 1.31 \\
1.27 & 10.55 & 1.38 & 348.5 & 1.00
\end{tabular}
\end{table}
We plot the binding energy per nucleon using Eq.~(\ref{eq:epn}) in
Fig.~\ref{fig:bepn}. The compressibility is $K = 348.5\text{ MeV}$
which is above the experimental value $K = 210 \pm 30 \text{
MeV}$, but well below the Walecka model \cite{Walecka:qa} value of
$560 \text{ MeV}$. We have performed the self-consistent
calculation to obtain the profile functions for zero density,
$0.5\rho_{0}$, $1.0\rho_{0}$ and $1.5\rho_{0}$ in
Fig.~\ref{fig:profile} (where $\rho_{0}=0.178\text{ fm}^{-3}$ is
nuclear density, consistent with the experimental value of
$0.167\pm 0.018\text{ fm}^{-3}$ \cite{Blaizot:tw}).

\begin{figure}
\centering
\includegraphics[scale=0.7]{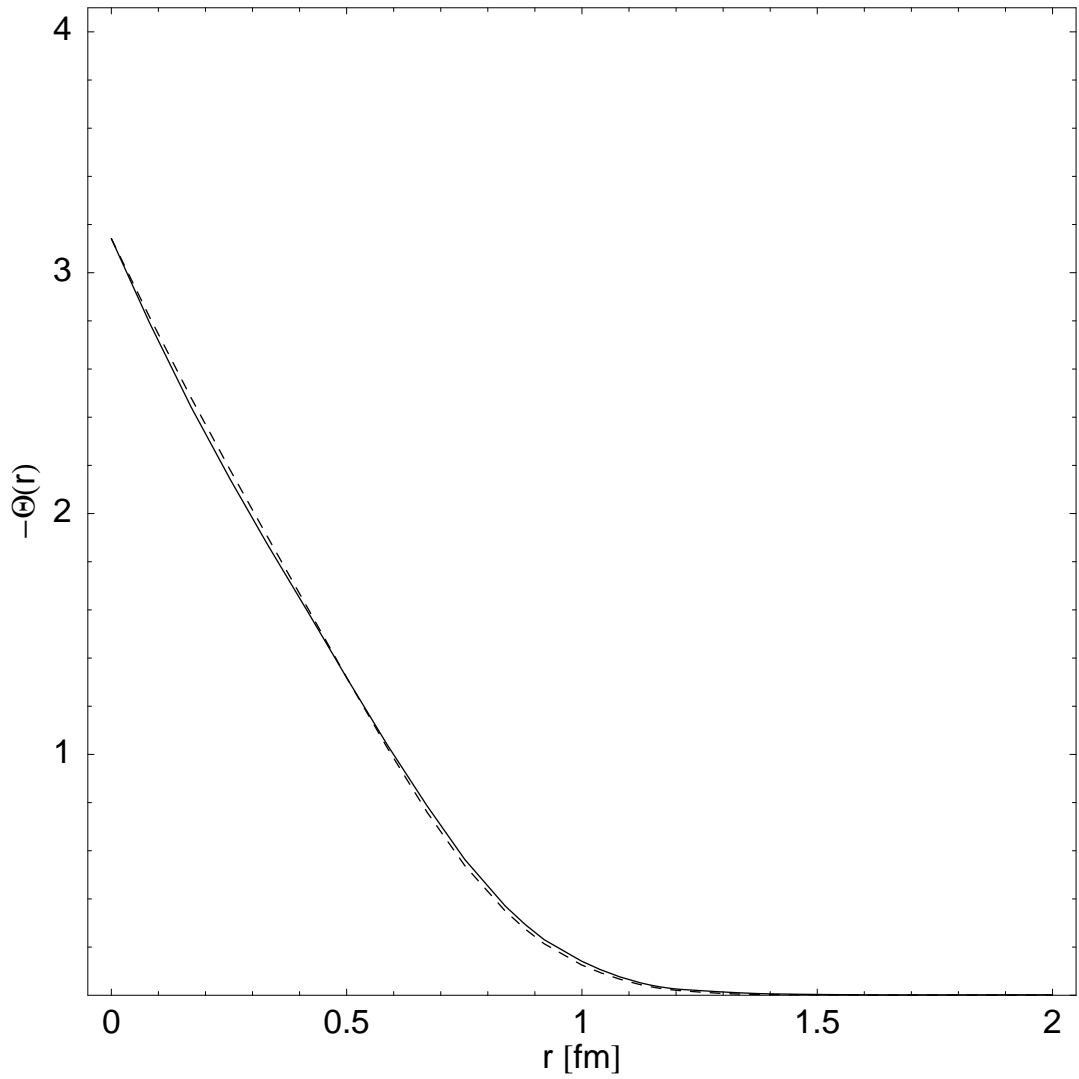}
\caption{Profile functions for the free nucleon using the valence
approximation of Eq.~(\ref{eq:rhos}) (solid) and a fully
self-consistent treatment (dashes).} \label{fig:profapprox}
\end{figure}

We need to address the valence-only approximation in
Eq.~(\ref{eq:rhos}) to the self-consistent profile solution.
Specifically, we are making the approximation that when the change
in the valence energy level $\Delta E^{v}$ is small, so is $\Delta
M_{N}$. We are also hypothesizing that the negative energy level
contributions to the scalar density are approximately constant in
space, and that the spatially varying part of the scalar density
is well approximated by the valence level. It is very important to
note that these approximations are only used in determining the
self-consistent profile, not in the wave functions used as input
into the quark distributions or electromagnetic form factors. We
can see the difference from a fully self-consistent calculation of
the profile function for a free nucleon in
Fig.~\ref{fig:profapprox}. The resulting profiles are virtually
identical.

\begin{figure}
\centering
\includegraphics[scale=0.7]{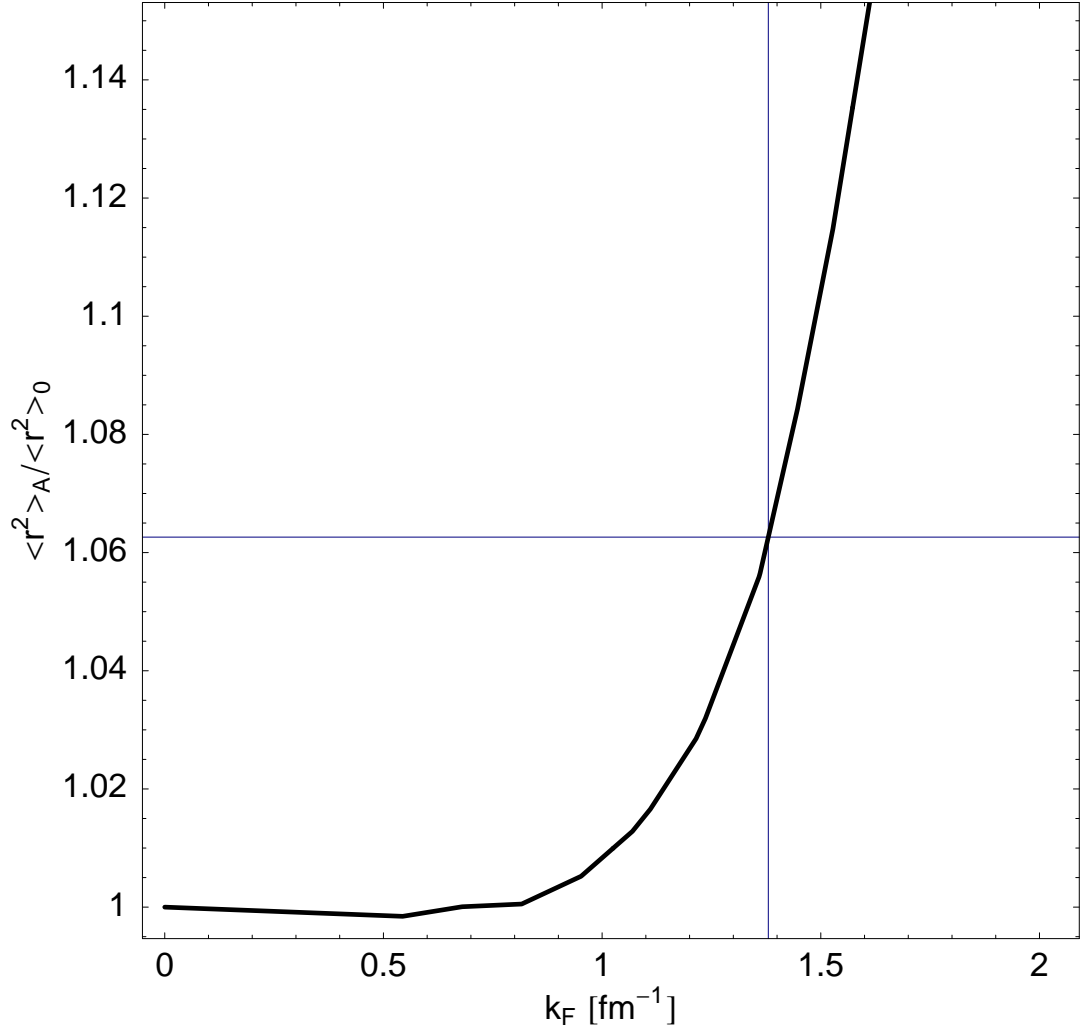}
\caption{The nucleon mean square radius as a function of Fermi
momentum. The 3.1\% increase at $k_{F} = 1.38\text{ fm}^{-1}$ is
indicated by the horizontal and vertical lines. Experiments
\cite{Mckeown:kn,Strauch:2002wu} limit nucleon root mean square
radius swelling to less than 6\%, or 1.12 on the vertical
axis.}\label{fig:radius}
\end{figure}

While the vacuum value of the condensate does not vary with the
Fermi momentum by definition, the effective condensate
Eq.~(\ref{eq:effcond}) falls $\sim 30 \%$ at nuclear density,
\textit{q.v.}~Eq.~(\ref{eq:qqmed}). This is consistent with the
model independent result \cite{Cohen:1991nk} that predicts a value
25-50\% below the vacuum value. The nucleon mean square radius is
plotted as a function of the Fermi momentum in
Fig.~\ref{fig:radius}. The influence of the nuclear medium on the
nucleon causes the root mean square radius of the baryon density
to increase by 3.1\% at $k_{F} = 1.38\text{ fm}^{-1}$. This
swelling is consistent with a $<6 \%$ increase ($\langle r^{2}
\rangle_{A}/\langle r^{2} \rangle_{0} < 1.12$ in
Fig.~\ref{fig:radius}) as constrained by quasi-elastic inclusive
electron-nucleus scattering data \cite{Mckeown:kn}.

\chapter{Quark Distributions}
\label{ch:dist}

One way to probe intermediate length scales and non-perturbative
physics is to examine the short distance structure of a large
object. The prime example is the European Muon Collaboration (EMC)
effect \cite{Aubert:1983xm} where the short distance ($\sim 5$
GeV, or $\sim 10^{-2}\text{ fm}$) structure of nuclei seen in deep
inelastic lepton-nucleus scattering differs from that of a
collection of free nucleons. This measurement showed that bound
nucleons are different than free ones, and implied that the medium
modifications could be significant for any nuclear observable
\cite{Geesaman:1995yd}.

Our primary concern is the depletion of the nuclear structure
function $F_{2}^{A}(x)$ in the valence quark regime $0.3\lesssim x
\lesssim 0.7$. While the general interpretation is that a valence
quark in a bound nucleon has less momentum than in a free one,
corresponding to some increased length scale, the specific
mechanism for this has eluded a universally accepted explanation
for 20 years
\cite{Geesaman:1995yd,Piller:1999wx,Arneodo:1992wf,Sargsian:2002wc}.
A popular explanation is the so-called `binding' effect which
originates from a possible mechanism in which mesons binding the
nucleus carry momentum. An important consequence is that the
mesonic presence would enhance the anti-quark content of the
nucleus \cite{Bickerstaff:1984ax,Ericson:1984vt}. Such an effect
has not been seen in Drell-Yan experiments \cite{Alde:im} in which
a quark in a proton beam annihilates with an antiquark in a
nuclear target producing a muon pair. Furthermore, relativistic
treatments, including the light-cone approach needed to obtain the
nucleon structure function, of the binding effect with
structureless hadrons fail
\cite{Miller:2001tg,Smith:2002ci,Birse:hu,Frankfurt:1985ui},
suggesting that modifications of the internal quark structure of
the nucleon are required to explain the deep inelastic scattering
data.

Any description of the EMC effect must be consistent with the
constraints set by both deep inelastic scattering and Drell-Yan
data. Thus a successful model must include antiquarks as well as
quarks, and show how the medium modifies both the valence and sea
quark distributions. Our purpose is to provide a mechanism for
that modification within the Chiral Quark-Soliton ($\chi$QS) model
\cite{Kahana:dx,Birse:1983gm,Diakonov:2000pa,Christov:1995vm}.
This phenomenological model has many desirable qualities: the
ability to describe a wide class of hadron observables with
surprising accuracy, the inclusion of antiquarks, positivity of
Generalized Parton Distributions, and a basis in QCD
\cite{Diakonov:2000pa}. Here we show how the model describes the
EMC effect, and satisfies the bounds on nuclear antiquark
enhancement provided by Drell-Yan experiments.


Polarized lepton-nucleus scattering experiments are another
important tool in hadronic physics. For example, in order to study
the spin structure function of the neutron, one must use nuclear
targets. It is already well known that there are significant
differences betweeen free and bound nucleons in the unpolarized
case (the EMC effect \cite{Aubert:1983xm}). It is reasonable to
assume that nuclear effects could appear in polarized quark
distributions. Our purpose here is to calculate the analogous
modification to the nucleon spin structure function
$g_{1}^{(p,n)}(x,Q^{2})$: a `polarized EMC effect'.

\section{Light-cone Correlation Function and Sum Rules}

If we insert a complete set of states into the light-cone
correlation function, we obtain
\begin{eqnarray}
q(x) & = & k^{+} \int \frac{dz}{2\pi}^{-} e^{i x k^{+}z^{-}}
\langle k | \overline{\psi}(0) \gamma^{+} \psi (z^{-})
| k \rangle \label{eq:lccf2a}\\
& = & N_{C} k^{+} \sum_{n} \int \frac{dz}{2\pi}^{-} e^{i x
k^{+}z^{-}} \langle k | p, n \rangle \gamma^{0} \gamma^{+} e^{-i
p_{n}^{+}z^{-}} \langle p, n | k \rangle. \label{eq:lccf2}
\end{eqnarray}
Evaluating Eq.~(\ref{eq:lccf2}) in the rest frame, with $k^{+} =
k^{-} = M_{N}[\Theta (r, k_{F})]$ from Eq.~(\ref{eq:mn}), and
writing the wave functions and plus and minus components
explicitly, we obtain
\begin{equation}
q(x)  =  N_{C} M_{N} \sum_{n} \int \frac{dz}{2\pi}^{0} e^{i x
M_{N} z^{0}} \psi_{n}^{\dag}(p) (1+ \gamma^{0} \gamma^{3}) e^{-i
(E_{n}+p^{3})z^{0}} \psi_{n}(p) \label{eq:lccf}
\end{equation}
We can put this in a form that is ideal for evaluation in the
finite KR basis
\begin{equation}
q(x) = N_{C} M_{N} \sum_{n} \langle n | (1 + \gamma^{0}
\gamma^{3}) \delta(E_{n} + p^{3} - x M_{N}) | n \rangle
\label{eq:me}
\end{equation}

In order to address the baryon and momentum sum rules, we will
follow Ref.~\cite{Diakonov:1996sr}. It is convenient to put the
light-cone correlator in the form of a functional trace using (in
the rest frame)
\[
\sum_{n} | n \rangle \langle n | e^{-i E_{n}z^{0}} = \int dE
\delta(E-H) e^{- i E z^{0} }
\]
where $H$ is the Dirac Hamiltonian Eq.~(\ref{eq:hamiltonian}).
With this substitution, we can write the moments of
Eq.~(\ref{eq:lccf}) as
\begin{equation}
\int dx x^{n-1} q(x) = N_{C} M_{N}^{1-n} \int dE \text{ Tr }
\left[ \delta(E-H) (E + p^{3})^{n-1} (1 + \gamma^{0} \gamma^{3})
\right]
\end{equation}
For $n = 1$, we obtain the baryon sum rule
\[
N_{C} \text{ Tr } \left[ \theta(E^{v} - H) - \theta(-H^{(0)})
\right] = N_{C} B
\]
since the $\theta$-function operator simply counts the occupied
levels in the nucleon, with the number of occupied levels in the
vacuum subtracted, leaving only the quarks in the valence level.
It should be noted here that the topological condition for the
profile function guarantees that only one energy level comes out
of the positive continuum, and hence that the winding number of
the soliton field is not the baryon number; it is only a
necessary, not sufficient, condition to obtain unit baryon number.

For $n = 2$, along with other terms that vanish by the hedgehog
symmetry of the profile function, we obtain the momentum sum rule
\begin{eqnarray} N_{C}
M_{N}^{-1} \text{ Tr } \left[ \theta(E^{v} - H)H -
\theta(-H^{(0)})H^{(0)} \right] & = & N_{C} M_{N}^{-1} \left[
E^{v} + \sum_{E_{n}<0} E_{n} - \sum_{E_{n}^{(0)}<0} E_{n}^{(0)}
\right]\nonumber\\
& = &  M_{N}^{-1} M_{N}[\Theta (r,k_{F})]\nonumber\\
& = & 1
\end{eqnarray}
The only details we have used are that we have a complete set of
basis states, a single valence energy level, and that the sum of
the energy eigenvalues of the Dirac Hamiltonian
Eq.~(\ref{eq:hamiltonian}) yield the nucleon mass
Eq.~(\ref{eq:mn}). Therefore, the calculation presented here is
independent of the addition of a non-zero vector potential. It is
important to note that both of these sum rules are automatically
satisfied as long as one remains in the same basis consistently
throughout the calculation, provided Eq.~(\ref{eq:lccf2a}) holds.

\section{Unpolarized Distributions}

\begin{figure}
\centering
\includegraphics[scale=0.7]{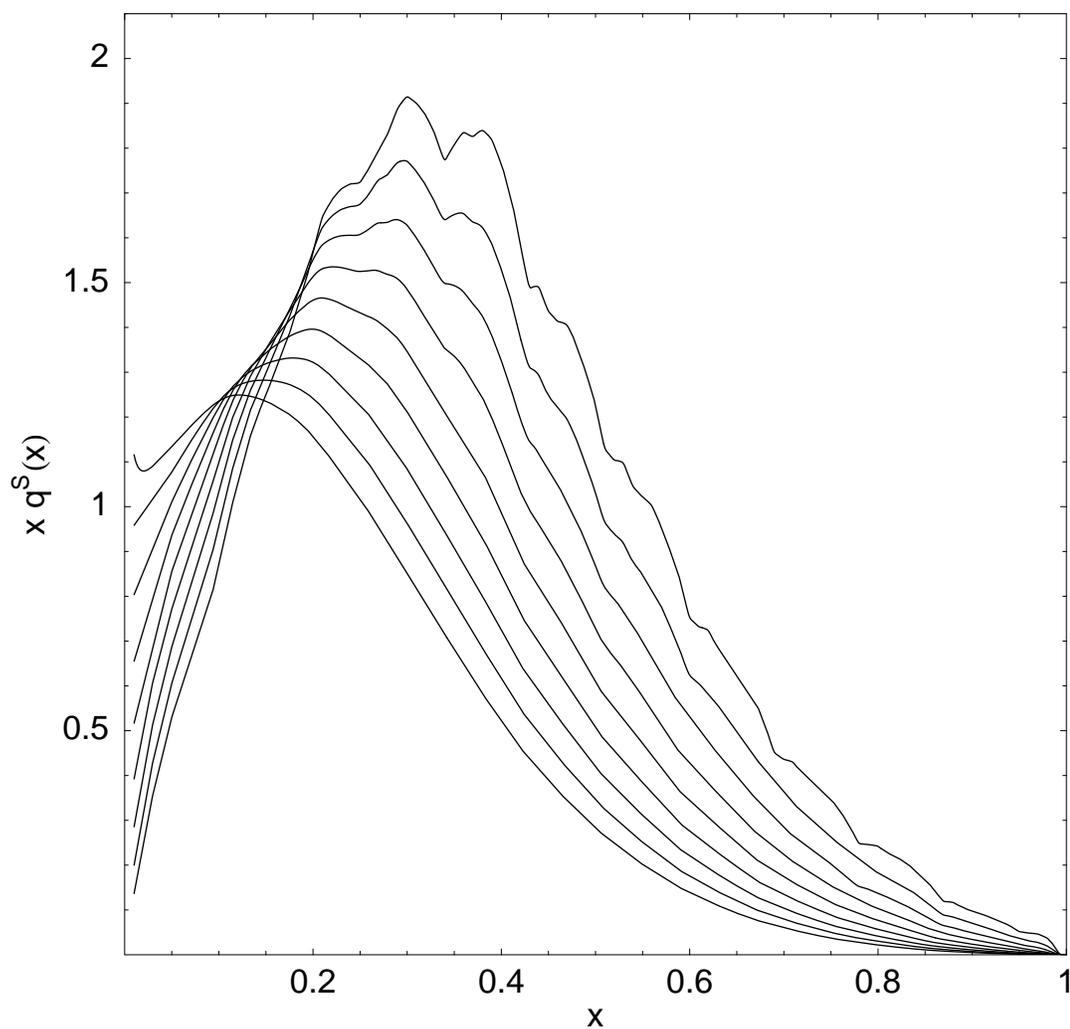}
\caption{The smoothing effect of evolution on the quark singlet
distribution. The curves are shown at scales $\log M_{PV}^{2}$
(highest curve at $x = 0.3$), and the DIS scale $\log Q^{2}$, with
$Q^{2} = 10 \text{ GeV}^{2}$ (lowest curve at $x = 0.3$).}
\label{fig:evolve}
\end{figure}

The isoscalar unpolarized distribution $q(x)= u(x)+d(x)$ is the
leading order term in $N_{C}$, with the isovector unpolarized
quark distribution $u(x)-d(x)$ smaller by a factor $ \sim 1/N_c$
and set to zero. The distributions are calculated using the KR
basis at $k_{F}=0$ and $k_{F}=1.38\text{ fm}^{-1}$ almost exactly
as in Ref.~\cite{Diakonov:1997vc} where the quark distribution is
given by the matrix element Eq.~(\ref{eq:me}) evaluated in the
finite basis
\begin{equation}
q(x) = N_{C} M_{N} \sum_{n} \sum_{i,j} c_{ni}^{*} c_{nj}
\frac{1}{2 p_{i}} \theta(p_{i} - |E_{n}-x M_{N}|)
\left[\delta_{ij} - \frac{E_{n}-x M_{N}}{p_{i}^{2}}
(\gamma^{0}\bm{p}\cdot\bm{\gamma})_{ij}\right] ,\label{eq:unpolme}
\end{equation}
with the Pauli-Villars regulated sum taken over occupied states.
The eigenvalues $E_{n}$ are determined from diagonalizing the
Hamiltonian, derived from the Lagrangian (\ref{eq:lagrangian}), in
the KR basis. The vector meson exchange is not explicit in
Eq.~(\ref{eq:unpolme}); it is implicit in the energy eigenvalues
$E_{n}$.  The antiquark distribution is given by $\bar{q}(x) =
-q(-x)$ where the sum is over unoccupied states
\cite{Diakonov:1997vc}. We use the exact sea wave functions, and
not the approximation used in Eq.~(\ref{eq:rhos}). The use of a
finite basis causes the distributions to be discontinuous. These
distributions are smooth functions of $x$ in the limit of infinite
momentum cutoff and box size, but numerical calculations are made
at finite values and leave some residual roughness. This is
overcome in Ref.~\cite{Diakonov:1997vc} by introducing a smoothing
function. We deviate from their procedure, and do not smooth the
results; instead we find the subsequent one-loop perturbative QCD
evolution \cite{Hagiwara:fs} to be sufficient. This can be seen in
Fig.~\ref{fig:evolve}.

\begin{figure}
\centering
\includegraphics[scale=0.7]{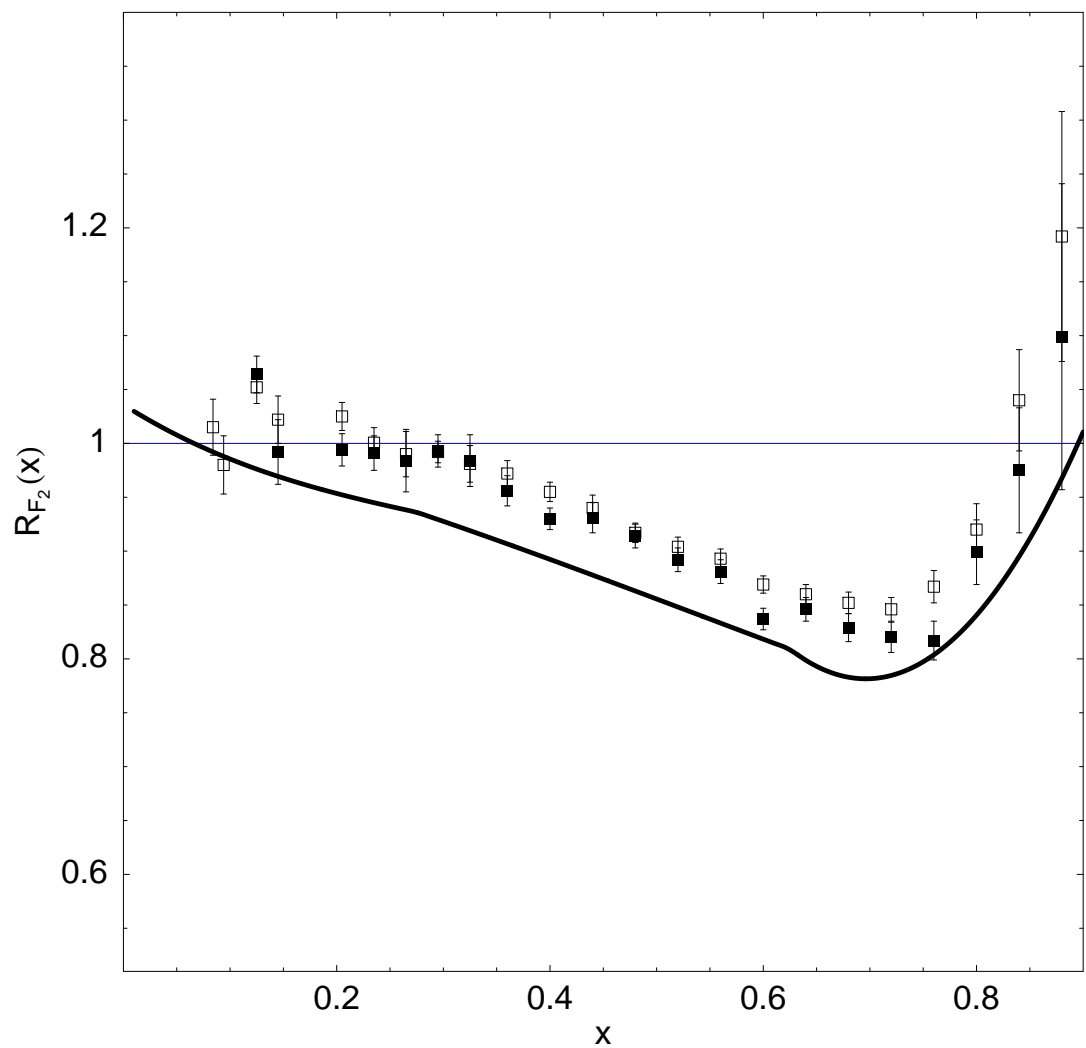}
\caption{The EMC ratio at a scale $Q^{2} = 10 \text{ GeV}^{2}$ for
nuclear matter. The data are for Iron (empty boxes) and Gold
(filled boxes) from SLAC-E139 \cite{Gomez:1993ri}.}\label{fig:emc}
\end{figure}

\begin{figure}
\centering
\includegraphics[scale=0.7]{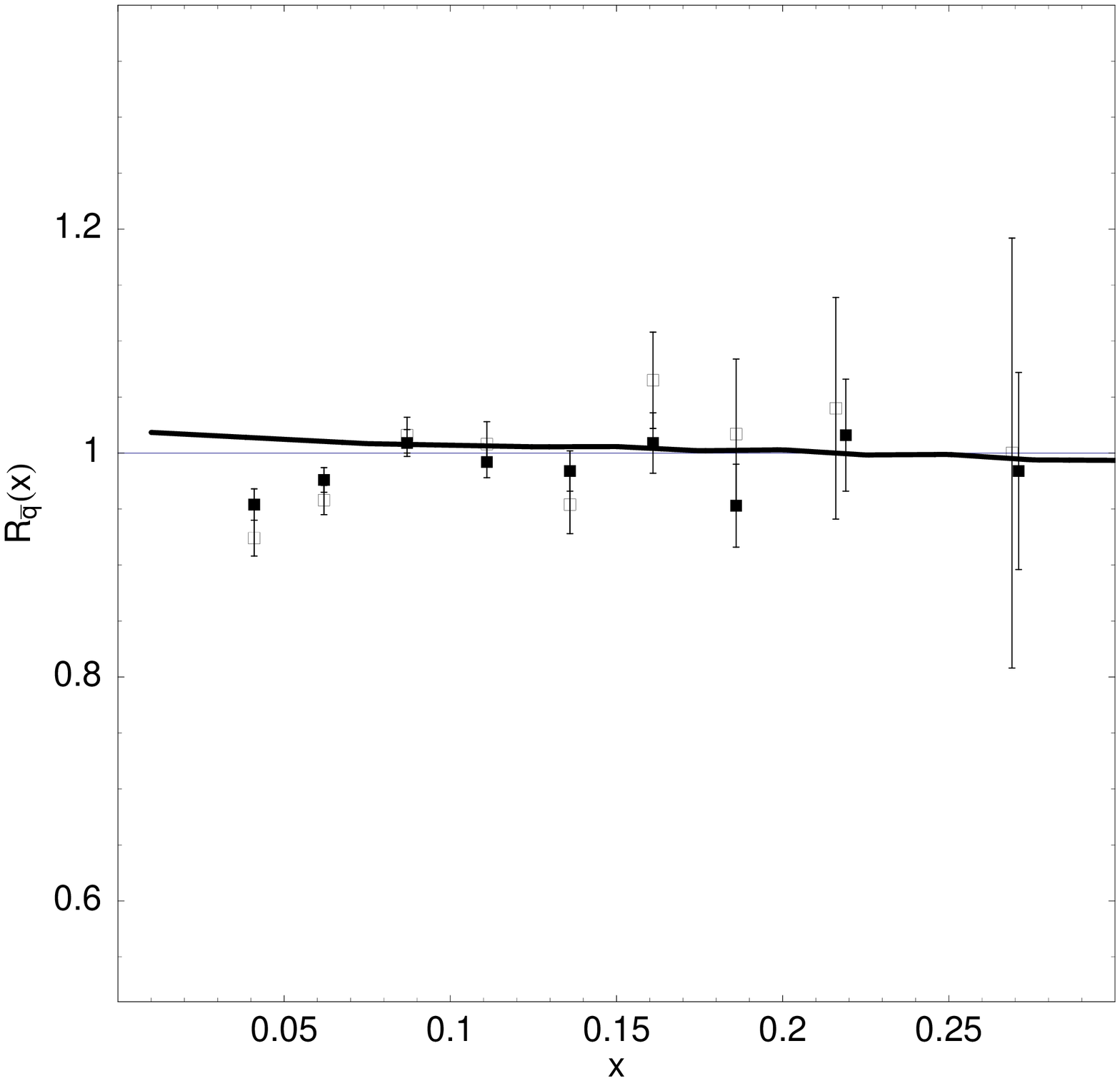}
\caption{The Drell-Yan ratio at a scale $Q^{2} = 10 \text{
GeV}^{2}$ for nuclear matter. The data are for Iron (empty boxes)
and Tungsten (filled boxes) from FNAL-E772
\cite{Alde:im}.}\label{fig:dy}
\end{figure}

These distributions are used as input at the model scale of $Q^{2}
= M_{PV}^{2}\simeq 0.34 \text{ GeV}^{2}$ for evolution to $Q^{2} =
10 \text{ GeV}^{2}$ in the case of the quark singlet distribution
$q^{S}(x)=q(x)+\bar{q}(x)\propto F_{2}^{N}(x)/x$ at leading order
in $N_{C}$. We define the generic ratio function to be
\begin{eqnarray}
R_{g}(x,Q^{2}) & = & \frac{f\otimes g(x,Q^{2},k_{F})}{A g(x,Q^{2},k_{F}=0)},\label{eq:ratio}\\
f\otimes g(x/y,Q^{2},k_{F}) & = & \int_{x}^{A} dy f(y)
g(x/y,Q^{2},k_{F}).\label{eq:conv}
\end{eqnarray}
In the following, we take the general function $g$ to be the
structure function $F_{2}^{(N)}$, the antiquark distribution
$\bar{q}$, or the spin structure function $g_{1}^{(p,n)}$. The
nucleon momentum distribution $f(y)$ following from a light-cone
approach, for any mean field theory of nuclear matter (see
Ref.~\cite{Miller:2001tg} and Eq.~(\ref{eq:pmdA}) in Appendix
\ref{ch:finite}) is
\begin{equation}
f(y) = \frac{3}{4\Delta_{F}^{3}} \theta(1+\Delta_{F}-y)
\theta(y-1+\Delta_{F}) \left[
\Delta_{F}^{2}-(1-y)^2\right],\label{eq:pmd}
\end{equation}
where $\Delta_{F} = k_{F}/\bar{M}_{N}$ and
$\bar{M}_{N}=M_{N}(0)-15.75 \text{ MeV}$. The distributions are
evolved to $Q^{2} = 10 \text{ GeV}^{2}$ for use in the EMC ratio
$R_{F_{2}}(x)$ and the Drell-Yan ratio $R_{\bar{q}}(x)$
Eq.~(\ref{eq:ratio}). The ratios are plotted in
Figs.~\ref{fig:emc} and \ref{fig:dy}, respectively. While the data
shown in Figs.~\ref{fig:emc} and \ref{fig:dy} are for large, but
finite, nuclei, our calculation reproduces the trend of both sets
of data. It falls slightly below the SLAC-E139 data
\cite{Gomez:1993ri} due to the higher density of nuclear matter.

\begin{figure}
\centering
\includegraphics[scale=0.7]{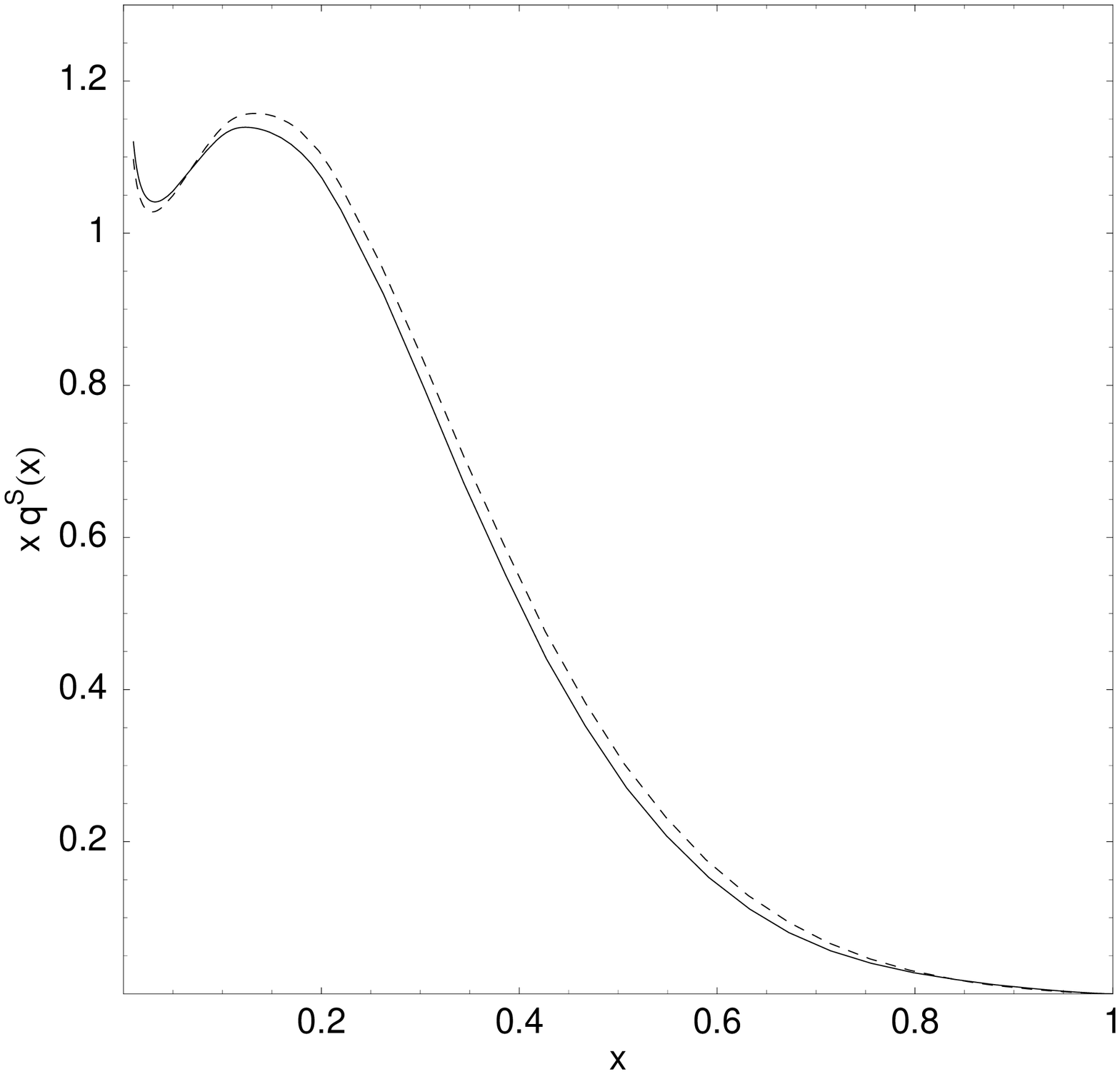}
\caption{The distribution $xq^{S}(x)$ in a free (dashed) and bound
(solid) nucleon at a scale $Q^{2} = 10 \text{ GeV}^{2}$.}
\label{fig:dists}
\end{figure}

In Fig.~\ref{fig:dists} we show the quark singlet distribution for
a free and bound nucleon at a scale $Q^{2}=10\text{ GeV}^{2}$.
There is a large depletion in the bound nucleon valence
distribution, accounting for nearly all of the depletion in
Fig.~\ref{fig:dists}, since the antiquark contribution is
virtually unchanged. If one uses only the valence contribution to
calculate the EMC ratio (\ref{eq:ratio}), produces too large an
effect. This large effect is comparable to that of the QMC model
impulse approximation calculation or the Guichon model
\cite{Saito:yw}, both of which only include valence quarks. This
valence effect is mitigated by a small enhancement in
$\bar{q}(x)$, consistent with the Drell-Yan data, so that the
singlet distribution has only a moderate depletion consistent with
the EMC effect.

A simple picture in terms of the uncertainty principle is
available. The influence of the nuclear medium on the nucleon
causes the root mean square radius of the baryon density to
increase by 3.1\%. This corresponds to a decreased momentum, and a
depletion of the bound structure function relative to the free
one. This swelling is consistent with a $<6 \%$ increase as
constrained by quasi-elastic inclusive electron-nucleus scattering
data \cite{Mckeown:kn}, and the recent polarization transfer
measurement \cite{Strauch:2002wu}.

We ignore the effects of shadowing, which occur when the virtual
photon striking the nucleus fluctuates into a quark-antiquark pair
over a distance $\sim 1/2 M_{N} x$ exceeding the inter-nucleon
separation. This causes a depletion in the structure function for
$x\leq 0.1$ and is relatively well understood
\cite{Geesaman:1995yd,Piller:1999wx,Arneodo:1992wf,Sargsian:2002wc}
and so we do not reiterate those results. Additionally, we ignore
contributions from quantum pion structure functions, which in this
model propagate through constituent quark loops, and would modify
the behavior at small $x$. These loops are suppressed by
${\mathcal O}(1/N_{C})$, and are not treated at leading order.

The present model provides a intuitive, qualitative treatment that
maintains consistency with all of the free nucleon properties
calculated by others \cite{Diakonov:2000pa,Christov:1995vm}. It
gives a reasonable description of nuclear saturation properties,
reproduces the EMC effect, and satisfies the constraints on the
nuclear sea obtained from Drell-Yan experiments with only two free
parameters: $g_{s}$ and $g_{v}$.

The central mechanism to explain the EMC effect is that the
nuclear medium provides an attractive scalar interaction that
modifies the nucleon wave function. This is also the dominant
mechanism in the QMC model approach to the EMC effect
\cite{Saito:yw} and also similar to the quark delocalization
approach \cite{Benesh:2003fk}. The improvements given here are the
explicit computation of the effects of the medium on the antiquark
distributions so that consistency with the Drell-Yan data could be
verified, and the reduction of the number of input parameters and
model assumptions. Our extension of the Chiral Quark-Soliton model
to nuclear matter provides a new, consistent way to calculate
possible medium modifications of a variety of observables that
could be measured in experiments.

\section{Polarized Quark Distribution Function}
\label{sec:pqdf}

The first discussion of nuclear effects in the polarized quark
distributions is in Ref.~\cite{Close:1987ay} in the context of
dynamical rescaling. A more recent calculation is in
Ref.~\cite{Cloet:2005rt}; it predicts dramatic effects for the
bound nucleon spin structure function.  We have previously shown
how the model describes nuclear saturation properties, reproduces
the EMC effect, and satisfies the bounds on unpolarized nuclear
antiquark enhancement provided by Drell-Yan experiments
\cite{Smith:2003hu}. Therefore, we expect the $\chi$QS model to
produce a reasonable result for the polarized distributions.

The polarized quark distribution for flavor $i$ is defined by the
difference between the quark distributions with spin parallel
$(\uparrow)$ and antiparallel $(\downarrow)$ to the nucleon
\begin{equation}
\Delta q_{i}(x,Q^{2}) = q_{i}^{\uparrow}(x,Q^{2}) - q_{i}^{\downarrow}(x,Q^{2}).
\end{equation}
The polarized antiquark distribution is defined analogously using
$\bar{q}_{i}^{\uparrow}$, and $\bar{q}_{i}^{\downarrow}$. The
isovector polarized distribution $\Delta q^{(T=1)}(x)=\Delta
u(x)-\Delta  d(x)$ is the leading order term in $N_{C}$, with the
isoscalar polarized quark distribution $\Delta q^{(T=0)}(x)=\Delta
u(x)+\Delta  d(x)$  smaller by a factor $ \sim 1/N_{C}$ and set to
zero. We will therefore suppress the isospin superscript in the
following. The distributions are calculated using the KR basis at
$k_{F}=0$ and $k_{F}=1.38\text{ fm}^{-1}$ (see
Refs.~\cite{Smith:2003hu,Smith:2004dn}) almost exactly as in
Ref.~\cite{Diakonov:1997vc} where the quark distribution is given
by the matrix element
\begin{eqnarray}
\Delta q(x) & = &  -\frac{1}{3} (2 T_{3})  N_{C} M_{N} \sum_{n}
\sum_{i,j} c_{ni}^{*} c_{nj} \frac{1}{2 p_{i}} \theta(p_{i} -
|E_{n}-x M_{N}|) \times \nonumber\\
&  & \Bigg[ \frac{E_{n}-x M_{N}}{p_{i}^{2}}
(\gamma_{5}\bm{p}\cdot\bm{\tau})_{ij} + \nonumber\\
&  & \left( -\frac{1}{2 p_{i}} + \frac{3}{2} \frac{(E_{n}-x
M_{N})^{2}}{p_{i}^{4}}\right) \sum_{k}
(\gamma^{0}\bm{p}\cdot\bm{\gamma})_{ik}
(\gamma_{5}\bm{p}\cdot\bm{\tau})_{kj} + \nonumber \\
&  & \left( \frac{1}{2} - \frac{1}{2} \frac{(E_{n}-x
M_{N})^{2}}{p_{i}^{2}}\right)(\gamma^{0}\bm{\tau}\cdot\bm{\gamma})_{ij}
\Bigg] ,\label{eq:polme}
\end{eqnarray}
with the Pauli-Villars regulated sum taken over occupied states.
The eigenvalues $E_{n}$ are determined from diagonalizing the
Hamiltonian, derived from the Lagrangian (\ref{eq:lagrangian}), in
the KR basis.

The antiquark distribution is given by $\Delta\bar{q}(x) = \Delta
q(-x)$ where the sum is over unoccupied states. The use of a
finite basis causes the distributions to be discontinuous. These
distributions are smooth functions of $x$ in the limit of infinite
momentum cutoff and box size, but numerical calculations are made
at finite values and leave some residual roughness. This is
overcome in Ref.~\cite{Diakonov:1997vc} by introducing a smoothing
function. We deviate from their procedure, and do not smooth the
results; instead we find that performing the one-loop perturbative
QCD evolution \cite{Hagiwara:fs} provides sufficient, but not
complete, smoothing. Some residual fluctuations due to the finite
basis remain visible in our results, and the size of these
fluctuations serve as a guide to the size of the error introduced
by the method.

These distributions are used as input at the model scale of $Q^{2}
= M_{PV}^{2}\simeq 0.34 \text{ GeV}^{2}$ for evolution to $Q^{2} =
10 \text{ GeV}^{2}$. The polarized structure function to leading
order in $N_{C}$ is given by
\begin{eqnarray}
g_{1}^{(p,n)} & = & \frac{1}{2} \sum_{i} e_{i}^{2} \big( \Delta q_{i} + \Delta \bar{q}_{i} \big) \\
g_{1}^{(p,n)} & = & \pm\frac{5}{18}  \Delta q^{NS}  + \mathcal{O}(N_{C}^{0}) \\
\Delta q^{NS} & = & \frac{3}{5} \big( \Delta u  - \Delta d +
\Delta \bar{u}  - \Delta \bar{d} \big) + \mathcal{O}(N_{C}^{0})
\end{eqnarray}
The nucleon momentum distribution is assumed to be the same as the
unpolarized case, Eq.~(\ref{eq:pmd}).

This is justified since the Hugenholtz-van Hove theorem follows
from the pressure vanishing for a stable nucleus (see Appendix
\ref{ch:finite}), which would remain valid for a polarized system.
Therefore, we would expect to still have a distribution sharply
peaked around $y\simeq 1$. The details of the functional form of
$f(y)$ have little effect on the ratio beyond the peak location.
We are already ignoring target polarization, which has the same
effect as the leading order correction (change in the
normalization) to the momentum distribution, with the sum on the
spin-dependent distributions yielding the unpolarized
distributions. Also, the spin-orbit force tends to average out in
nuclear matter, leading to the same distribution for all nucleons.
The nuclear matter momentum distribution we use is fairly close to
those calculated for Lithium isotopes \cite{Saito:2001gv}.

\begin{figure}
\centering
\includegraphics[scale=0.7]{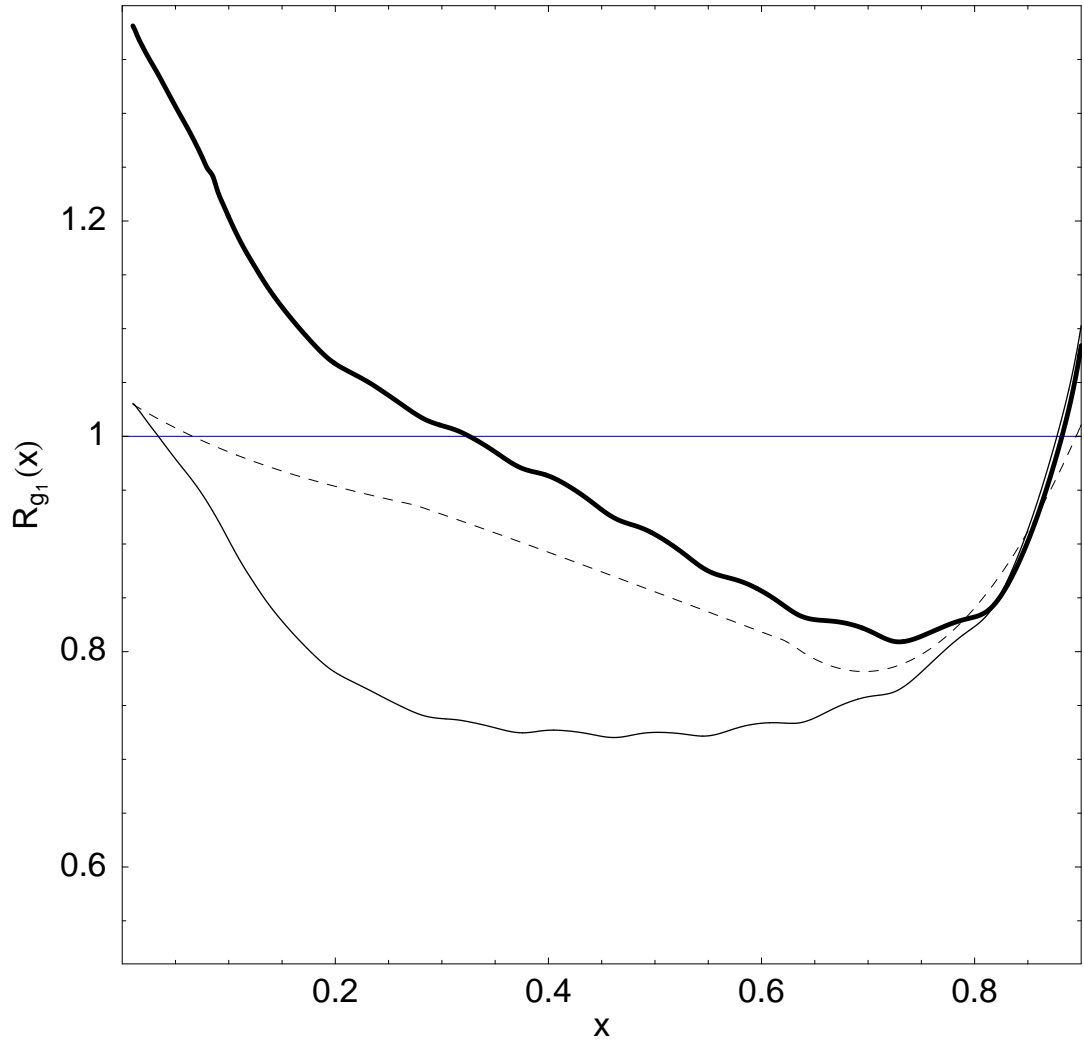}
\caption{The ratio Eq.~(\ref{eq:ratio}) at scale
$Q^{2} = 10 \text{ GeV}^{2}$  for nuclear
matter. The heavy line is the full calculation for nuclear matter.
The light line is the effect calulated using only medium modifications
to the `valence' energy level as decribed in the text.}\label{fig:polEMC}
\end{figure}

We show the ratio Eq.~(\ref{eq:ratio}) in Fig.~\ref{fig:polEMC}
using a `valence'-like distribution, as well as for the full
distribution. The latter includes all medium modifications, while
the former distribution uses the medium modified energy level
eigenstate, but the same free nucleon sea quark distribution for
both the free and bound nucleon. This was done in order to compare
our results with the model in Ref.~\cite{Cloet:2005rt}, which only
has valence quarks at the model scale. The single energy level
actually has a contribution to the polarized antiquark
distribution, so it alone cannot be considered a true valence spin
structure function. However, this contribution is small, so we
effectively reproduce the result of a valence quark model,
especially in the region $x\gtrsim 0.3$.
\begin{figure}
\centering
\includegraphics[scale=0.7]{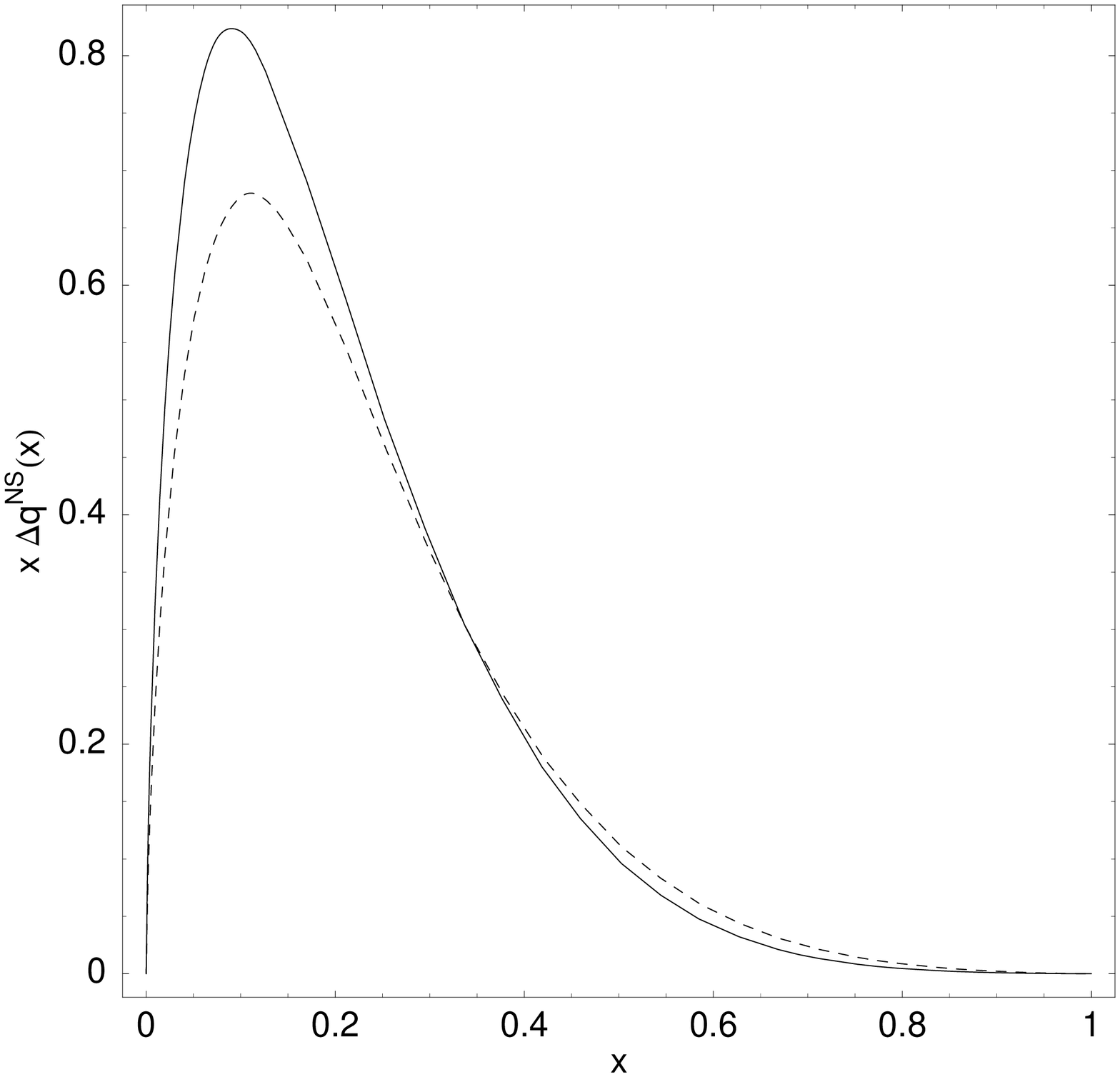}
\caption{The distribution $\Delta q^{NS}(x)$ in a free (dashed)
and bound (solid) nucleon at a scale $Q = 10 \text{ GeV}^{2}$.}
\label{fig:poldists}
\end{figure}

In Fig.~\ref{fig:polEMC}, one can see that there is a large
depletion in the polarized `valence' quark distribution . This
produces a large depletion in the isovector axial coupling
$g_{A}^{(T=1)}$ of $17.8$\%. This large effect is comparable to
that of the calculation in Ref.~\cite{Cloet:2005rt} which only
includes valence quarks at the model scale. This valence effect is
mitigated by a large enhancement in the sea quark contribution, so
that the full polarized distribution has only a moderate depletion
in the region $0.3\lesssim x \lesssim 0.7$ of the same size as the
EMC effect in unpolarized nuclear structure functions. There is a
large enhancement for $x \lesssim 0.3$ due to the sea quarks. This
large enhancement is very different from the small effect
calculated in the unpolarized case \cite{Smith:2003hu}, and seen
in unpolarized Drell-Yan experiments \cite{Alde:im}. This would
suggest that one might see a significant enhancement in a
polarized Drell-Yan experiment, even after including shadowing
corrections (which we address later).

The larger sensitivity to the lower components of the wave
functions, and the smaller size of the relevant distribution
$\Delta q(x)$ are the primary sources for the greater sea quark
enhancement in the polarized case than in the unpolarized case.
This model gives a smaller value for the isovector axial coupling
$g_{A}^{(T=1)} = 1.11$ than experiment. This is to be expected at
leading order in $N_{C}$ as there are important corrections due to
rotational modes of the soliton. These corrections should be
relatively minor when considering the ratio Eq.~(\ref{eq:ratio}).

The axial coupling $g_{A}^{(T=1)}$ is enhanced by 9.8\% in the
nuclear medium. This is in accord with an earlier finding of a
$\sim 25$\% enhancement for $g_{A}$ in a different soliton model
by Birse \cite{Birse:1993nr}. There, the effect is also seen as a
competition between enhancement and depletion.

In order to address the medium modification of the Bjorken sum rule
\cite{Bjorken:1966jh,Bjorken:1969mm}
\begin{equation}
\lim_{Q^{2}\rightarrow\infty} \int_{0}^{1} dx \:  g_{1}^{(p)}(x,Q^{2}) -
g_{1}^{(n)}(x,Q^{2}) = \frac{g_{A}}{6}\label{eq:BjSR}
\end{equation}
as an integral of the experimentally observed nuclear
distribution, one must account for the effects of shadowing. This
occurs when the virtual photon striking the nucleus fluctuates
into a quark-antiquark pair over a distance $\sim 1/2 M_{N} x$
exceeding the inter-nucleon separation. This causes a depletion in
the structure function for $x\lesssim 0.1$ and is relatively well
understood \cite{Piller:1999wx,Arneodo:1992wf,Sargsian:2002wc}.
Shadowing in the polarized case is expected to be larger than in
the unpolarized case by roughly a factor of 2 simply from the
combinatorics of multiple scattering (see
\textit{e.g.}~Ref.~\cite{Guzey:1999rq}).

The enhancement at $x \sim 0.1-0.2$ in Fig.~\ref{fig:polEMC} is
comparable to that seen by Guzey and Strikman \cite{Guzey:1999rq};
they assume that the combined effects of shadowing, enhancement,
and target polarization lead to the empirical value of the nuclear
Bjorken sum rule for $^{3}$He and $^{7}$Li. Shadowing effects
become large for $x\lesssim 0.05$, but we ignore them as well as
target polarization; such precision is not necessary for our
relatively qualitative analysis. One needs $\sim 10$ times the
shadowing observed in the unpolarized case for Lead in order to
counter the enhancement at $x \sim 0.1-0.2$, and give the same
value for the Bjorken sum rule (\ref{eq:BjSR}) in matter and free
space. This assumes that shadowing is the only effect neglected at
small $x$ in our calculation of the unpolarized quark distribution
\cite{Smith:2003hu}.

\begin{figure}
\centering
\includegraphics[scale=0.7]{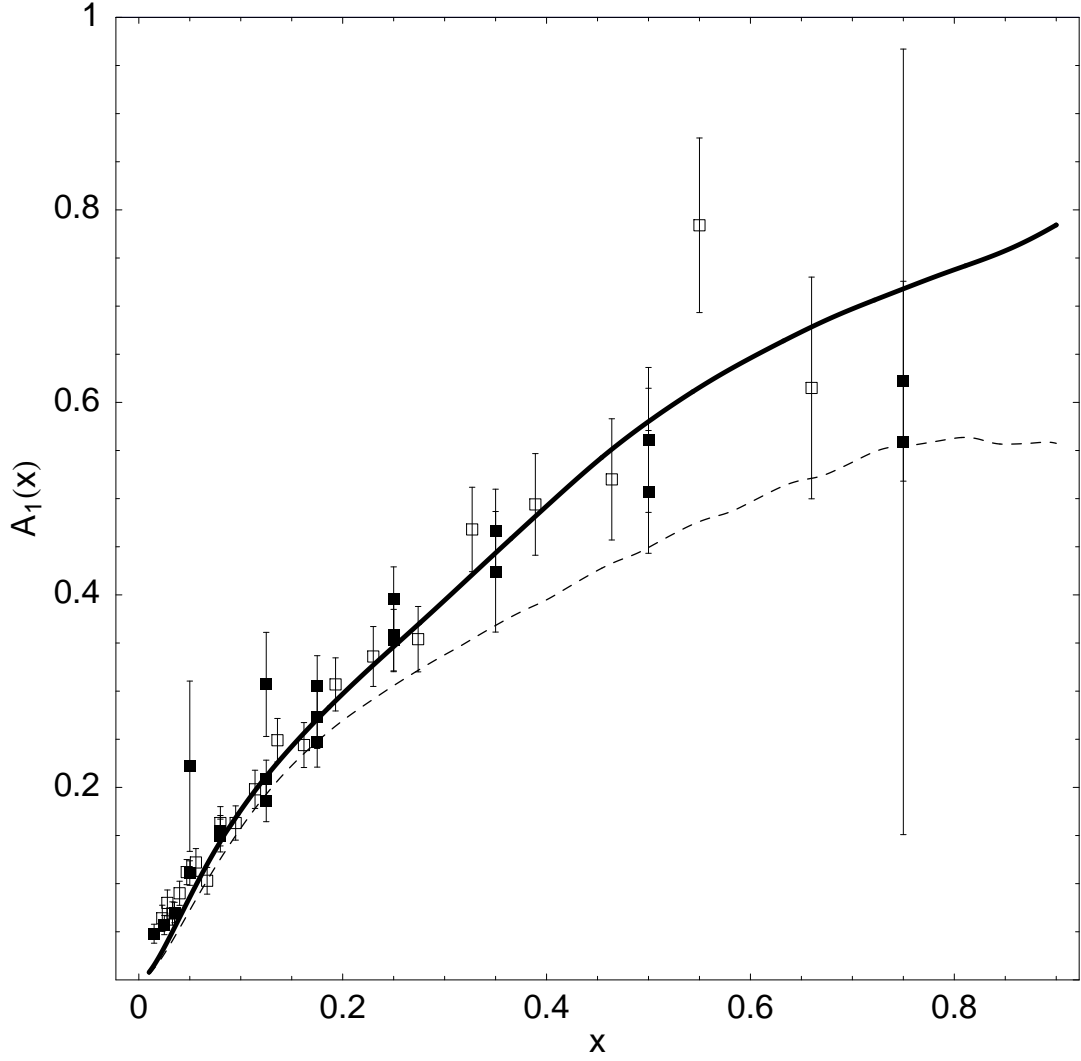}
\caption{The asymmetry $A_{1}^{(p)}$ Eq.~(\ref{eq:A1}) at scale
$Q^{2} = 10 \text{ GeV}^{2}$. The heavy line is for nuclear
matter. The dashed line is for the free proton. The data are from
SLAC \cite{Anthony:2000fn} (filled) for $Q^{2}\sim 1 - 40 \text{
GeV}^{2}$ and HERMES \cite{Airapetian:1998wi} (empty) for
$Q^{2}\sim 1 - 20 \text{ GeV}^{2}$ The free curve falls slightly
below the data due to the lower value of $g_{A}$ calculated in the
large $N_{C}$ limit. }\label{fig:polEMCA1}
\end{figure}

We also present, in Fig.~\ref{fig:polEMCA1}, the results for the
spin asymmetry
\begin{equation}
A_{1}^{(p)}(x,Q^{2}) = \frac{\sum_{i}e_{i}^{2}  [\Delta
q_{i}(x,Q^{2})+\Delta \bar{q}_{i}(x,Q^{2})]}{\sum_{i}e_{i}^{2}
[q_{i}(x,Q^{2})+\bar{q}_{i}(x,Q^{2})]} \label{eq:A1}.
\end{equation}
The nuclear asymmetry $A_{1}^{(A)}$ is defined by replacing the
polarized and unpolarized quark distributions, represented
generically as $q$, with $f \otimes q$ in Eq.~(\ref{eq:conv}). We
find that for the free case, the calculation falls slightly below
the data due to the smaller value of $g_{A}$ in the large $N_{C}$
limit, and that the size of the medium modification is of the same
order as the experimental error for the free proton
\cite{Anthony:2000fn,Airapetian:1998wi}.

The central mechanism to explain the EMC effect is that the
nuclear medium provides an attractive scalar interaction that
modifies the nucleon wave function. We see this again in the
polarized case. This is also the dominant mechanism in the
Nambu-Jona-Lasinio (NJL) type quark-diquark model of Cloet
\textit{et al} \cite{Cloet:2005rt}, and the soliton model of Birse
\cite{Birse:1993nr}.

It is worth noting that the model of Ref.~\cite{Cloet:2005rt} uses
proper time infrared and ultraviolet cutoffs. These cutoffs are
known \cite{Diakonov:1997vc} in the $\chi$QS model to ruin the
equivalence of sums using occupied and unoccupied states,
producing unphysical behavior in the antiquark distributions. This
lack of equivalence can be traced to a violation of causality,
therefore the NJL model of Ref.~\cite{Cloet:2005rt}, like the MIT
bag model used in the QMC model, probably could not be used to
calculate medium effects on sea quarks.

The present model provides a intuitive, qualitative treatment that
maintains consistency with all of the free nucleon properties
calculated by others \cite{Diakonov:2000pa,Christov:1995vm}. It
provides reasonable description of nuclear saturation properties,
reproduces the EMC effect, and satisfies the constraints on the
nuclear sea obtained from Drell-Yan experiments with only two
parameters for the nuclear physics ($g_{s}$ and $g_{v}$) fixed by
the binding energy and density of nuclear matter. Therefore, we
expect the results presented here to manifest themselves in future
experiments with polarized nuclei.

\chapter{Electromagnetic Form Factors}
\label{ch:emff}


Recent polarization transfer experiments at TJNAF
\cite{Strauch:2002wu} observed a difference in the electromagnetic
form factors of a proton bound in a Helium nucleus compared to a
free one. The form factors encode the structure information of the
proton observed in elastic scattering. There is extensive work on
the medium modifications of electromagnetic properties of the
nucleon in the literature (for example, see
Refs.~\cite{RuizArriola:ex,Frank:1995pv,Yakhshiev:2002sr,Lu:1998tn}
). This includes effective Lagrangians as well as models that
include the quark substructure of hadrons. While in principle
these effects could be couched in terms of effective field theory
operators, it is our thesis that such results may be more
transparent, physically intuitive or straightforward to calculate
when viewed as a change in the internal structure of the hadrons.

We will use the $\chi$QS model to calculate the electromagnetic
form factors. The overall procedure is similar to the Quark-Meson
Coupling model (QMC) \cite{Lu:1998tn}, which uses the MIT bag
model for the nucleon. The bag model does not include sea quarks.
It is a confining model, whereas the $\chi$QS model is not.
Additionally, the QMC model calculation, when coupled with a
Relativistic Distorted Wave Impulse Approximation (RDWIA)
calculation \cite{Udias:1999tm} or a Relativistic
Multiple-Scattering Glauber Approximation (RMSGA) calculation
\cite{Ryckebusch:2003fc,Lava:2004mp}, improves the agreement
between theory the TJNAF data \cite{Strauch:2002wu}. With our
study, we hope to reinforce the interpretation of the medium
effect in terms of quark degrees of freedom, as well as provide an
alternate model when the accuracy of the data is improved.

The procedure is much the same as that in Chapter \ref{ch:dist}
for the quark distributions, except that we are looking at the
spatial structure of the nucleon in elastic scattering (after a
Fourier transform) instead of the momentum distributions of the
constituents observed in inelastic scattering. In fact,
Generalized Parton Distributions (GPD) contain both form factors
and quark distributions in the same framework. We will begin our
description of electromagnetic form factors unconventionally
starting with GPDs.

The leading twist GPD is given by the off-forward (incoming
nucleon has a different momentum $k$ than the outgoing one $k'$)
matrix element analogous to Eq.~(\ref{eq:pdf1})
\begin{equation}
F(x, \xi, Q^{2}) =  k^{+} \int \frac{dz}{2\pi}^{-} e^{i x
k^{+}z^{-}} \langle k' | \overline{\psi}(0) \gamma^{+} \psi
(z^{-}) | k \rangle
\end{equation}
which can be written in terms of two functions $H$ and $E$
\begin{equation}
F(x, \xi, Q^{2}) =  \overline{u}(k') \left[ \gamma^{+}
H(x,\xi,Q^{2}) + \frac{i \sigma^{+\mu}q_{\mu}}{2 M_{N} }
E(x,\xi,Q^{2}) \right] u(k)
\end{equation}
The functions $u(k)$ and $\overline{u}(k')$ are the incoming and
outgoing nucleon spinors. The other variables are defined by
\begin{eqnarray}
q & \equiv & k-k' \\
Q^{2} & \equiv & -\bm{q}^{2}\\
\xi & \equiv & \frac{q^{+}}{(k+k')^{+}}
\end{eqnarray}
The parton distributions of Chapter \ref{ch:dist} are given by the
forward limit $k=k'$, so that
\begin{eqnarray}
q(x) & = & H(x,0,0) \\
\Delta q(x) & = & E(x,0,0)
\end{eqnarray}
The electromagnetic form factors are integrals over $x$ of
functions $H$ and $E$
\begin{eqnarray}
\int_{-1}^{1} dx H(x,\xi,Q^{2}) & = & F_{1}(Q^{2}) \\
\int_{-1}^{1} dx E(x,\xi,Q^{2}) & = & F_{2}(Q^{2})
\end{eqnarray}
where the $F_{1,2}$ are the usual Dirac and Pauli form
factors\footnote{There are not to be confused with the structure
functions $F_{1,2}^{(N)}(x)$ discussed Chapters \ref{ch:intro} and
\ref{ch:dist}.}. The dependence on $\xi$ drops out as a special
case of polynomiality: the moments of $H$ and $E$ must be
polynomials in $\xi$. This property is rooted in Lorentz
invariance, and is preserved by the $\chi$QS model
\cite{Schweitzer:2002nm}. The Dirac/Pauli form factors can be
rewritten as the Sachs form factors
\begin{eqnarray}
G_{E}(Q^{2}) & = & F_{1}(Q^{2}) - \frac{Q^{2}}{4 M_{N}} F_{2}(Q^{2})\\
G_{M}(Q^{2}) & = & F_{1}(Q^{2}) + F_{2}(Q^{2})
\end{eqnarray}

\begin{figure}
\centering
\includegraphics[scale=0.7]{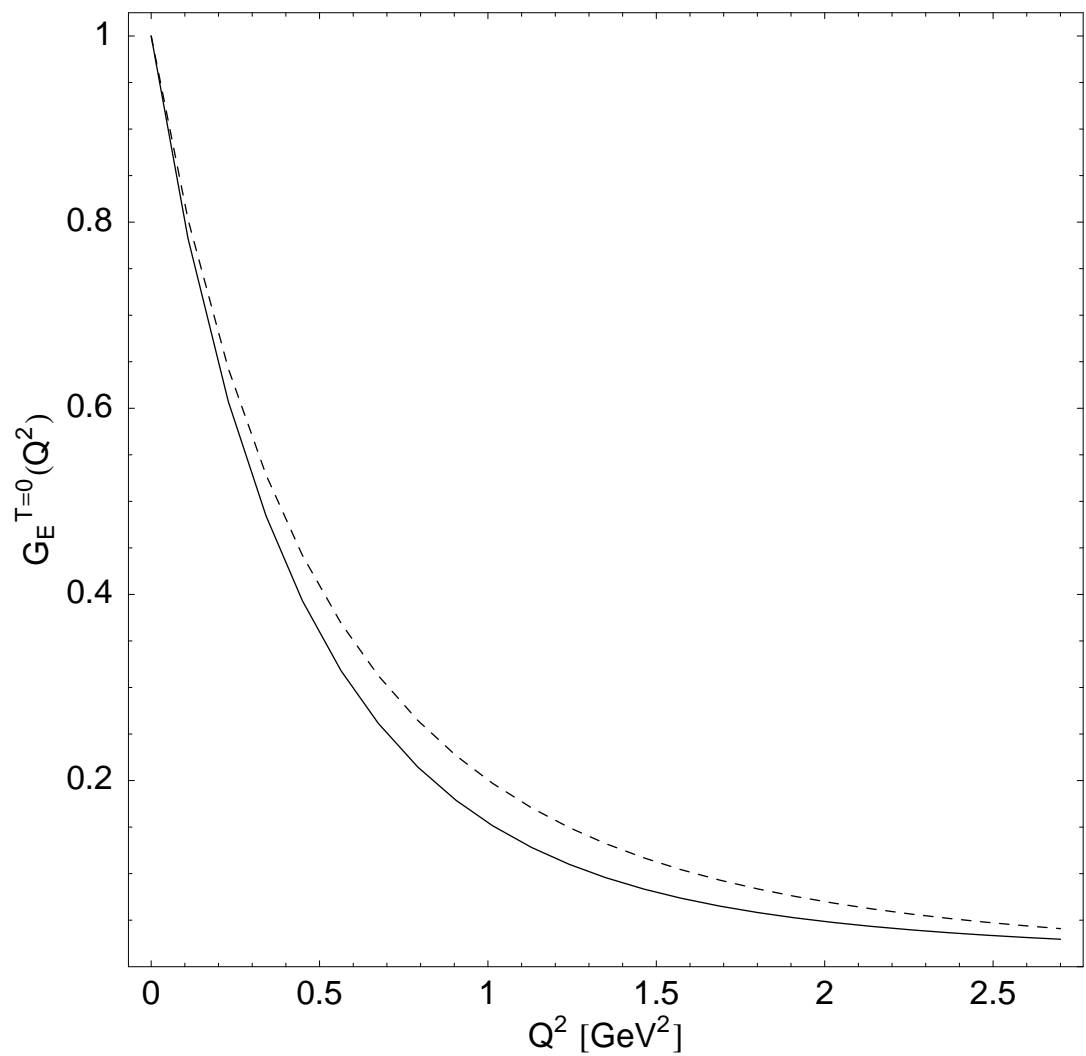}
\caption{The isoscalar electric form factor at nuclear density
(solid) and at zero density (dashes).} \label{fig:eff}
\end{figure}

\begin{figure}
\centering
\includegraphics[scale=0.7]{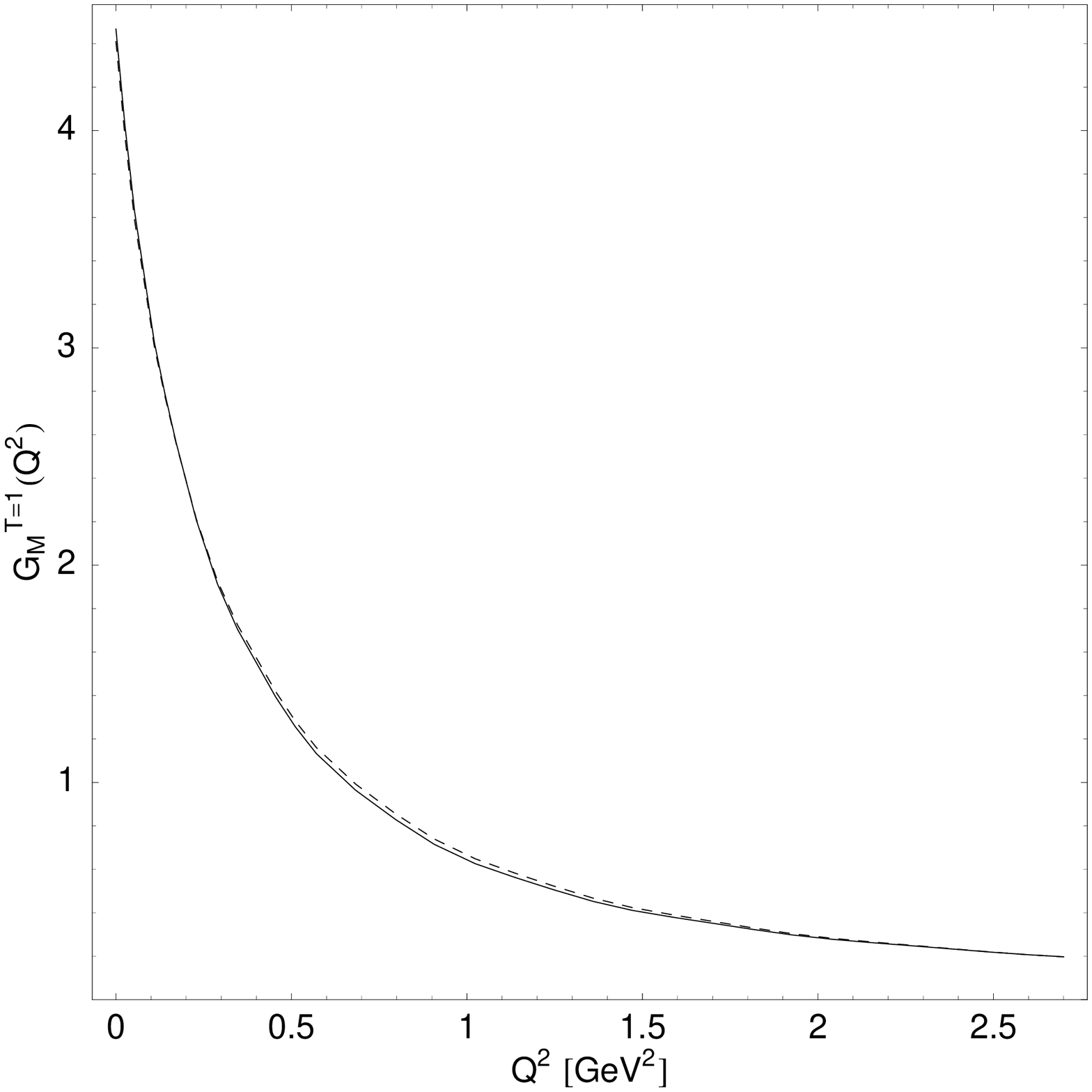}
\caption{The isovector magnetic form factor at nuclear density
(solid) and at zero density (dashes).} \label{fig:mff}
\end{figure}

\begin{figure}
\centering
\includegraphics[scale=0.7]{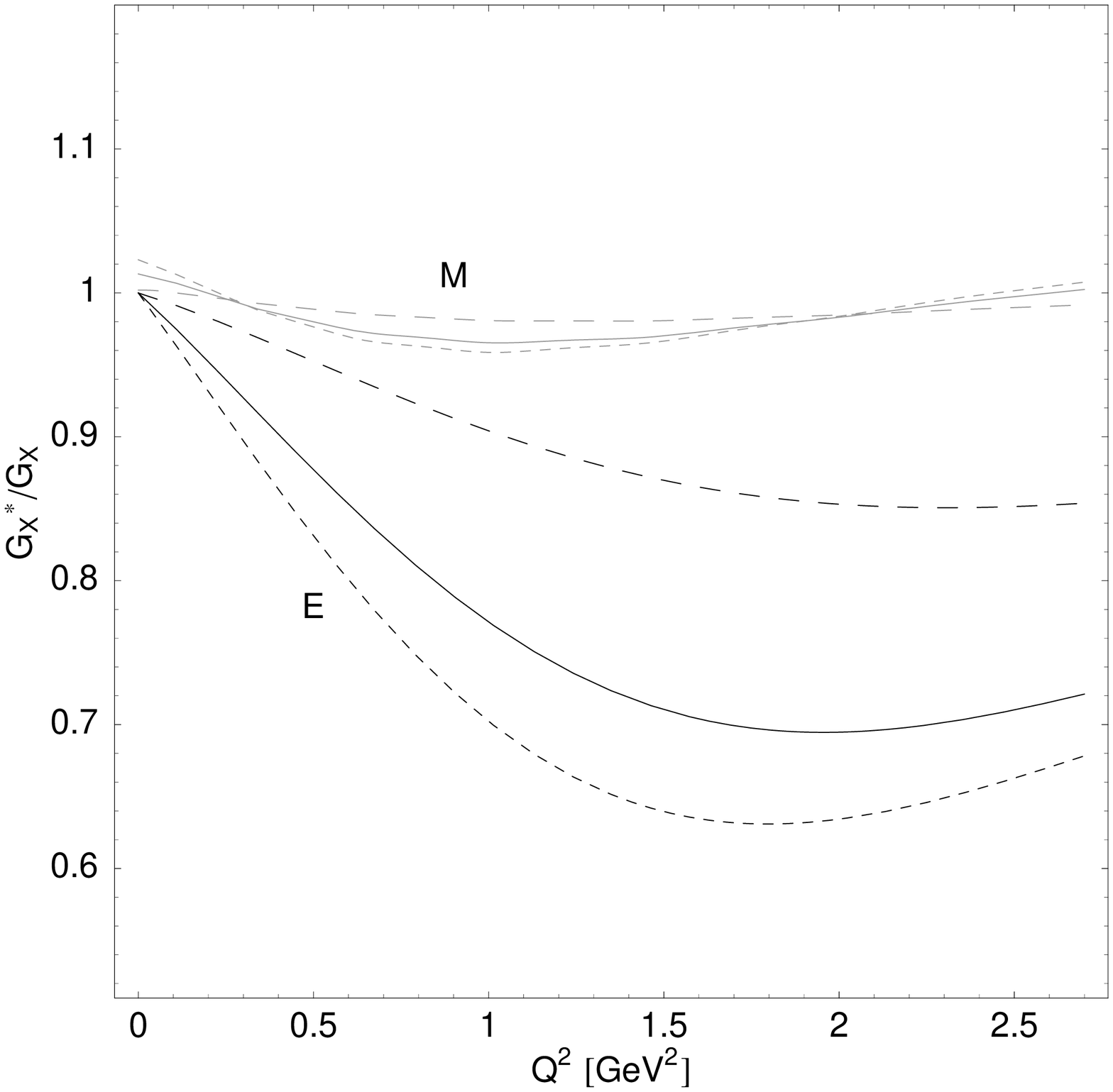}
\caption{The electric (lower three curves) and magnetic (upper
three curves) form factor ratios in Eq.~(\ref{eq:ffr}) for
$0.5\rho_{0}$ (long dashes), $1.0\rho_{0}$ (solid) and
$1.5\rho_{0}$ (short dashes).} \label{fig:ffr}
\end{figure}

\begin{figure}
\centering
\includegraphics[scale=0.7]{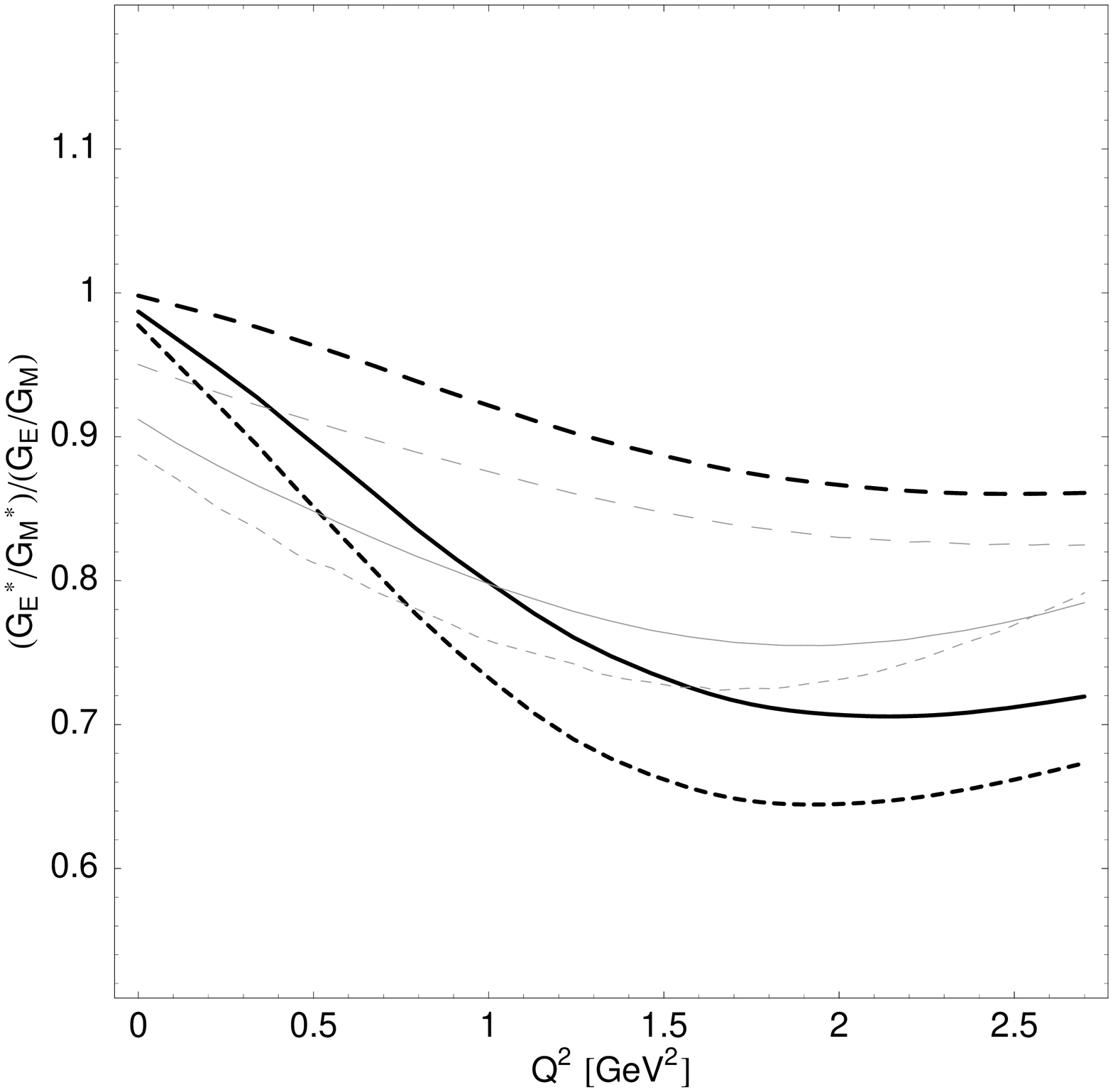}
\caption{The double ratio Eq.~(\ref{eq:ffdr}) of the electric to
magnetic form factors in nuclear matter and in the vacuum from the
$\chi$QS model (heavy) and the QMC model \cite{Lu:1998tn} (light).
Three densities are shown: $0.5\rho_{0}$ (long dashes),
$1.0\rho_{0}$ (solid) and $1.5\rho_{0}$ (short dashes).}
\label{fig:ffdr}
\end{figure}

These electromagnetic form factors can be evaluated in terms of
the quark wave functions, and are derived in
Ref.~\cite{Christov:1995hr}. The formulae are reproduced here,
with a Pauli-Villars regulator, for convenience. To leading order
in $N_{C}$, we have only the isoscalar electric and isovector
magnetic form factors ($G_{X}^{T=0,1} = G_{X}^{p} \pm G_{X}^{n}$)
\begin{subequations}\label{eq:GEM}
\begin{eqnarray}
G_{E}^{T=0}(q^{2}) & = & \frac{N_{C}}{3} \int d\bm{r}\:
e^{i\bm{q}\cdot\bm{r}} \Bigg\{ \sum_{E_{n}\leq E^{v}}
\psi_{n}^{\dag}(\bm{r})\psi_{n}(\bm{r})
-\sum_{E_{n}^{(0)}\leq 0} \psi_{n}^{(0)\dag}(\bm{r})\psi_{n}^{(0)}(\bm{r}) \Bigg\}\label{eq:GE}\\
G_{M}^{T=1}(q^{2}) & = & \frac{N_{C} M_{N}}{3} \varepsilon^{jkl}
\frac{iq^{j}}{|q^{2}|} \int d\bm{r}\: e^{i\bm{q}\cdot\bm{r}}
\Bigg\{ \sum_{E_{n}\leq E^{v}} \psi_{n}^{\dag}(\bm{r})
\gamma^{0}\gamma^{k}\tau^{l} \psi_{n}(\bm{r})\Bigg\} \nonumber\\
& & - PV \label{eq:GM}.
\end{eqnarray}
\end{subequations}
In the nuclear medium, Eqs.~(\ref{eq:GEM}) acquire a dependence on
the Fermi momentum through the wave functions as described in
Chapter \ref{ch:np}. We use Eqs.~(\ref{eq:GE}) and (\ref{eq:GM})
to calculate the form factors, which we present in
Figs.~\ref{fig:eff} and \ref{fig:mff}.
We also present the results in terms of the ratios
\begin{equation}
\frac{G_{E,M}^{T=0,1}(Q^{2},k_{F})}{G_{E,M}^{T=0,1}(Q^{2},0)}
\equiv \frac{G_{X}^{*}(Q^{2})}{G_{X}(Q^{2})}\label{eq:ffr},
\end{equation}
where $X \text{ is } E (T=0) \text{ or } M (T=1)$, and the double
ratio
\begin{equation}
\frac{G_{E}^{*}(Q^{2})/G_{M}^{*}(Q^{2})}{G_{E}(Q^{2})/G_{M}(Q^{2})}\label{eq:ffdr}.
\end{equation}
These ratios are plotted in Figs.~\ref{fig:ffr} and \ref{fig:ffdr}
for $0.5\rho_{0}$, $1.0\rho_{0}$ and $1.5\rho_{0}$.

The electric form factor is dominated by the valence contribution
and shows a dramatic effect, while the magnetic form factor has
equally important contributions from the valence and the sea. The
latter shows almost no change in nuclear matter; it shows only a
1.3\% enhancement of the magnetic moment at nuclear density, and a
2.3\% enhancement at 1.5 times nuclear density. These are
consistent with the constraints of a $<2 \%$ increase set by
Ref.~\cite{Ericson:1986zz}. In contrast, the QMC model
\cite{Lu:1998tn} predicts an enhancement of roughly 5\% for Oxygen
and 10\% for Lead. The influence of the nuclear medium on the
nucleon also causes the root mean square radius of the baryon
density to increase by 3.1\% in the $\chi$QS model. This swelling
is consistent with a $<6 \%$ increase as constrained by
quasi-elastic inclusive electron-nucleus scattering data
\cite{Mckeown:kn}. The effect in the electric form factor
calculated here is comparable to that of the QMC model; the main
difference from that calculation lies in the lack of enhancement
in the magnetic form factor, specifically the practically
unchanged value of the magnetic moment.

While both form factors use the same wave functions, the isovector
magnetic form factor includes an extra weighting by a factor of
the angular momentum of the state (relative to the electric form
factor) due to the $\gamma^{k}$ in Eq.~(\ref{eq:GM}). This extra
factor is not only responsible for making the regularization of
Eq.~(\ref{eq:GM}) necessary, but for making the sea contribution
as important as the valence. In the $\chi$QS model, the orbital
angular momentum carried by the sea is comparable to the orbital
angular momentum carried by the valence quarks
\cite{Wakamatsu:1990ud} (the sum of which make up about 60\% of
the total angular momentum of the nucleon state, with the
remainder belonging to the intrinsic spin of the constituent
quarks).

Conversely, the isoscalar electric form factor (which is finite,
after the vacuum subtraction) does not have as large of a
contribution from the sea. The valence level is the most important
piece, even at $Q^{2}>0$, since the $Q^{2}$ dependence in the form
factors arises from the wave functions \cite{Christov:1995hr}. The
negative Dirac continuum wave functions largely cancel in the
vacuum subtraction in Eq.~(\ref{eq:GE}).

The magnetic form factors are sensitive to the tail of the quark
wave functions, and the mere existence of a tail is due to the
lack of confinement. This is one reason for the discrepancy
between the current results and the QMC model \cite{Lu:1998tn},
but the primary source is due to the resistance to change of the
sea. The former accounts for only a few percent of the difference;
it is the latter that is our most important result. We see that
the role of antiquarks is again prevalent as in our previous work
\cite{Smith:2003hu}.

The double ratio obtained in Fig.~\ref{fig:ffdr} has the same
trend as the QMC model \cite{Lu:1998tn}, but differs in the
details. Since we obtain a similar double ratio, we expect to have
similar results if we compare these results with the polarization
transfer data \cite{Strauch:2002wu}. This requires one to take the
final state and relativistic effects into account through the use
of the RDWIA \cite{Udias:1999tm} or the RMSGA \cite{Lava:2004mp},
which accounts for a few percent of the discrepancy between the
results for bound and free protons. A RMSGA calculation for the
Helium reaction studied in Ref.~\cite{Strauch:2002wu} has been
done with these $\chi$QS model results \cite{Lava:private}, and it
delivers remarkably similar results to the same calculation done
with the QMC model \cite{Lava:2004mp}. The $\chi$QS model predicts
a smaller deviation than the QMC model from a Relativistic Plane
Wave Impulse Approximation (RPWIA) calculation, which is taken as
a baseline in Ref.~\cite{Strauch:2002wu}. While it slightly
worsens the agreement with the data at $Q^{2}\lesssim 1$, the
differences are of the same order of magnitude as the current
experimental error, and both models under predict the observed
deviation from a RPWIA calculation. At higher $Q^{2}$, the two
models produce nearly identical results for Helium.

We ignore potentially important corrections that follow from
integrating out the rotational and translational zero modes of the
soliton that are suppressed by $1/N_{C}$. These corrections break
the $N-\Delta$ degeneracy, and improve the agreement of the vacuum
form factors with experiment \cite{Christov:1995hr}. More relevant
to the calculation presented here, the rotational corrections do
not affect the $Q^2$ dependence, but instead affect the
normalization of the form factors \cite{Christov:1995hr}. However,
there is no reason at that level to continue to ignore quantum
fluctuations of the the pion field (quark loops, also suppressed
by $1/N_{C}$) or center of mass corrections, and treat the soliton
as a purely self-consistent, static mean field. We will save this
difficult problem for the future.

We have calculated the electric and magnetic form factors at
leading order in $N_{C}$ at nuclear density using the $\chi$QS
model. Our results help validate the apparent success of the QMC
model in describing the polarization transfer experiment
\cite{Strauch:2002wu,Lu:1998tn}, and provide a counterpoint
consistent with constraints on the nucleon radius and magnetic
moment in nuclei to be distinguished when finer resolution becomes
available in the data. In fact, the difference between the
$\chi$QS model double ratio and the QMC model \cite{Lu:1998tn} is
roughly the size as the current experimental error. Specifically,
data on the bound nucleon magnetic form factor at low $Q^{2}$,
particularly the magnetic moment, could serve to determine the
role of sea quarks in nuclei.

\chapter{Summary}
\label{ch:summary}

The Chiral Quark-Soliton model provides a intuitive, qualitative
treatment that maintains consistency with all of the free nucleon
properties calculated by others
\cite{Diakonov:2000pa,Christov:1995vm}. It gives a reasonable
description of nuclear saturation properties, reproduces the EMC
effect, and satisfies the constraints on the nuclear sea obtained
from Drell-Yan experiments with only two free parameters: $g_{s}$
and $g_{v}$.

The central mechanism behind these calculated effects is that the
nuclear medium provides an attractive scalar interaction that
modifies the nucleon wave function. This is the same mechanism
involved in other calculations
\cite{Saito:yw,Benesh:2003fk,Cloet:2005rt,Birse:1993nr}. The major
difference between this calculation and others is the inclusion of
antiquarks at the model scale, and we see that they play an
important role in the modifications, maintaining not only
consistency with constraints such as those imposed by Drell-Yan
experiments \cite{Alde:im} and the theoretical and experimental
understanding of magnetic moments \cite{Ericson:1986zz}, but also
positivity and baryon and momentum sum rules. The latter is deeply
connected with the inclusion of antiquarks and a regulator which
does not interfere with the completeness of the set of quark
states or the equivalence of summing over occupied and unoccupied
states in the spectrum.

Dynamical rescaling, a moderately successful attempt at
simultaneous phenomenological description of the EMC effect and
related Drell-Yan experiments, automatically maintains the baryon
and momentum sum rules due to the structure of the renormalization
group equations. A change in the nucleon size in nuclei was
believed to be the scale driving the effect. These are all
properties of the $\chi$QS model, with the nucleon size swelling
3.1\% in the nuclear medium. Although there is no special scale
(one could point to the relative sizes of the nuclear scalar
density and chiral condensate), we have a mechanism in the scalar
interaction of the quarks in a nucleon with the medium. It may be
possible that the spectrum of quark states in a bound nucleon is
the exactly the same as in a free nucleon at a smaller model scale
$M_{PV}$, the Pauli-Villars mass, and hence one would have a
`derivation' of dynamical rescaling. It also may just be a
coincidence that any model that maintains the momentum and baryon
sum rules with some swelling of the nucleon describes the EMC
effect and Drell-Yan data within the accuracy of current
experiments.

The Wilson coefficients $c$ in Eq.~(\ref{eq:ope}) in the OPE are
dependent on the starting renormalization scale. Nuclear
interactions are effective QCD interactions which renormalize a
free nucleon into a bound one. In the handbag diagram
Fig.~\ref{fig:parton}, all three quarks, not just the spectator
quarks, in the intermediate state are renormalized by QCD from the
same starting scale, and hence all feel the effects of nuclear
interactions. The picture presented in Chapter \ref{ch:dist} is
consistent with this viewpoint, while other approaches
\cite{Cloet:2005rt,Saito:yw} treat the struck quark as if it had
the same normalization point as one in a free nucleon.

We have also calculated the electric and magnetic form factors at
nuclear density. Our results help validate the apparent success of
the QMC model in describing the polarization transfer experiment
\cite{Strauch:2002wu,Lu:1998tn}, and provide a counterpoint to be
distinguished when finer resolution becomes available in the data.
In fact, the difference between the $\chi$QS model double ratio
and the QMC model \cite{Lu:1998tn} is roughly the size as the
current experimental error.

These successful descriptions of the medium modification of quark
distributions and form factors give weight to the model and point
to future experiments. For example, data on the bound nucleon
magnetic form factor at low $Q^{2}$, particularly the magnetic
moment, where the difference between the QMC and $\chi$QS models
is the largest, may be able to differentiate between the two
models. We also expect the results for the polarized quark
distribution to manifest themselves in future experiments with
polarized nuclei. These experiments could serve to help determine
the role of both quarks and antiquarks in nuclei.

\nocite{*}
\bibliographystyle{plain}
\bibliography{mythesis}

\begin{thebibliography}{100}

\bibitem{Airapetian:1998wi}
A.~Airapetian et~al.
\newblock Measurement of the proton spin structure function $g_{1}^{(p)}$ with
  a pure hydrogen target.
\newblock {\em Phys.~Lett.~B}, 442:484, 1998.

\bibitem{Alde:im}
D.M. Alde et~al.
\newblock Nuclear dependence of dimuon production at 800 {GeV}. {FNAL}-772
  experiment.
\newblock {\em Phys.~Rev.~Lett.}, 64:2479, 1990.

\bibitem{Alkofer:1994ph}
R.~Alkofer, H.~Reinhardt, and H.~Weigel.
\newblock Baryons as chiral solitons in the {N}ambu-{J}ona-{L}asinio model.
\newblock {\em Phys.~Rept.}, 265:139, 1996.

\bibitem{Altarelli:1977zs}
G.~Altarelli and G.~Parisi.
\newblock Asymptotic freedom in parton language.
\newblock {\em Nucl. Phys. B}, 126:298, 1977.

\bibitem{Anthony:2000fn}
P.L. Anthony et~al.
\newblock Measurements of the ${Q}^{2}$ dependence of the proton and neutron
  spin structure functions $g_{1}^{(p)}$ and $g_{1}^{(n)}$.
\newblock {\em Phys.~Lett.~B}, 493:19, 2000.

\bibitem{Arneodo:1992wf}
M.~Arneodo.
\newblock Nuclear effects in structure functions.
\newblock {\em Phys.~Rept.}, 240:301, 1994.

\bibitem{Aubert:1983xm}
J.J. Aubert et~al.
\newblock The ratio of the nucleon structure functions $\textit{F}_{2}^{(N)}$
  for \text{Iron} and \text{Deuterium}.
\newblock {\em Phys.~Lett.~B}, 123:275, 1983.

\bibitem{Barmin:2003vv}
V.V. Barmin et~al.
\newblock Observation of a baryon resonance with positive strangeness in
  ${K}^{+}$ collisions with {Xe} nuclei.
\newblock {\em Phys.~Atom.~Nucl.}, 66:1715, 2003.

\bibitem{Benesh:2003fk}
C.J. Benesh, T.~Goldman, and G.J. Stephenson.
\newblock Valence quark distribution in {A} = 3 nuclei.
\newblock {\em Phys.~Rev.~C}, 68:045208, 2003.

\bibitem{Benhar:1999up}
O.~Benhar, V.R. Pandharipande, and I.~Sick.
\newblock Density dependence of the {EMC} effect.
\newblock {\em Phys.~Lett.~B}, 469:19, 1999.

\bibitem{Berger:1986dr}
E.L. Berger.
\newblock Nuclear effects in massive lepton pair production.
\newblock {\em Nucl.~Phys.~B}, 267:231, 1986.

\bibitem{Bickerstaff:1984ax}
R.P. Bickerstaff, M.C. Birse, and G.A. Miller.
\newblock Disentangling explanations of deep inelastic lepton nucleus
  scattering by lepton pair production.
\newblock {\em Phys.~Rev.~Lett.}, 53:2532, 1984.

\bibitem{Birse:1993hu}
M.~C. Birse.
\newblock Relativistic mean fields and the {EMC} effect.
\newblock {\em Phys. Lett.}, B299:186--188, 1993.

\bibitem{Birse:1993nr}
M.C. Birse.
\newblock The axial charge of a nucleon in matter.
\newblock {\em Phys.~Lett.~B}, 316:472, 1993.

\bibitem{Birse:hu}
M.C. Birse.
\newblock Relativistic mean fields and the {EMC} effect.
\newblock {\em Phys.~Lett.~B}, 299:186, 1993.

\bibitem{Birse:1983gm}
M.C. Birse and M.K. Banerjee.
\newblock A chiral soliton model of nucleon and delta.
\newblock {\em Phys.~Lett.~B}, 136:284, 1984.

\bibitem{Bjorken:1968dy}
J.~D. Bjorken.
\newblock Asymptotic sum rules at infinite momentum.
\newblock {\em Phys. Rev.}, 179:1547--1553, 1969.

\bibitem{Bjorken:1966jh}
J.D. Bjorken.
\newblock Applications of the chiral ${U}(6) \times {U}(6)$ algebra of current
  densities.
\newblock {\em Phys.~Rev.}, 148:1467, 1966.

\bibitem{Bjorken:1969mm}
J.D. Bjorken.
\newblock Inelastic scattering of polarized leptons from polarized nucleons.
\newblock {\em Phys.~Rev.~D}, 1:1376, 1970.

\bibitem{Blaizot:tw}
J.P. Blaizot.
\newblock Nuclear compressibilities.
\newblock {\em Phys.~Rept.}, 64:171, 1980.

\bibitem{Blunden:1999gq}
P.G. Blunden, M.~Burkardt, and G.A. Miller.
\newblock Light-front nuclear physics: Mean field theory for finite nuclei.
\newblock {\em Phys.~Rev.~C}, 60:055211, 1999.

\bibitem{Blunden:1996kc}
P.G. Blunden and G.A. Miller.
\newblock Quark-meson coupling model for finite nuclei.
\newblock {\em Phys.~Rev.~C}, 54:359, 1996.

\bibitem{lcrev4}
S.J. Brodsky and G.P. Lepage.
\newblock Exclusive processes in quantum chromodynamics.
\newblock In A.~Mueller, editor, {\em Perturbative Quantum Chromodynamics}.
  World Scientific, Singapore, 1989.

\bibitem{lcrev1}
S.J. Brodsky, H.C. Pauli, and S.S. Pinsky.
\newblock Quantum chromodynamics and other field theories on the light cone.
\newblock {\em Phys.\ Rept.}, 301:299, 1998.

\bibitem{lcrev3}
M.~Burkardt.
\newblock Light front quantization.
\newblock {\em Adv.~Nucl.~Phys.}, 23:1, 1996.

\bibitem{Christov:1995vm}
C.V. Christov et~al.
\newblock Baryons as non-topological chiral solitons.
\newblock {\em Prog.~Part.~Nucl.~Phys.}, 37:91, 1996.

\bibitem{Christov:1995hr}
C.V. Christov, A.Z. Gorski, K.~Goeke, and P.V. Pobylitsa.
\newblock Electromagnetic form-factors of the nucleon in the chiral quark
  soliton model.
\newblock {\em Nucl.~Phys.~A}, 592:513, 1995.

\bibitem{Cloet:2005rt}
I.C. Cloet, W.~Bentz, and A.W. Thomas.
\newblock Spin-dependent structure functions in nuclear matter and the
  polarized {EMC} effect.
\newblock {\em arXiv:nucl-th/0504019}, 2005.

\bibitem{Close:1984zn}
F.~E. Close, R.~L. Jaffe, R.~G. Roberts, and Graham~G. Ross.
\newblock Change of confinement scale in nuclei: Predictions for structure
  functions confront electroproduction data.
\newblock {\em Phys. Rev.}, D31:1004, 1985.

\bibitem{Close:1987ay}
F.E. Close, R.G. Roberts, and G.G. Ross.
\newblock Factorization scale independence, the connection between alternative
  explanations of the {EMC} effect and {QCD} predictions for nuclear
  properties.
\newblock {\em Nucl.~Phys.~B}, 296:582, 1988.

\bibitem{Cohen:1991nk}
T.D. Cohen, R.J. Furnstahl, and D.K. Griegel.
\newblock Quark and gluon condensates in nuclear matter.
\newblock {\em Phys.~Rev.~C}, 45:1881, 1992.

\bibitem{complications}
J.R. Cooke and G.A. Miller.
\newblock Deuteron binding energies and form factors from light front field
  theory.
\newblock {\em Phys.~Rev.~C}, 66:034002, 2002.

\bibitem{deGroot:yb}
J.G. de~Groot et~al.
\newblock {QCD} analysis of charged current structure functions.
\newblock {\em Phys.~Lett.~B}, 82:456, 1979.

\bibitem{Diakonov:2002fq}
D.~Diakonov.
\newblock Instantons at work.
\newblock {\em Prog. Part. Nucl. Phys.}, 51:173--222, 2003.

\bibitem{Diakonov:2005eq}
D.~Diakonov.
\newblock Relativistic mean field approximation to baryons.
\newblock {\em Eur.~Phys.~J.~A}, 24{S}1:3, 2005.

\bibitem{Diakonov:1997vc}
D.~Diakonov et~al.
\newblock Unpolarized and polarized quark distributions in the large-${N}_{C}$
  limit.
\newblock {\em Phys.~Rev.~D}, 56:4069, 1997.

\bibitem{Diakonov:1983hh}
D.~Diakonov and V.~Y. Petrov.
\newblock Instanton based vacuum from {F}eynman variational principle.
\newblock {\em Nucl. Phys.}, B245:259, 1984.

\bibitem{Diakonov:1987ty}
D.~Diakonov, V.~Y. Petrov, and P.~V. Pobylitsa.
\newblock A chiral theory of nucleons.
\newblock {\em Nucl. Phys.}, B306:809, 1988.

\bibitem{Diakonov:2000pa}
D.~Diakonov and V.Y. Petrov.
\newblock Nucleons as chiral solitons.
\newblock In M.~Shifman, editor, {\em At the Frontier of Particle Physics,
  Vol.~1}, page 359. World Scientific, Singapore, 2001.

\bibitem{Diakonov:1996sr}
D.~Diakonov, V.Y. Petrov, P.~Pobylitsa, M.~V. Polyakov, and C.~Weiss.
\newblock Nucleon parton distributions at low normalization point in the large
  ${N}_{C}$ limit.
\newblock {\em Nucl. Phys.}, B480:341--380, 1996.

\bibitem{Diakonov:1997mm}
D.~Diakonov, V.Y. Petrov, and M.V. Polyakov.
\newblock Exotic anti-decuplet of baryons: Prediction from chiral solitons.
\newblock {\em Z.~Phys.~A}, 359:305, 1997.

\bibitem{Dokshitzer:1977sg}
Y.~L. Dokshitzer.
\newblock Calculation of the structure functions for deep inelastic scattering
  and $e^{+} e^{-}$ annihilation by perturbation theory in quantum
  chromodynamics. ({I}n {R}ussian).
\newblock {\em Sov. Phys. JETP}, 46:641--653, 1977.

\bibitem{Dzierba:2004db}
A.~R. Dzierba, C.~A. Meyer, and A.~P. Szczepaniak.
\newblock Reviewing the evidence for pentaquarks.
\newblock {\em hep-ex/0412077}, 2004.
\newblock Talk at 1st Meeting of the APS Topical Group on Hadronic Physics
  (GHP2004).

\bibitem{Ericson:1984vt}
M.~Ericson and A.W. Thomas.
\newblock Evidence for an enhanced nuclear sea from the proton - nucleus
  {D}rell-{Y}an process.
\newblock {\em Phys.~Lett.~B}, 148:191, 1984.

\bibitem{Ericson:1986zz}
T.~E.~O. Ericson and A.~Richter.
\newblock Limits on the scaling of nuleon magnetic moments in nuclei.
\newblock {\em Phys. Lett.}, B183:249, 1987.

\bibitem{Frank:1995pv}
M.R. Frank, B.K. Jennings, and G.A. Miller.
\newblock The role of color neutrality in nuclear physics--{M}odifications of
  nucleonic wave functions.
\newblock {\em Phys.~Rev.~C}, 54:920, 1996.

\bibitem{Frankfurt:1985ui}
L.L. Frankfurt and M.I. Strikman.
\newblock On the normalization of nucleus spectral function and the {EMC}
  effect.
\newblock {\em Phys.~Lett.~B}, 183:254, 1987.

\bibitem{Frankfurt:nt}
L.L. Frankfurt and M.I. Strikman.
\newblock Hard nuclear processes and microscopic nuclear structure.
\newblock {\em Phys.~Rept.}, 160:235, 1988.

\bibitem{Gasser:1983yg}
J.~Gasser and H.~Leutwyler.
\newblock Chiral perturbation theory to one loop.
\newblock {\em Ann. Phys.}, 158:142, 1984.

\bibitem{Geesaman:1995yd}
D.F. Geesaman, K.~Saito, and A.W. Thomas.
\newblock The nuclear {EMC} effect.
\newblock {\em Ann.~Rev.~Nucl.~Part.~Sci.}, 45:337, 1995.

\bibitem{lcrev2}
S.D. Glazek, editor.
\newblock {\em Theory of hadrons and light-front {QCD}}.
\newblock World Scientific, Singapore, 1994.

\bibitem{Gomez:1993ri}
J.~Gomez et~al.
\newblock Measurement of the {A}-dependence of deep inelastic electron
  scattering.
\newblock {\em Phys.~Rev.~D}, 49:4348, 1994.

\bibitem{Gribov:1972ri}
V.~N. Gribov and L.~N. Lipatov.
\newblock Deep inelastic $e$-$p$ scattering in perturbation theory.
\newblock {\em Sov. J. Nucl. Phys.}, 15:438--450, 1972.

\bibitem{Gross:1973id}
D.~J. Gross and F.~Wilczek.
\newblock Ultraviolet behavior of non-{A}belian gauge theories.
\newblock {\em Phys. Rev. Lett.}, 30:1343--1346, 1973.

\bibitem{Gross:1992pi}
F.~Gross and S.~Liuti.
\newblock Role of nuclear binding in the {EMC} effect.
\newblock {\em Phys.~Rev.~C}, 45:1374, 1992.

\bibitem{Guzey:1999rq}
V.~Guzey and M.~Strikman.
\newblock Nuclear effects in $g_{1}^{(A)}(x,{Q}^{2})$ at small $x$ in deep
  inelastic scattering on $^{7}\text{Li}$ and $^{3}\text{He}$.
\newblock {\em Phys.~Rev.~C}, 61:014002, 2000.

\bibitem{Hagiwara:fs}
K.~Hagiwara et~al.
\newblock Review of particle physics.
\newblock {\em Phys.~Rev.~D}, 66:010001, 2002.

\bibitem{lcrev9}
A.~Harindranath.
\newblock An introduction to light-front dynamics for pedestrians.
\newblock In J.P. Vary and F.~W$\text{\"o}$lz, editors, {\em Light-Front
  Quantization and Non-Perturbative {QCD}}. Int.~Inst.~of Theoretical and
  Applied Physics, Ames, IA, 1997.

\bibitem{Hirai:1997gb}
M.~Hirai, S.~Kumano, and M.~Miyama.
\newblock Numerical solution of ${Q}^{2}$ evolution equations for polarized
  structure functions.
\newblock {\em Comput. Phys. Commun.}, 108:38, 1998.

\bibitem{cjht}
C.J. Horowitz.
\newblock {TIMORA}.
\newblock In K.~Langanke, J.A. Maruhn, and S.E. Koonin, editors, {\em
  Computational Nuclear Physics. Vol.~1: Nuclear Structure}. Springer-Verlag,
  New York, 1991.

\bibitem{Horowitz:1981xw}
C.J. Horowitz and B.D. Serot.
\newblock Selfconsistent {H}artree description of finite nuclei in a
  relativistic quantum field theory.
\newblock {\em Nucl.~Phys.~A}, 368:503, 1981.

\bibitem{HvH}
N.M. Hugenholtz and L.~van Hove.
\newblock A theorem on the single particle energy in a {F}ermi gas with
  interaction.
\newblock {\em Physica}, 24:363, 1958.

\bibitem{Ioffe:kw}
B.L. Ioffe.
\newblock Calculation of baryon masses in quantum chromodynamics.
\newblock {\em Nucl.~Phys.~B}, 188:317, 1981.
\newblock \textit{Erratum-ibid.}~191:591,1981.

\bibitem{Ji:2004gf}
X.~Ji.
\newblock Generalized parton distributions.
\newblock {\em Ann. Rev. Nucl. Part. Sci.}, 54:413--450, 2004.

\bibitem{lcrev5}
X.-D. Ji.
\newblock Quantum field theory in light front coordinates.
\newblock {\em Comments Nucl.~Part.~Phys.}, 21:123, 1992.

\bibitem{Jung:1988jw}
H.~Jung and G.A. Miller.
\newblock Nucleonic contribution to lepton nucleus deep inelastic scattering.
\newblock {\em Phys.~Lett.~B}, 200:351, 1988.

\bibitem{Kahana:1984be}
S.~Kahana and G.~Ripka.
\newblock Baryon density of quarks coupled to a chiral field.
\newblock {\em Nucl.~Phys.~A}, 419:462, 1984.

\bibitem{Kahana:dx}
S.~Kahana, G.~Ripka, and V.~Soni.
\newblock Soliton with valence quarks in the chiral invariant sigma model.
\newblock {\em Nucl.~Phys.~A}, 415:351, 1984.

\bibitem{Kubota:1999hx}
T.~Kubota, M.~Wakamatsu, and T.~Watabe.
\newblock On the chiral quark soliton model with {P}auli-{V}illars
  regularization.
\newblock {\em Phys.~Rev.~D}, 60:014016, 1999.

\bibitem{Lava:private}
P.~Lava, J.~Ryckebusch, and S.~Strauch.
\newblock {P}rivate communication.

\bibitem{Lava:2004mp}
P.~Lava, J.~Ryckebusch, B.~Van~Overmeire, and S.~Strauch.
\newblock Polarization transfer in $^{4}\text{He}(\vec{e},e'\vec{p})$ and
  $^{16}\text{O}(\vec{e},e'\vec{p})$ in a relativistic {G}lauber model.
\newblock {\em Phys.~Rev.~C}, 70:014605, 2005.

\bibitem{Lu:1997mu}
D.H. Lu, A.W. Thomas, K.~Tsushima, A.G. Williams, and K.~Saito.
\newblock In-medium electron nucleon scattering.
\newblock {\em Phys.~Lett.~B}, 417:217, 1998.

\bibitem{Lu:1998tn}
D.H. Lu, K.~Tsushima, A.W. Thomas, A.G. Williams, and K.~Saito.
\newblock Electromagnetic form factors of the bound nucleon.
\newblock {\em Phys.~Rev.~C}, 60:068201, 1999.

\bibitem{Mcallister:1956ng}
R.~W. Mcallister and R.~Hofstadter.
\newblock Elastic scattering of 188-{M}e{V} electrons from the proton and the
  alpha particle.
\newblock {\em Phys. Rev.}, 102:851--856, 1956.

\bibitem{Mckeown:kn}
R.D. Mckeown.
\newblock Precise determination of the nucleon radius in $^{3}\text{He}$.
\newblock {\em Phys.~Rev.~Lett.}, 56:1452, 1986.

\bibitem{Miller:1997cr}
G.A. Miller.
\newblock Light front treatment of nuclei: Formalism and simple applications.
\newblock {\em Phys.~Rev.~C}, 56:2789, 1997.

\bibitem{previous}
G.A. Miller.
\newblock Light front quantization: A technique for relativistic and realistic
  nuclear physics.
\newblock {\em Prog.~Part.~Nucl.~Phys.}, 45:83, 2000.

\bibitem{Miller:1999ap}
G.A. Miller and R.~Machleidt.
\newblock Infinite nuclear matter on the light front: Nucleon-nucleon
  correlations.
\newblock {\em Phys.~Rev.~C}, 60:035202, 1999.

\bibitem{Miller:2001tg}
G.A. Miller and J.R. Smith.
\newblock Return of the {EMC} effect.
\newblock {\em Phys.~Rev.~C}, 65:015211, 2002.
\newblock \textit{Erratum-ibid.}~66:049903,2002.

\bibitem{Miyama:1995bd}
M.~Miyama and S.~Kumano.
\newblock Numerical solution of ${Q}^{2}$ evolution equations in a brute force
  method.
\newblock {\em Comput. Phys. Commun.}, 94:185--215, 1996.

\bibitem{Nakano:2003qx}
T.~Nakano et~al.
\newblock Evidence for a narrow ${S} = +1$ baryon resonance in photo-production
  from neutron.
\newblock {\em Phys.~Rev.~Lett.}, 91:012002, 2003.

\bibitem{lcrev8}
R.J. Perry.
\newblock Hamiltonian light front field theory and quantum chromodynamics.
\newblock In V.~E.~Herscovitz \textit{et al}, editor, {\em Hadron Physics 94:
  Topics on the structure and interactions of hadronic systems}. World
  Scientific, Singapore, 1994.

\bibitem{Piller:1999wx}
G.~Piller and W.~Weise.
\newblock Nuclear deep-inelastic lepton scattering and coherence phenomena.
\newblock {\em Phys.~Rept.}, 330:1, 2000.

\bibitem{Politzer:1973fx}
H.~D. Politzer.
\newblock Reliable perturbative results for strong interactions?
\newblock {\em Phys. Rev. Lett.}, 30:1346--1349, 1973.

\bibitem{Praszalowicz:1989dh}
M.~Praszalowicz and G.~Valencia.
\newblock Quark models and chiral lagrangians.
\newblock {\em Nucl. Phys. B}, 341:27--49, 1990.

\bibitem{RuizArriola:2004en}
E.~Ruiz~Arriola, P.~O. Bowman, and W.~Broniowski.
\newblock Landau-gauge condensates from the quark propagator on the lattice.
\newblock {\em Phys. Rev.}, D70:097505, 2004.

\bibitem{RuizArriola:ex}
E.~Ruiz~Arriola, C.V. Christov, and K.~Goeke.
\newblock Medium effects on nucleon properties.
\newblock {\em Phys.~Lett.~B}, 225:22, 1989.

\bibitem{Rutherford:1911}
E.~Rutherford.
\newblock The scattering of alpha and beta particles by matter and the
  structure of the atom.
\newblock {\em Phil. Mag.}, 21:669--688, 1911.

\bibitem{Ryckebusch:2003fc}
J.~Ryckebusch, D.~Debruyne, P.~Lava, S.~Janssen, B.~Van~Overmeire, and
  T.~Van~Cauteren.
\newblock Relativistic formulation of {G}lauber theory for ${A}(e,e' p)$
  reactions.
\newblock {\em Nucl.~Phys.~A}, 728:226, 2003.

\bibitem{Saito:yw}
K.~Saito and A.W. Thomas.
\newblock A microscopic understanding of the structure functions of finite
  nuclei.
\newblock {\em Nucl.~Phys.~A}, 574:659, 1994.

\bibitem{Saito:1994ki}
K.~Saito and A.W. Thomas.
\newblock A quark-meson coupling model for nuclear and neutron matter.
\newblock {\em Phys.~Lett.~B}, 327:9, 1994.

\bibitem{Saito:2001gv}
K.~Saito, M.~Ueda, K.~Tsushima, and A.~W. Thomas.
\newblock Structure functions of unstable lithium isotopes.
\newblock {\em Nucl. Phys. A}, 705:119--152, 2002.

\bibitem{Sargsian:2002wc}
M.M. Sargsian et~al.
\newblock Hadrons in the nuclear medium.
\newblock {\em J.~Phys.~G}, 29:R1, 2003.

\bibitem{Schweitzer:2002nm}
P.~Schweitzer, S.~Boffi, and M.~Radici.
\newblock Polynomiality of off-forward distribution functions in the chiral
  quark soliton model.
\newblock {\em Phys. Rev. D}, 66:114004, 2002.

\bibitem{Serot:1986ey}
B.D. Serot and J.D. Walecka.
\newblock The relativistic nuclear many body problem.
\newblock {\em Adv.~Nucl.~Phys.}, 16:1, 1986.

\bibitem{Serot:1997xg}
B.D. Serot and J.D. Walecka.
\newblock Recent progress in quantum hadrodynamics.
\newblock {\em Int.~J.~Mod.~Phys.~E}, 6:515, 1997.

\bibitem{Sick:1992pw}
I.~Sick and D.~Day.
\newblock The {EMC} effect of nuclear matter.
\newblock {\em Phys.~Lett.~B}, 274:16, 1992.

\bibitem{Smith:2002ci}
J.R. Smith and G.A. Miller.
\newblock Return of the {EMC} effect: Finite nuclei.
\newblock {\em Phys.~Rev.~C}, 65:055206, 2002.

\bibitem{Smith:2003hu}
J.R. Smith and G.A. Miller.
\newblock Chiral solitons in nuclei: Saturation, {EMC} effect, and
  {D}rell-{Y}an experiments.
\newblock {\em Phys.\ Rev.\ Lett.}, 91:212301, 2003.

\bibitem{Smith:2004dn}
J.R. Smith and G.A. Miller.
\newblock Chiral solitons in nuclei: Electromagnetic form factors.
\newblock {\em Phys.~Rev.~C}, 70:065205, 2004.

\bibitem{Soper:1971wn}
D.E. Soper.
\newblock Massive quantum electrodynamics in the infinite momentum frame.
\newblock {\em Phys.~Rev.~D}, 4:1620, 1971.

\bibitem{Strauch:2002wu}
S.~Strauch et~al.
\newblock Polarization transfer in the $^{4}\text{He}(\vec{e},e'
  \vec{p})^{3}\text{H}$ reaction up to $\textit{Q}^{2} = 2.6 \text{
  (GeV/c)}^{2}$.
\newblock {\em Phys.~Rev.~Lett.}, 91:052301, 2003.

\bibitem{'tHooft:1976fv}
G.~'t~Hooft.
\newblock Computation of the quantum effects due to a four- dimensional
  pseudoparticle.
\newblock {\em Phys. Rev.}, D14:3432--3450, 1976.

\bibitem{'tHooft:1976up}
G.~'t~Hooft.
\newblock Symmetry breaking through {B}ell-{J}ackiw anomalies.
\newblock {\em Phys. Rev. Lett.}, 37:8--11, 1976.

\bibitem{Thomas:1998eu}
A.W. Thomas, D.H. Lu, K.~Tsushima, A.G. Williams, and K.~Saito.
\newblock Recent results from {QMC} relevant to {TJNAF}.
\newblock {\em arXiv:nucl-th/9807027}, 1998.
\newblock Invited talk at TJNAF.

\bibitem{Udias:1999tm}
J.M. Udias, J.A. Caballero, E.~Moya~de Guerra, J.E. Amaro, and T.W. Donnelly.
\newblock Quaiselastic scattering from relativistic bound nucleons:
  Transverse-longitudinal response.
\newblock {\em Phys.~Rev.~Lett.}, 83:5451, 1999.

\bibitem{Wakamatsu:1990ud}
M.~Wakamatsu and H.~Yoshiki.
\newblock A chiral quark model of the nucleon.
\newblock {\em Nucl.~Phys.~A}, 524:561, 1991.

\bibitem{Walecka:qa}
J.D. Walecka.
\newblock A theory of highly condensed matter.
\newblock {\em Annals Phys.}, 83:491, 1974.

\bibitem{Witten:1979}
E.~Witten.
\newblock Baryons in the $1/{N}$ expansion.
\newblock {\em Nucl. Phys. B}, 160:57, 1979.

\bibitem{Yakhshiev:2002sr}
U.T. Yakhshiev, U.G. Meissner, and A.~Wirzba.
\newblock Electromagnetic form factors of bound nucleons revisited.
\newblock {\em Eur.~Phys.~J.~A}, 16:569, 2003.

\bibitem{Yan:1973qf}
T.-M. Yan.
\newblock Quantum field theories in the infinite momentum frame 3.
  {Q}uantization of coupled spin one fields.
\newblock {\em Phys.~Rev.~D}, 7:1760, 1973.

\bibitem{lcrev6}
W.-M. Zhang.
\newblock Light front dynamics and light front {QCD}.
\newblock {\em Chinese J.\ Phys.}, 32:717, 1994.

\end{thebibliography}

\appendix
\raggedbottom\sloppy

\chapter{The Connection to Chiral Perturbation Theory and Nonperturbative QCD}
\label{ch:npqp}

In this Appendix, we perform a low-momentum expansion of the
$\chi$QS model Lagrangian (\ref{eq:lagrangian}) with a non-zero
current quark mass, and compare it to the Lagrangian of Gasser and
Leutwyler. This has been done before in
Ref.~\cite{Praszalowicz:1989dh}, for example; we present a
slightly different approach here for completeness. We also fit
quenched lattice data for the dynamical quark mass to a sum of
propagators (as appearing in Pauli-Villars regulation of
divergences).

\section{The Chiral Lagrangian}

We will first derive Eq.~(\ref{eq:GLp2}) with current quark mass
$m_{0} \neq 0$. Then we will proceed to higher order in the
derivative expansion.

\subsection{The $\mathcal{O}(p^{2})$ Lagrangian}

The constituent quark model Lagrangian in Euclidean space with
(anti)quark fields $\psi^{\dag},\psi$ that we use is
\begin{equation}
\mathcal{L} =  \psi^{\dag} ( i \partial \!\!\!\!\!\:/\, + i
\hat{M} U^{\gamma_{5}} + i m_{0}) \psi - \psi^{\dag} ( i \partial
\!\!\!\!\!\:/\, + i \hat{M}) \psi \label{eq:lagrangian-appendix}
\end{equation}
where
\begin{equation}
U^{\gamma_{5}} \equiv
U\frac{1-\gamma_{5}}{2}+U^{\dag}\frac{1+\gamma_{5}}{2}
\end{equation}
The constituent quark mass, $\hat{M}$, is a function of momentum,
falling to zero as the momentum goes to infinity. The second term
represents the vacuum subtraction. One can integrate over the
constituent quark fields in the Euclidean partition function to
obtain the effective action
\begin{equation}
\tilde{S}[U] = - N_{C} \Big\{ \log \det ( i \partial
\!\!\!\!\!\:/\, + i M U^{\gamma_{5}} + i m_{0}) -\log \det ( i
\partial \!\!\!\!\!\:/\, + i M) \Big\}
\end{equation}
We wish to look at the real part of this effective action.
\begin{equation}
\text{Re }\tilde{S}[U] = - \frac{N_{C}}{2}  \log \det \Big[1 -
\frac{1}{-\partial^{2}+\hat{M}^{2}}(\hat{M}
\partial \!\!\!\!\!\:/\, U^{\gamma_{5}} - m_{0} \hat{M} [U+U^{\dag}]) \Big]
\end{equation}
To simplify our calculation we introduce the `regulated'
derivative and scalar source operators
\begin{subequations}
\label{eq:operators}
\begin{eqnarray}
\tilde{\nabla}^{\mu} & \equiv &
\frac{\hat{M}}{-\partial^{2}+\hat{M}^{2}}\; \partial^{\mu}\\
\tilde{\chi} & \equiv & 2 m_{0}
\frac{\hat{M}}{-\partial^{2}+\hat{M}^{2}}
\end{eqnarray}
\end{subequations}
Note that $\tilde{\nabla}^{\mu}U$ is leading $\mathcal{O}(p)$ and
$\tilde{\chi}$ is leading $\mathcal{O}(p^{2})$ from the current
quark mass. After rewriting our effective action in terms of this
new notation, and using $\log \det = \text{Tr} \log$ where
$\text{Tr}$ means a functional as well as Dirac and flavor trace,
we obtain
\begin{equation}
\text{Re} \tilde{S}[U] = - \frac{N_{C}}{2} \text{Tr} \log \Big[1 -
\tilde{\nabla} \!\!\!\!\!\!\:/\, U^{\gamma_{5}} -
\frac{1}{2}(\tilde{\chi}^{\dag}U+U^{\dag}\tilde{\chi}) \Big]
\end{equation}
Then one can expand in derivatives of the field
$\tilde{\nabla}^{\mu}U$ and current quark mass, and obtain the
momentum expansion of chiral perturbation theory. One essentially
expands
\[
\log (1-\varepsilon) = -\varepsilon -\frac{1}{2}\varepsilon^{2}
-\frac{1}{3}\varepsilon^{3} + \cdots
\]

To order $\mathcal{O}(p^{2})$, after taking the Dirac trace, we
obtain the chiral Lagrangian in leading order in $N_{C}$
\begin{equation}
\mathcal{L}^{(N_{C},2)} = N_{C} \text{tr} \tilde{\nabla}^{\mu}U
\tilde{\nabla}_{\mu}U^{\dag} + N_{C}
\text{tr}(\tilde{\chi}^{\dag}U+U^{\dag}\tilde{\chi})
\end{equation}
Where the lowercase trace is over flavor indices, and functional
traces over the operators defined in Eq.~(\ref{eq:operators}).

\subsection{The $\mathcal{O}(p^{4})$ Lagrangian}

To order $\mathcal{O}(p^{4})$ and leading order in $N_{C}$, we
obtain the operators
\begin{eqnarray}
\mathcal{L}^{(N_{C},4)} & = & \frac{N_{C}}{2} \text{tr} \Big(
\tilde{\nabla}^{\mu}U\tilde{\nabla}^{\nu}U^{\dag}\tilde{\nabla}_{\nu}U\tilde{\nabla}_{\mu}U^{\dag}
+\tilde{\nabla}^{\mu}U\tilde{\nabla}_{\mu}U^{\dag}\tilde{\nabla}^{\nu}U\tilde{\nabla}_{\nu}U^{\dag}\Big)\nonumber\\
& & - \frac{N_{C}}{2} \text{tr} \Big(
\tilde{\nabla}^{\mu}U\tilde{\nabla}^{\nu}U^{\dag}\tilde{\nabla}_{\mu}U\tilde{\nabla}_{\nu}U^{\dag}
\Big)\nonumber\\
& & + N_{C} \text{tr}  \Big( \tilde{\nabla}^{\mu}U
\tilde{\nabla}_{\mu}U^{\dag}
(\tilde{\chi}^{\dag}U+U^{\dag}\tilde{\chi}) \Big)\nonumber\\
& & + \frac{N_{C}}{2} \text{tr}  \Big(
(\tilde{\chi}^{\dag}U+U^{\dag}\tilde{\chi})^{2} \Big)
\end{eqnarray}

We will use the identity, for $N_{f} = 3$
\begin{equation}
\text{tr} (ABAB) = -2 \text{tr} (A^{2}B^{2}) + \frac{1}{2}
\text{tr} (A^{2})\text{tr}(B^{2}) + \text{tr} (AB)^{2}
\end{equation}
to put this Lagrangian in the form of Gasser and Leutwyler
\cite{Gasser:1983yg}
\begin{eqnarray}
\mathcal{L}^{(N_{C},4)} & = & \frac{N_{C}}{4} (\text{tr}
\tilde{\nabla}^{\mu}U\tilde{\nabla}_{\mu}U^{\dag})^{2}
+\frac{N_{C}}{2}
\text{tr}(\tilde{\nabla}^{\mu}U\tilde{\nabla}^{\nu}U^{\dag})\text{tr}(\tilde{\nabla}_{\mu}U\tilde{\nabla}_{\nu}U^{\dag})
\nonumber\\
& & - N_{C} \text{tr}
(\tilde{\nabla}^{\mu}U\tilde{\nabla}_{\mu}U^{\dag}\tilde{\nabla}^{\nu}U\tilde{\nabla}_{\nu}U^{\dag})
\nonumber\\
& & + N_{C} \text{tr}  \Big( \tilde{\nabla}^{\mu}U
\tilde{\nabla}_{\mu}U^{\dag}
(\tilde{\chi}^{\dag}U+U^{\dag}\tilde{\chi}) \Big)\nonumber\\
& & + \frac{N_{C}}{2} \text{tr}
(\tilde{\chi}^{\dag}U\tilde{\chi}^{\dag}U+U^{\dag}\tilde{\chi}U^{\dag}\tilde{\chi})
\end{eqnarray}
This means that to leading order in $N_{C}$, the constituent quark
Lagrangian Eq.~(\ref{eq:lagrangian-appendix}) gives the Gasser and
Leutwyler coefficients $L_{1}$, $L_{2}$, $L_{3}$, $L_{5}$, and
$L_{8} \neq 0$ and $L_{4} = L_{6} = L_{7} = 0$. When one takes the
functional trace over the operator, one gets
\begin{equation}
L_{i} = \lambda_{i} N_{C} \int \frac{d^{4}k}{(2\pi)^{2}}
\Bigg(\frac{\hat{M}(k)}{k^{2}+\hat{M}(k)^{2}}\Bigg)^{\alpha_{i}}\label{eq:li}
\end{equation}
where $\lambda_{i}$ and $\alpha_{i}$ are given in Table
\ref{table:integrals}.
\begin{table}\centering
\caption{Constants appearing in the integrals for $L_{i}$.}
\label{table:integrals}
\begin{tabular}{cccccc}
$i =$ & 1 & 2 & 3 & 5 & 8 \\
\hline $\lambda_{i}$ & $\frac{1}{4}$ & $\frac{1}{2}$ & 1 & 1 & $\frac{1}{2}$ \\
$\alpha_{i}$ & 4 & 4 & 4 & 3 & 2
\end{tabular}
\end{table}

\section{The Quark Propagator}

\begin{figure}\centering
\includegraphics[scale=0.7]{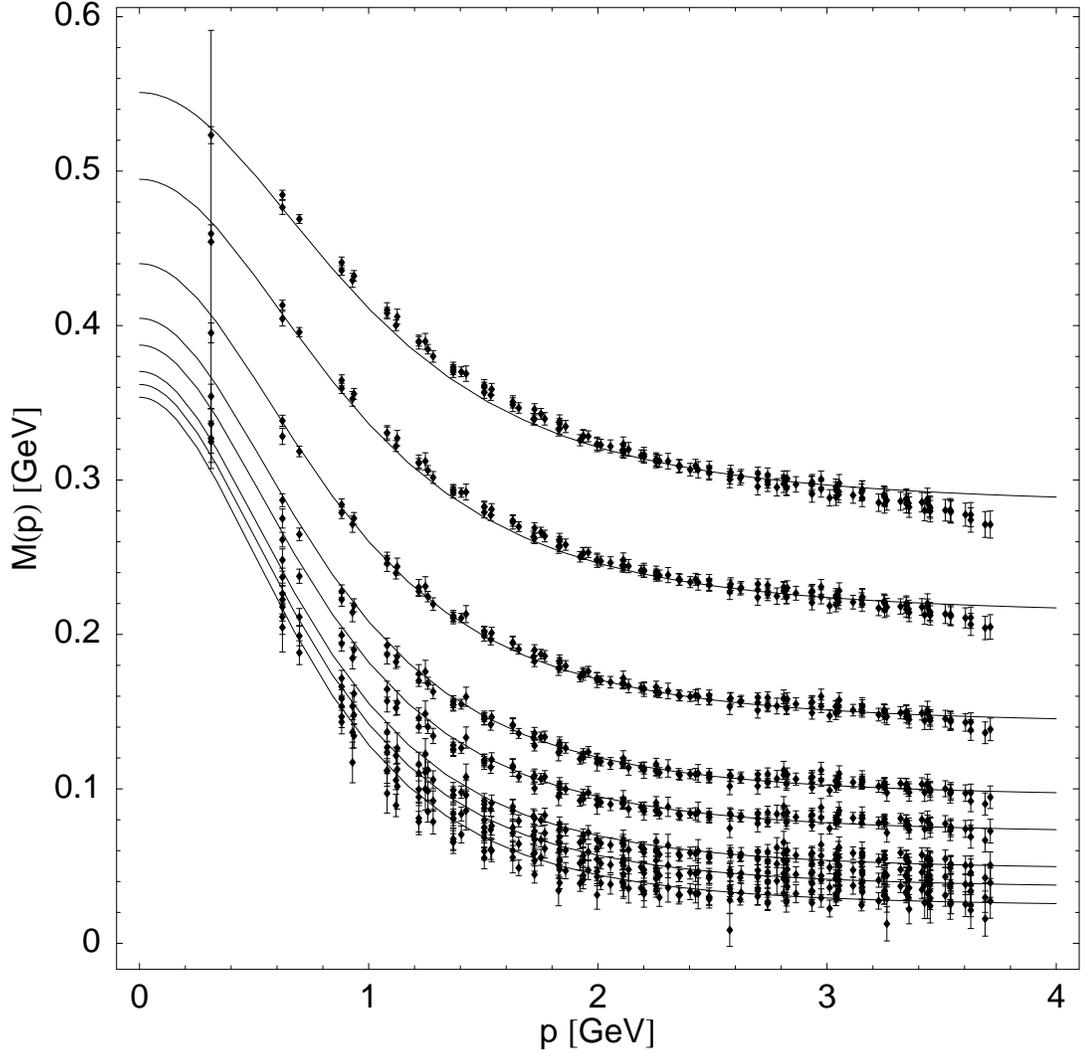}
\caption{Fits of Eq.~(\ref{eq:fit}) for $n=3$ to the quenched
lattice data \cite{RuizArriola:2004en} for the dynamical quark
mass $M(p)$ at current quark masses $m_{q} a = 0.012$, 0.018,
0.024, 0.036, 0.048, 0.072, 0.108, and 0.144, where $a$ is the
lattice spacing.}\label{fig:mp_lat}
\end{figure}

The (quenched) lattice data \cite{RuizArriola:2004en} gives the
quark propagator in terms of functions $Z(k)$, the wave function
renormalization, and $M(k)$, the constituent quark mass. We fit
this data with a sum of propagators as the parameterization (much
like what appears in Pauli-Villars regularization)
\begin{equation}
\frac{Z(k)}{ik\!\!\!\!\!\:/\, + i M(k)} \rightarrow \sum_{i=0}^{n}
Z_{i}\frac{\mu_{0}}{\mu_{i}}\frac{1}{ik\!\!\!\!\!\:/\, + i \mu_{i}
+i m_{0}}\label{eq:fit}
\end{equation}
Regularizing the divergent integrals appearing in the momentum
expansion (specifically the quadratically divergent chiral
condensate and the logarithmically divergent pion decay constant
in $\mathcal{L}^{(2)}$), along with the restriction that
$Z(k\rightarrow\infty)\rightarrow 1$, requires
\begin{subequations}
\begin{eqnarray}
\sum_{i=0}^{n} Z_{i}\frac{\mu_{0}}{\mu_{i}} & = & 1\\
\sum_{i=0}^{n} Z_{i} & = & 0\\
\sum_{i=0}^{n} Z_{i}\mu_{i} & = & 0
\end{eqnarray}
\end{subequations}
We have taken $n=3$, so we have 5 parameters to fit the lattice
data. A fit of the momentum dependent quark mass $M(p)$ with
Eq.~(\ref{eq:fit}) for $n=3$ to the quenched lattice data for
$m_{q} \simeq 0.14 \text{ GeV}$ are shown in
Fig.~\ref{fig:mp_lat}. The same parameters do a good job of
reproducing the lattice data for several other quark masses, so
the extrapolation to the chiral limit shown in Fig.~\ref{fig:mp}
is expected to allow a good qualitative comparison with the
instanton model and the single Pauli-Villars subtraction used in
the self-consistent calculations.
\begin{figure}\centering
\includegraphics[scale=0.7]{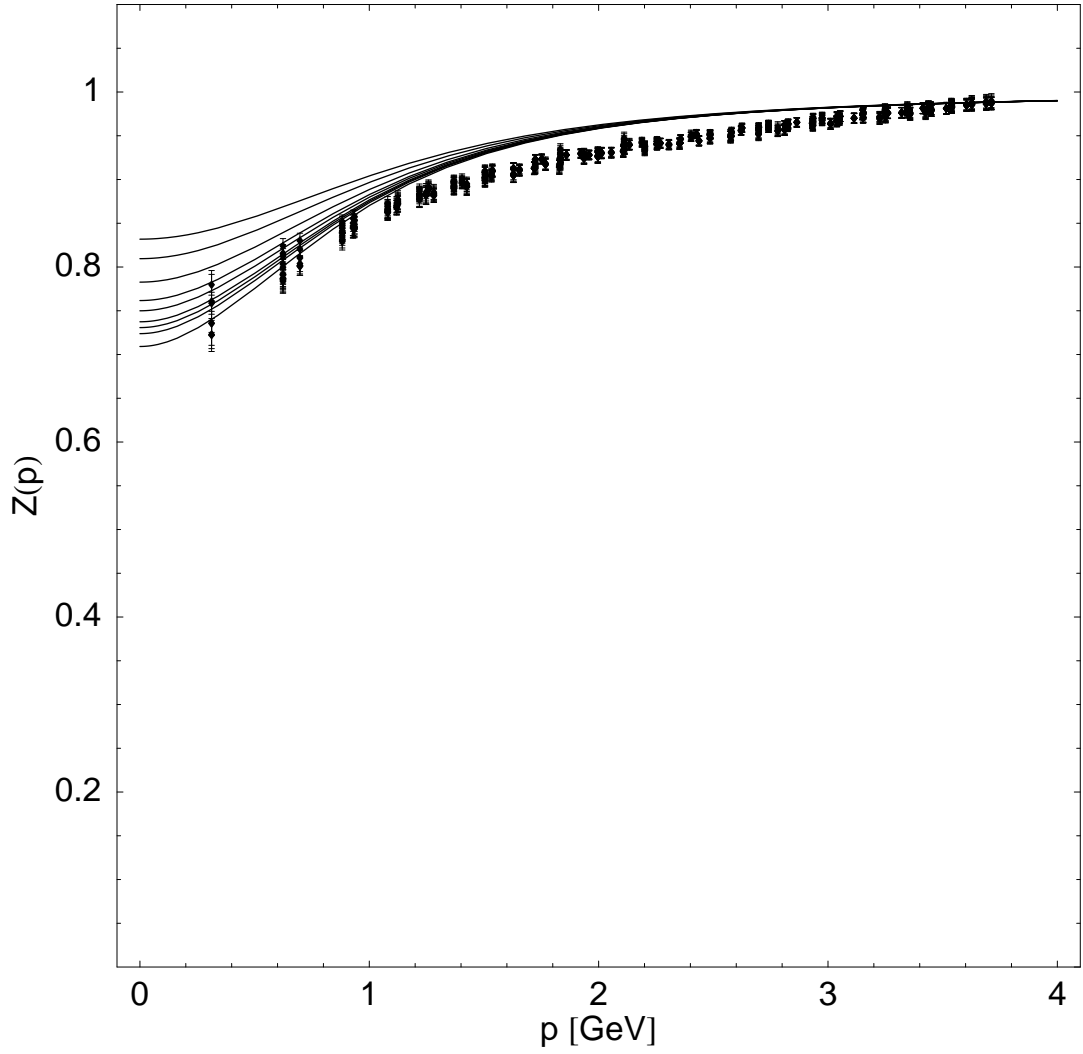}
\caption{Fits of Eq.~(\ref{eq:fit}) for $n=3$ to the quenched
lattice data for the wavefunction renormalization function $Z(p)$
at various quark masses. The data falls slightly below the fits.
This can be attributed to an effect of the finite volume of the
lattice \cite{RuizArriola:2004en}. }\label{fig:zp_lat}
\end{figure}

\chapter{The Kahana-Ripka Basis}
\label{ch:KR}

The Kahana-Ripka basis is defined as the following set of
functions inside a spherical box with radius $L$:
\begin{equation}
\phi_{nGjlMst} (\bm{r}) = N_{n,l}
j_{l}(p_{n}r)\Xi^{M}_{Gjlst}(\bm{n}) \label{eq:basisfunction}
\end{equation}
where $j_{l}(k_n r)$ is a spherical Bessel function. The momenta
$p_{n}$ are defined by
\[
j_{G}(p_{n} L) = 0
\]
where $p_{n}<\Lambda$ and $\Lambda$ is the basis momentum cut-off,
and $G$ is the grand spin, defined below in Eq.~(\ref{eq:grand}).
The normalization $N_{n,l}$ is defined by a property of the zeros
of spherical Bessel functions. If $a$ and $a'$ are two zeros of
$j_{l}(x)$
\[
j_{l}(a) = j_{l}(a') = 0
\]
then
\begin{eqnarray}
\int_{0}^{1} dy \, y^{2} j_{l}(a y) j_{l}(a' y) & = & \int_{0}^{1}
dy \, y^{2} j_{l\pm 1}(a y) j_{l \pm 1}(a' y)\\
& = & \delta_{a,a'} \frac{1}{2} [j_{l\pm 1}(a)]^{2}
\end{eqnarray}
The quantum number $n$ is a good quantum number for the free Dirac
Hamiltonian Eq.~(\ref{eq:H0}). The Dirac Hamiltonian in the
$\chi$QS model Eq.~(\ref{eq:hamiltonian}) derived from the
Lagrangian Eq.~(\ref{eq:lagrangian}) commutes with the grand spin
$G$, the sum of the spin, angular momentum and isospin operators
\begin{equation}
G = J + T = L + S + T, \label{eq:grand}
\end{equation}
therefore we construct
eigenstates $\Xi$ of $G$. The functions $\Xi$ are
spinor-isospinor-spherical harmonics, defined by the sum of
Clebsch-Gordan coefficients $C_{j_{1}m_{1}j_{2}m_{2}}^{jm}$,
spherical harmonics $Y$ and (iso)spinors $\chi$
\[
\Xi^{M}_{Gjlst}(\bm{n}) = \sum_{j_{3}} \sum_{t_{3}} \sum_{l_{3}}
\sum_{s_{3}} C^{GM}_{jj_{3}tt_{s}} C^{jj_{3}}_{ll_{3}ss_{s}}
Y_{l}^{l_{3}}(\bm{n}) \chi_{s_{3}} \chi_{t_{3}}
\]
The functions Eq.~(\ref{eq:basisfunction}) are organized into
eight (four in the case of $G=0$) basis functions: four with
parity $P = (-1)^{l} = (-1)^{G}$ and four with $P = (-1)^{l} =
(-1)^{G+1}$. The eight can also be separated into two groups of
four with $l = G+1, G \: (\times 2)$ and $G-1$. There are only
four functions for $G = 0$ since $l = G - 1 $ and $j = G - 1/2$
cannot be satisfied. The $\Xi$ are eight element vectors,
corresponding to the four of the Dirac spinor and the two of the
isospinor. For an explicit example, the basis states with $G^{P} =
0^{+}$ are
\[
\phi_{n0\frac{1}{2}0}(\bm{r}) = \frac{1}{\sqrt{8\pi}} N_{n,0}
j_{0}(p_{n} r) \left(
\begin{array}{c}
\left( \begin{array}{c} 0\\ 0 \end{array} \right)\vspace{1mm}\\
\left( \begin{array}{c} 0\\ 0 \end{array} \right)\vspace{1mm}\\
\left( \begin{array}{c} 0\\ -1 \end{array} \right)\vspace{1mm}\\
\left( \begin{array}{c} 1\\ 0 \end{array} \right)
\end{array} \right)
\]
\[
\phi_{n0\frac{1}{2}1}(\bm{r}) = \frac{1}{\sqrt{8\pi}} N_{n,1}
j_{1}(p_{n} r) \left(
\begin{array}{c}
\left( \begin{array}{c} e^{-i\phi}\sin \theta\\ -\cos \theta \end{array} \right)\vspace{1mm}\\
\left( \begin{array}{c} -\cos \theta \\ -e^{-i\phi}\sin \theta \end{array} \right)\vspace{1mm}\\
\left( \begin{array}{c} 0\\ 0 \end{array} \right)\vspace{1mm}\\
\left( \begin{array}{c} 0\\ 0 \end{array} \right)
\end{array} \right)
\]
with $N_{n,l} = \sqrt{2}/j_{l+1}(p_{n}L)$. They satisfy
$l=G=j-1/2$ and $l=G+1=j+1/2$ respectively.

An explicit example of the momenta for a Kahana-Ripka basis with
cutoff $\Lambda = 600$ MeV and box size $L = 5$ fm is in Table
\ref{tab:krbasis}. These represent a finite basis of 1576
functions. The largest basis used to calculate the numerical
results presented here has cutoff $\Lambda = 5000$ MeV and box
size $L = 6$ fm, for which there are $\sim 1.7 \times 10^{6}$
functions. Symmetry considerations immediately reduce this number
to $\sim 20,000$ depending on the specific matrix elements
calculated.
\begin{table}
\centering \caption{The momenta for each grand spin $G$ for the
1576 Kahana-Ripka basis functions, with cut-off and box size
$\Lambda = 600 $ MeV and $L = 5$ fm, respectively. The number of
basis functions for each $G$ for $\Lambda = 600$ MeV (I), and
$\Lambda = 5000$ MeV (II) are shown in the far right two columns.}
\label{tab:krbasis}
\begin{tabular}{cccccrr}
$G$ & $p_{n=1}$ & $p_{2}$ & $p_{3}$ & $p_{4}$ & Total states &\\
    & [MeV] & [MeV] & [MeV] & [MeV] & (I) & (II) \\
\hline
0  & 123.984 & 247.968 & 371.953 & 495.937 & 16  & 192 \\
1  & 177.334 & 304.880 & 430.336 & 555.128 & 96  & 1128\\
2  & 227.457 & 358.938 & 486.330 &         & 120 & 1880\\
3  & 275.782 & 411.116 & 540.598 &         & 168 & 2576\\
4  & 322.928 & 461.939 & 593.546 &         & 216 & 3312\\
5  & 369.321 & 511.729 &         &         & 176 & 3960\\
6  & 414.893 & 560.700 &         &         & 208 & 4680\\
7  & 460.049 &         &         &         & 120 & 5280\\
8  & 504.793 &         &         &         & 136 & 5984\\
9  & 549.194 &         &         &         & 152 & 6536\\
10 & 593.302 &         &         &         & 168 & 7224\\
\hline
  &  &  & &  & 1576 & 42752\\
\end{tabular}
\end{table}

\chapter{DGLAP Evolution}
\label{ch:evolve}

Quantum Chromodynamics tells us how to calculate parton
distributions at another scale $Q^{2}$ if they are given at a
scale $Q_{0}^{2}$ through the
Dokshitzer-Gribov-Lipatov-Altarelli-Parisi (DGLAP) evolution
equations \cite{Dokshitzer:1977sg,Gribov:1972ri,Altarelli:1977zs}.
In Chapter \ref{ch:dist}, we make use of QCD evolution to not only
obtain the (un)polarized distributions at $Q^{2} = 10\text{
GeV}^{2}$ where DIS takes place, but to smooth out the effect of
the finite KR basis. First, we must write down the polarized
distribution in its nonsinglet form, and the unpolarized
distribution in its singlet form; these are more convenient for
the application of the DGLAP equations.

The unpolarized singlet distribution is given in terms of the
distributions of up and down quarks as
\begin{eqnarray}
q^{S}(x) & = & \sum_{i} q_{i}(x)+\bar{q}_{i}(x)\\
& = & u(x)+d(x)+\bar{u}(x)+\bar{d}(x) \label{eq:qS}.
\end{eqnarray}
The polarized nonsinglet distribution is
\begin{eqnarray}
\Delta q^{NS}(x) & = & \sum_{i}
\frac{e_{i}^{2}-\langle e_{i}^{2} \rangle}{\langle e_{i}^{2} \rangle} \left(\Delta q_{i}(x)+\Delta \bar{q}_{i}(x)\right)\\
& = & \frac{3}{5}\left(\Delta u(x)- \Delta d(x)+ \Delta \bar{u}(x)
- \Delta \bar{d}(x)\right) + \mathcal{O}(N_{C}^{0})
\label{eq:qNS}.
\end{eqnarray}
Note that Eqs.~(\ref{eq:qS}) and (\ref{eq:qNS}) are normalized to
the baryon number and axial coupling, respectively, which are both
$\mathcal{O}(N_{C})$. The combinations $u-d$ and $\Delta u +
\Delta d$ are normalized to the isospin and spin, respectively;
these are $\mathcal{O}(N_{C}^{0})$. The gluon distribution is
denoted $g(x,Q^{2})$.

If we define
\[
t\equiv -\frac{2}{\beta_{0}} \log
\frac{\alpha_{s}(Q^{2})}{\alpha_{s}(Q_{0}^{2})}
\]
and
\[
p\otimes q (x)= \int_{x}^{1} \frac{dy}{y} p(x/y) q(y)
\]
then the DGLAP equations are given by
\begin{eqnarray}
\frac{d}{dt}\Delta q^{NS}(x,t) & = & P_{qq} \otimes \Delta
q^{NS}(x,t)\\
\frac{d}{dt} \bm{q}(x,t) & = & \bm{P} \otimes \bm{q}(x,t)
\end{eqnarray}
where
\[
\bm{q} = \left( \begin{array}{c} q^{S}\\ g \end{array} \right)
\qquad \bm{P} = \left( \begin{array}{cc} P_{qq} & 2 N_{f} P_{qg}  \\
P_{gq} & P_{gg} \end{array} \right)
\]
The splitting functions $P_{ij}$ to leading order in $\alpha_{s}$
are
\begin{eqnarray}
P_{qq}(x) & = & C_{F} \left[
\frac{1+x^{2}}{(1-x)_{+}}+\frac{3}{2}\delta(1-x) \right] \\
P_{qg}(x) & = & T_{R} \left[
x^{2}+(1-x^{2}) \right] \\
P_{gq}(x) & = & C_{F} \left[
\frac{1+(1-x)^{2}}{x} \right] \\
P_{gg}(x) & = & 2 C_{G} \left[ \frac{x}{(1-x)_{+}} + \frac{1-x}{x}
+ x(1-x)
\left(\frac{11}{12}-\frac{1}{3}\frac{N_{f}T_{R}}{C_{G}}\right)\delta(1-x)
\right]
\end{eqnarray}
where the plus-distribution is defined by its integral with an
arbitrary function $f$(x)
\[
\int_{0}^{1} dx \frac{f(x)}{(1-x)_{+}} = \int_{0}^{1} dx
\frac{f(x)-f(1)}{(1-x)}
\]

The DGLAP evolution equations are solved here with the brute force
method following Refs.~\cite{Miyama:1995bd,Hirai:1997gb}, with
$N_{t} = 200$ points in the scale variable $t$, $\Delta t_{j} =
t_{j+1}-t_{j}$, and $N_{x} = 1000$ points in $x$, $\Delta x_{k} =
x_{k}-x_{k-1}$. We then have, for the nonsinglet polarized
distribution
\[
\Delta q^{NS}(x_{i},t_{j+1}) = \Delta q^{NS}(x_{i},t_{j}) + \Delta
t_{j} \sum_{k=i}^{N_{x}} \frac{\Delta x_{k}}{x_{k}}
P_{qq}(x_{i}/x_{k}) q^{NS}(x_{k},t_{j}).
\]
and similarly for the singlet unpolarized distribution.

\chapter{The Failure of Relativistic Nuclear Mean Field Theory}
\label{ch:finite}

We have shown \cite{Miller:2001tg,Smith:2002ci} that the use of
the relativistic mean field approximation, and the assumption that
the structure of the nucleon is not modified by effects of the
medium, to describe infinite nuclear matter leads to no
appreciable binding effect. The failure was encapsulated in terms
of the Hugenholtz-van Hove theorem \cite{HvH} which states that
the average nuclear binding energy per nucleon is equal to the
binding energy of a nucleon at the top of the Fermi sea. The light
front version of this theorem is obtained from the requirement
that, in the nuclear rest frame, the expectation values of the
total plus and minus momentum are equal. The original version of
the theorem was obtained in a non-relativistic theory in which
nucleons are the only degrees of freedom. Here, the mesons are
important and the theory is relativistic, but the theorem still
holds. This theorem can be shown to restrict \cite{Miller:2001tg}
the plus momentum carried by nucleons to be the mass of the
nucleus, which in turn implies that the probability for a nucleon
to have a plus momentum $k^{+}$ is narrowly peaked about
$k^{+}=M_{A}/A=\bar{M}_{N}$. Thus the only binding effect arises
from the average binding energy, which is much smaller than the
average separation energy. Therefore dynamics beyond the
relativistic mean field approximation must be invoked to explain
the EMC effect.

We briefly outline our procedure. In Sections \ref{sec:green} and
\ref{sec:subtle} we present the covariant deep inelastic
scattering formalism of Ref.~\cite{Jung:1988jw} and derive its
representation in terms of nucleon single particle wave functions.
The plus momentum distribution follows from this representation in
Section \ref{sec:derivation} where we also derive a relativistic
version of the Hugenholtz-van Hove theorem. Then we present the
results of analytic and numerical calculations in Section
\ref{sec:numbers}, the latter giving an $A$ dependence of the
ratio function contrary to experimental results. This demonstrates
that the use of the relativistic mean field approximation,
combined with the assumption that the nuclear medium does not
modify the structure of the nucleon, cannot describe the EMC
effect. The reasons for the subtle differences between the results
for finite nuclei and nuclear matter are detailed in Section
\ref{sec:scalar}.

\section{Nucleon Green's Function for Finite Nuclei}
\label{sec:green}

The use of a manifestly covariant formulation to derive the
expression for the structure function leads to a convolution
formula Eqs.~(\ref{eq:conv-intro}) and (\ref{eq:fn})
\cite{Jung:1988jw}. The function $\chi(k,P)$ is (proportional to)
the connected part of the nuclear expectation value of the nucleon
Green's function, and the trace is over the Dirac and isospin
indices. We begin with the covariant plus momentum distribution
function. Eq.~(\ref{eq:fn}) where we identify
\begin{eqnarray}
\chi^{A}(k,P) & \equiv & -i \int d^{4}x \int d^{4}y e^{-ik\cdot
(x-y)} G^C(x,y),\label{eq:chi}
\end{eqnarray}
\begin{figure}
\centering
\includegraphics[scale=1.0]{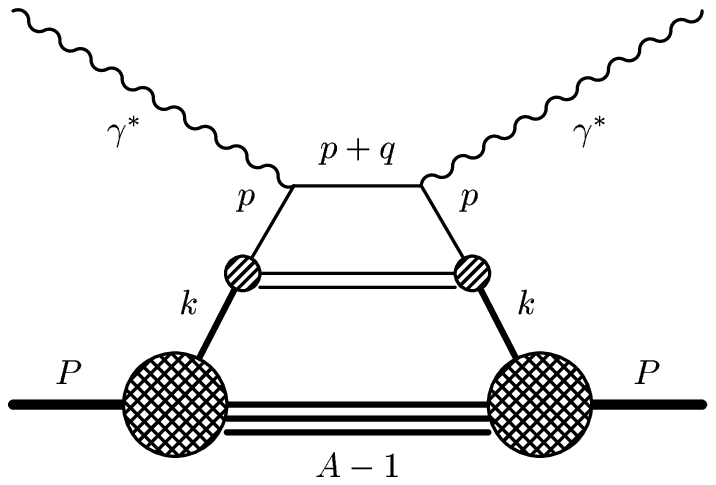}
\caption{Feynman diagram for deep inelastic lepton-nucleus
scattering. A nucleus of momentum $P$ is struck by a virtual
photon of momentum $q$. We label nucleon momentum $k$, and quark
momentum $p$.} \label{fig:dis}
\end{figure}
\begin{figure}
\centering
\includegraphics[scale=1.0]{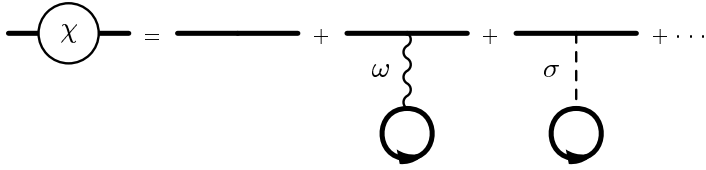}
\caption{The function $\chi$ in a mean field theory with scalar
($\sigma$) and vector ($\omega$) mesons.}
\end{figure}
where $G^{C}(x,y)$ is the connected part of the nucleon Green's
function:
\begin{eqnarray}
iG(x,y) & \equiv &  \langle P | T^{+} \{\Psi'(x) \overline{\Psi}'
(y)\} |P \rangle. \label{eq:green}
\end{eqnarray}
This result is directly determined from the Feynman diagram in
Fig.~\ref{fig:dis} following Ref.~\cite{Jung:1988jw}, but with
$\chi$ having a different normalization. So far this is
independent of the particular relativistic mean field model, but
for concreteness we use a Quantum Hadrodynamics (QHD) Lagrangian
\cite{Serot:1997xg,Serot:1986ey}, specifically QHD-I as in
Ref.~\cite{Blunden:1999gq}, where the nucleon fields, $\Psi'$,
that appear in Eq.~(\ref{eq:green}) are those appearing in the
Lagrangian. Light front quantization requires that the plus
component of all vector potential fields vanishes, and this is
obtained by using the Soper-Yan transformation
\cite{Soper:1971wn,Yan:1973qf}
\begin{equation}
\Psi'(x)\equiv e^{-ig_{v}\Lambda(x)}\Psi(x),\qquad
\partial^{+}\Lambda(x)=V^{+}(x). \label{eq:soperyan}
\end{equation}
to define the nucleon field operator $\Psi$ for various models
\cite{Miller:2001tg}. This transformation allows the use of the
eigenmode expansion for the $\Psi$ fields which have been obtained
previously in Ref.~\cite{Blunden:1999gq}
\begin{eqnarray}
\Psi(x) & = & \sum_{\alpha}  \left[ a_{\alpha} \Psi_{\alpha} (x) +
b_{\alpha}^{\dag} \varphi_{\alpha} (x) \right]\nonumber \\
& = & \sum_{\alpha}  \left[ a_{\alpha} \Psi_{\alpha} (\bm{x})
e^{-i p^{-}_{\alpha} x^{+}/2 }+ b_{\alpha}^{\dag} \varphi_{\alpha}
(\bm{x}) e^{i p^{-}_{\alpha} x^{+}/2 } \right],
\label{eq:mode-exp}
\end{eqnarray}
where $a_{\alpha}$ and $b^{\dag}_{\alpha}$ are (anti-)nucleon
annihilation operators and we define $z \equiv -x^{-}/2$ with
$\partial^{+} = 2\partial_{-} = -\partial_{z} $ and
$(\bm{x}_{\bot},z) \equiv \bm{x}$ which allows us to treat the
minus and perpendicular co\"{o}rdinates on equal footing. The
$\Psi_{\alpha}$ and $\varphi_{\alpha}$ are co\"{o}rdinate space
4-component spinor solutions to the light front Dirac equation
with eigenvalues $p^{-}_{\alpha}/2 = M_{N}-\varepsilon_{\alpha}$.
To simplify the analysis we will temporarily ignore
electromagnetic effects, but we will include them in the final
numerical results. The light front mode equations in QHD-I are
obtained by minimizing the $P^{-}$ operator (light front
Hamiltonian) with the constraint \cite{Blunden:1999gq} that
$P^{+}=P^{-}$. The result is
\begin{eqnarray}
-i\partial_{z}\Psi_{\alpha}^{-}(\bm{x}) & = &
\left[\bm{\alpha}_{\bot}\cdot(\bm{p}_{\bot}-g_{v}\bar{\bm{V}}_{\bot})+\beta(M_{N}+g_{s}\phi)\right]
\Psi_{\alpha}^{+}(\bm{x}) \label{eq:dirac1} \\
p^{-}_{\alpha}\Psi_{\alpha}^{+}(\bm{x}) & = &
\left[-i\partial_{z}+2g_{v}\bar{V}^{-}\right]\Psi_{\alpha}^{+}(\bm{x})
\nonumber\\* & & +
\left[\bm{\alpha}_{\bot}\cdot(\bm{p}_{\bot}-g_{v}\bar{\bm{V}}_{\bot})+\beta(M_{N}+g_{s}\phi)\right]
\Psi_{\alpha}^{-}(\bm{x}), \label{eq:dirac2}
\end{eqnarray}
with
\begin{eqnarray}
\Lambda_{\pm}\Psi_{\alpha} & = &
\frac{1}{2}\gamma^{0}\gamma^{\pm}\Psi_{\alpha} =
\Psi_{\alpha}^{\pm}\\*
\partial^{+}\bar{V}^{\mu} & = & \partial^{+}V^{\mu} -
\partial^{\mu}V^{+}.
\end{eqnarray}
Using standard manipulations \cite{Serot:1986ey} and defining
$\varepsilon_{F}$ as the energy of the highest occupied state, we
find the Green's function to be

\begin{eqnarray}
G(x,y) & = & \sum_{\alpha} \Psi_{\alpha} (\bm{x})
\overline{\Psi}_{\alpha}
(\bm{y})e^{-ig_{v}[\Lambda(\bm{x})-\Lambda(\bm{y})]} \int
\frac{dk^{-}}{2\pi} e ^{-ik^{-}(x^{+}-y^{+})/2} \Bigg[
\frac{1}{k^{-}-p^{-}_{\alpha} + i\varepsilon} \nonumber\\
 & & + 2\pi i \delta(k^{-}-p^{-}_{\alpha})
\theta(\varepsilon_{F}-\varepsilon_{\alpha}) \Bigg]\nonumber\\
& \equiv & G^{D}(x,y) + G^{C}(x,y),
\end{eqnarray}

where the superscripts $D$ and $C$ represent the disconnected and
connected parts of the nucleon Green's function, respectively. The
connected part is relevant to deep inelastic scattering and is
given by
\begin{equation}
G^{C}(x,y) = i \sum_{\alpha\in F} \Psi_{\alpha} (\bm{x})
\overline{\Psi}_{\alpha} (\bm{y})
e^{-ig_{v}[\Lambda(\bm{x})-\Lambda(\bm{y})]}
e^{-ip^{-}_{\alpha}(x^{+}-y^{+})/2}, \label{eq:cgreen}
\end{equation}
where the sum is over occupied levels $\alpha$ in the Fermi sea
$F$. We now substitute Eq.~(\ref{eq:cgreen}) into
Eq.~(\ref{eq:chi}), first defining $(\bm{k}_{\bot},k^{+}) \equiv
\bm{k}$ where $\bm{k}\cdot \bm{x} = \bm{k}_{\bot}\cdot
\bm{x}_{\bot} + k^{+}z  = \bm{k}_{\bot}\cdot \bm{x}_{\bot} -
k^{+}x^{-}/2 $, $d\bm{x}=d^{2}\bm{x}_{\bot}dz$,
$d\bm{k}=d^{2}\bm{k}_{\bot}dk^{+}$ and
\begin{equation}
\Psi''_{\alpha}(\bm{k}) \equiv \int d\bm{x}
e^{-i\bm{k}\cdot\bm{x}}e^{-ig_{v}\Lambda(\bm{x})}\Psi_{\alpha}(\bm{x}).
\label{eq:tranpsi}
\end{equation}
We find
\begin{equation}
\chi^{A}(k,P) = (2\pi)^{2}\sum_{\alpha\in F}
\Psi''_{\alpha}(\bm{k})\overline{\Psi}''_{\alpha}(\bm{k})
\delta(k^{-}-p^{-}_{\alpha}). \label{eq:new-chi}
\end{equation}
The motivation for the `double-prime' notation is the subject of
the next section.

\section{Wave function Subtleties}
\label{sec:subtle}

It would be useful to express $\chi^{A}(k,P)$ in terms of
solutions of the ordinary Dirac equation, because one may use a
standard computer program \cite{cjht}. If we use
Eq.~(\ref{eq:tranpsi}) we find that these `double-primed' fields
satisfy another version of the mode equations
Eq.~(\ref{eq:dirac1}) and Eq.~(\ref{eq:dirac2}) following from an
application of the Soper-Yan transformation
Eq.~(\ref{eq:soperyan}), and are given by
\begin{eqnarray}
\left[-i\partial_{z}-g_{v}V^{+}\right] \Psi_{\alpha}''^{-} & = &
\left[\bm{\alpha}_{\bot}\cdot(\bm{p}_{\bot} -g_{v}\bm{V}_{\bot}) +
\beta(M_{N}+g_{s}\phi)\right] \Psi_{\alpha}''^{+}
\label{eq:dirac1-primed} \\*
\left[i\partial_{z}+g_{v}V^{+}-2g_{v}\bar{V}^{-}+p^{-}_{\alpha}\right]
 \Psi_{\alpha}''^{+} & = &
\left[\bm{\alpha}_{\bot}\cdot(\bm{p}_{\bot} -g_{v}\bm{V}_{\bot}) +
\beta(M_{N}+g_{s}\phi)\right] \Psi_{\alpha}''^{-}.
\label{eq:dirac2-primed}
\end{eqnarray}
If one multiplies Eq.~(\ref{eq:dirac1-primed}) by $\gamma^{+}$ and
Eq.~(\ref{eq:dirac2-primed}) by $\gamma^{-}$ and adds the two
equations, using $V^{+}=V^{-}=\bar{V}^{-}=V^{0}$, one obtains
\begin{equation}
\left[-\gamma^{3}(i\partial_{z}+p^{-}_{\alpha}/2)+\gamma^{0}
(p^{-}_{\alpha}/2-g_{v}V^{0}) \right] \Psi_{\alpha}''(\bm{x}) =
\left[ \bm{\gamma}_{\bot}\cdot\bm{p}_{\bot} + M_{N}+g_{s}\phi
\right]\Psi_{\alpha}''(\bm{x}) \label{eq:almost-dirac}
\end{equation}
Which looks nearly like the ordinary Dirac equation which the
exception of the $\gamma^{3} p^{-}_{\alpha}/2$ term. To remove
this term, set
\begin{equation}
\widetilde{\Psi}_{\alpha}(\bm{x}) \equiv e^{-ip^{-}_{\alpha}z/2}
\Psi''_{\alpha} (\bm{x})
\end{equation}
and substitute into Eq.~(\ref{eq:almost-dirac}), so that
\begin{equation}
\gamma^{0} (p^{-}_{\alpha}/2-g_{v}V^{0})
\widetilde{\Psi}_{\alpha}(\bm{x}) = \left[ \bm{\gamma} \cdot
\bm{p} + M_{N}+g_{s}\phi \right] \widetilde{\Psi}_{\alpha}(\bm{x})
\label{eq:dirac-form}
\end{equation}

which looks superficially like the ordinary Dirac equation for the
$\Psi'$ fields which appear in the Lagrangian. There is a subtle
difference; since the light front energy is given by $k^{2} =
(\bm{k}_{\bot}^{2}+M_{N}^{2})/k^{+}$, Eq.~(\ref{eq:almost-dirac})
has support for $k^{+}>0$. This means Eq.~(\ref{eq:dirac-form})
has support for $k^{3}>-p^{-}_{\alpha}/2$ which in turn implies
that it can only be considered the ordinary Dirac equation with a
momentum cutoff at $p^{-}_{\alpha}/2\simeq M_{N}$. This
restriction is nearly superfluous since the probability that a
nucleon is carrying $|k^{3}|>M_{N}$ is suppressed by a factor of
order $e^{-M_{N}^{2}R^{2}}$ with $R$ being the nuclear radius.
This allows us to effectively identify $\widetilde{\Psi}_{\alpha}
\simeq \Psi'$ which gives the approximate relationships between
the (equal time) Lagrangian fields $\Psi'$, the Soper-Yan
transformed light front fields $\Psi$ and the untransformed light
front fields $\Psi''$ that appear in Eq.~(\ref{eq:new-chi}). We
have
\begin{eqnarray}
\Psi'_{\alpha} (\bm{x}) & \simeq & e^{-ip^{-}_{\alpha}z/2}
\Psi''_{\alpha}(\bm{x}) \label{eq:prime-doubleprime}\\* & = &
e^{-ip^{-}_{\alpha}z/2} e^{-ig_{v}\Lambda(\bm{x})} \Psi_{\alpha}
(\bm{x}) \label{eq:prime-noprime}
\end{eqnarray}
Eq.~(\ref{eq:prime-noprime}) is the approximate relationship
between the $\Psi$ and $\Psi'$ fields in
Ref.~\cite{Blunden:1999gq}. We now are ready to derive a
representation of Eq.~(\ref{eq:fn}) in terms of these nucleon wave
functions.

\section{Derivation of the Plus Momentum Distribution}
\label{sec:derivation}

\subsection{General Case}
\label{sec:generalcase}

In Ref.~\cite{Blunden:1999gq}, it was determined that a plus
momentum distribution in QHD-I is given by
\begin{equation}
f(k^{+}) = 2 \sum_{\alpha\in F} \int d^{2}\bm{x}_{\bot} \left|
\Psi^{+}_{\alpha}(\bm{x}_{\bot},k^{+}) \right| ^{2}.
\label{eq:flf}
\end{equation}
This distribution peaks at $k^{+}/\bar{M}_{N}\equiv y \simeq 0.8$
for $^{16}$O, (with smaller values for heavier nuclei) but is  not
the distribution obtained from the covariant formalism of Section
\ref{sec:green}. The connection between this $f(y)$ and the
covariant $f(y)$ was made in Ref.~\cite{Miller:2001tg}; it was
determined that, in the limit of infinite nuclear matter,  the
relationship between $f(y)$ and $f(y)$ is  simply a shift in the
argument by the vector meson potential:
\begin{equation}
f(y) = f(y+g_{v}V^{+}/\bar{M}_{N}). \label{eq:shift-nm}
\end{equation}
This shift arises from the use of the Soper-Yan transformation
Eq.~(\ref{eq:soperyan}) where the $\Psi'$ fields are those
appearing in the Lagrangian and are used to determine $f(y)$,
whereas the $\Psi$ fields are used to determine $f(y)$. In finite
nuclei, this relationship is somewhat more complicated since the
vector meson potential is no longer a constant over all space. We
start with Eq.~(\ref{eq:new-chi}), and see that
\begin{eqnarray}
\text{Tr}\gamma^{+}\chi^{A}(k,P) & = &  (2\pi)^{2} \sum_{\alpha\in
F} \text{Tr} \left[\gamma^{+}\Psi''_{\alpha}(\bm{k})
\overline{\Psi}''_{\alpha}(\bm{k})\right]
\delta(k^{-}-p^{-}_{\alpha})\nonumber\\* & = & 8\pi^{2}
\sum_{\alpha\in F} \left| \Psi''^{+}_{\alpha}(\bm{k}) \right|^{2}
\delta(k^{-}-p^{-}_{\alpha}).\nonumber
\end{eqnarray}
Substituting into Eq.~(\ref{eq:fn}) we obtain
\begin{eqnarray}
f(y) & = & \frac{2}{(2\pi)^{2}} \sum_{\alpha\in F} \int d\bm{k}\,
\delta(y-k^{+}/\bar{M}_{N}) \left|\Psi''^{+}_{\alpha}(\bm{k})
\right|^{2}. \label{eq:fny}
\end{eqnarray}
Use of Parseval's identity and integrating over $k^{+}$ gives us
our main result:
\begin{eqnarray}
f(y) & = & 2\bar{M}_{N} \sum_{\alpha\in F} \int d^{2}\bm{x}_{\bot}
\left| \Psi''^{+}_{\alpha}(\bm{x}_{\bot}, \bar{M}_{N}y)
\right|^{2}, \label{eq:fn-general}
\end{eqnarray}
so the plus momentum distribution is related to Fourier transform
of the $\Psi''$ wave functions. One can see the similarity to
Eq.~(\ref{eq:flf}); the difference lies entirely in
Eq.~(\ref{eq:tranpsi}). It should be emphasized that this result
does not depend on the approximation in Section \ref{sec:subtle}.

We shall use $f(y)$ to compute the nuclear structure function
$F_{2}^{(A)}(x)$ in Section \ref{sec:numbers}, but first we derive
a version of the Hugenholtz-van Hove theorem valid for finite
nuclei. To do that, multiply Eq.~(\ref{eq:fny}) by $y$ and
integrate
\begin{eqnarray}
\langle y \rangle & \equiv & \int dy y f(y)\nonumber\\* & = &
\frac{2}{(2\pi)^{2}} \sum_{\alpha\in F}\int d\bm{k}
\frac{k^{+}}{\bar{M}_{N}} \left| \Psi''^{+}_{\alpha}(\bm{k})
\right|^{2},
\end{eqnarray}
Now remove the plus projections and re-express $\Psi''$ and its
complex conjugate in co\"{o}rdinate spaces $\bm{x}$ and $\bm{x'}$.
One can then integrate over $\bm{k}$ yielding a delta function
$\delta (\bm{x}-\bm{x'})$ which allows integration over $\bm{x'}$
yielding
\begin{eqnarray}
\langle y \rangle & = & \frac{1}{\bar{M}_{N}} \sum_{\alpha\in
F}\int d\bm{x} \Psi''^{\dag}_{\alpha}(\bm{x}) \gamma^{0}
\gamma^{+} i\partial^{+} \Psi_{\alpha}''(\bm{x}) \nonumber
\end{eqnarray}
We wish to look at the $\Psi$ fields in order to understand our
result in the context of Ref.~\cite{Blunden:1999gq}, so we need to
perform the Soper-Yan transformation Eq.~(\ref{eq:soperyan}) and
use $X^{\dag}\gamma^{0}\equiv\overline{X}$
\begin{eqnarray}
\langle y \rangle & = & \frac{1}{\bar{M}_{N}} \sum_{\alpha\in
F}\int d\bm{x} \overline{\Psi}_{\alpha}(\bm{x}) \gamma^{+}
[i\partial^{+}+g_{v}V^{+}(\bm{x})] \Psi_{\alpha}(\bm{x}) \nonumber
\end{eqnarray}
If we explicitly put in the the nuclear state vectors, we can
perform the sum on $\alpha$ by inserting creation and annihilation
operators; we can add the time dependence for free since it is
unaffected by $\partial^{+}$ and cancels with both fermion fields,
and the vector potential is static. We have effectively undone the
substitution Eq.~(\ref{eq:mode-exp}) and now have an expectation
value of an operator
\begin{eqnarray}
\langle y \rangle & = & \frac{1}{M_{A}} \int d\bm{x} \langle P |
\overline{\Psi} \gamma^{+} [i\partial^{+}+g_{v}V^{+}] \Psi | P
\rangle \label{eq:avg-y-op}
\end{eqnarray}
The second term of Eq.~(\ref{eq:avg-y-op}) was essentially
included by Birse \cite{Birse:hu} via a kinematic argument; here
it follows from a fully covariant light front treatment. Using the
vector meson field equation in QHD-I
\[
\partial_{\mu}V^{\mu+}+m_{v}^{2}V^{+}=g_{v}\overline{\Psi}
\gamma^{+}\Psi,
\]
integrating by parts, and anti-symmetrizing one can re-express the
second term of Eq.~(\ref{eq:avg-y-op})
\begin{eqnarray}
\langle y \rangle & = & \frac{1}{M_{A}} \int d\bm{x} \langle P |
\overline{\Psi}\gamma^{+} i\partial^{+} \Psi +
m_{v}^{2}V^{+}V^{+}+V^{+\mu}V_{\mu}^{\:\:+} | P \rangle \nonumber
\\*
& = & \frac{1}{M_{A}} \int d\bm{x} \langle P| T^{++} -
\partial^{+} \phi \partial^{+}\phi | P \rangle \nonumber\\*
& = & \frac{1}{M_{A}}\left( P^{+}-P^{+}_{s} \right)\nonumber\\* &
= & 1 - \frac{P^{+}_{s}}{M_{A}} \simeq 1 \label{eq:HvH}
\end{eqnarray}
where $T^{++}$ is one component of the canonical energy momentum
tensor, $P^{+}_{s}$ is the plus momentum of the scalar meson
fields, and $P^{+}$ is the total nuclear plus momentum. The result
Eq.~(\ref{eq:HvH}) constitutes an analog of the Hugenholtz-van
Hove theorem \cite{HvH} for finite relativistic systems; the
equality becomes exact in the nuclear matter limit, where the
scalar meson contribution vanishes, as shown in our previous work
\cite{Miller:2001tg}. This means that we may anticipate that the
binding effect will again be small. The vector operator `mixing'
and the scalar meson contribution will be elaborated on in a more
general context in Section \ref{sec:scalar}.

It is also worthwhile to explicitly evaluate the expression
Eq.~(\ref{eq:fn-general}) for $f(y)$ in the limit of infinite
nuclear matter. In this case, $V^{0}=V^{+}=V^{-}$ are constant and
$\bm{V}_{\bot}=0$, so we find
\begin{eqnarray}
\Lambda(z,\bm{x}_{\bot}) & = & \int_{z}^{\infty}dz'
V^{0}(z',\bm{x}_{\bot})\nonumber\\
& = & -V^{0}z\nonumber\\
& = & -V^{+}z,
\end{eqnarray}
so that Eq.~(\ref{eq:tranpsi}) becomes
\begin{eqnarray}
\Psi''^{+}_{\alpha}(\bm{k}) & = & \int d\bm{x}
e^{-i\bm{k}_{\bot}\cdot\bm{x}_{\bot}} e^{-i(k^{+}-g_{v}V^{+})z}
\Psi^{+}_{\alpha}(\bm{x}) \nonumber\\
& = & \Psi^{+}_{\alpha}(\bm{k}_{\bot}, k^{+}-g_{v}V^{+}).
\end{eqnarray}
Therefore Eq.~(\ref{eq:fn-general}) becomes
\begin{eqnarray}
f(y) & = & 2\bar{M}_{N} \sum_{\alpha\in F} \int d^{2}\bm{x}_{\bot}
\left|\Psi^{+}_{\alpha}(\bm{x}_{\bot}, \bar{M}_{N}y-g_{v}V^{+})
\right|^{2},
\end{eqnarray}
which is simply the expression~(\ref{eq:flf}) modified by a shift
in the argument of $g_{v}V^{+}/\bar{M}_{N}$. Thus  we find
Eq.~(\ref{eq:shift-nm}) is satisfied in the nuclear matter limit.
It is important to stress that all that is recovered here is the
shift in the argument and not any particular form of the plus
momentum distribution which arises from the specific model used
for the wavefunctions.

\subsection{Nuclear Matter} \label{sec:fy_nm}

In nuclear matter, we can derive an explicit form of $f(y)$ valid
for all mean field theories with $k_{F}$ and $M_{N}$ as the only
input parameters. The manifestly covariant form of the single
nucleon Green's function has been known for a long time
\cite{Serot:1986ey}, and its use (in the nucleus rest frame) leads
to the result
\begin{eqnarray}
\chi(k,P)&=&-i\left(\gamma\cdot(k-g_{v}V)+M_{N}^* \right)\nonumber\\
&\times&\left[\frac{1}{(k-V)^2-{M_{N}^*}^2+i\epsilon} +
\frac{i\pi}{E^*(k)}\delta(k^0-E^*(k)-g_{v}V^0))\theta(k_F-\vert{\bf
k}\vert)\right], \label{eq:gf}
\end{eqnarray}
where
\begin{equation}
E^*(k)\equiv\sqrt{{M_{N}^*}^2+{\bf k}^2}.
\end{equation}
The general form of the Green's function depends on a vector
potential $V=(V^0,{\bf 0})$ for a nucleus at rest, and the
effective mass $M_{N}^*$ which includes the effects of
interactions on the nucleon mass. The values of $V$ and $M_{N}^*$
depend on the specific Lagrangian employed, but the form of the
Green's function is general. Recall also that ${V}^-=V^+=V^0$ for
the expectation values of vector meson fields in the nucleus rest
frame.

The result Eq.~(\ref{eq:gf}) was first obtained using the
conventional equal time approach, but the very same can also be
obtained from the light front formalism. In that case it is
necessary to include the effects of the instantaneous part of the
nucleon Green's function and those of the instantaneous meson
exchange.

The next step is to insert the connected part (second term) of
Eq.~(\ref{eq:gf}) into Eq.~(\ref{eq:fn}) for $f(y)$. This gives,
after taking the trace and using the delta function to integrate
over $k^0$, the result
\begin{eqnarray}
f(y)=\frac{4}{(2\pi)^3\rho_B} \int d^2k_\perp\;dk^3 \frac{E^*(k)
+k^3}{E^*(k)} \delta(y- \frac{E^*(k)+g_{v}V^++k^3}{\bar{M}_{N}}
)\theta(k_F-\vert{\bf k}\vert) .\label{eq:mid}
\end{eqnarray}
The integration is simplified by using the transformation
\begin{eqnarray}
k^+\equiv E^*(k)+k^3,\label{eq:k3}
\end{eqnarray}
which makes a connection with light front variables. It is an
exercise in geometry to show that the Fermi volume can be
re-expressed in terms of $k^+$ using
\begin{equation}
k_\perp^2+(k^{+}-E^{*}(\bm{k}_{\perp},k^{+}))^2\le k_F^2,\quad
E_F^*\equiv \sqrt{k_F^2+{M_{N}^*}^2},\label{eq:fermisphere}
\end{equation}
\begin{equation}
E^{*}(\bm{k}_{\perp},k^{+}) = \frac{1}{2}\left( k^{+} +
\frac{k_{\perp}^{2}+{M_{N}^{*}}^{2}}{k^{+}} \right)
\end{equation}
so that Eq.~(\ref{eq:mid}) becomes
\begin{eqnarray}
f(y)=\frac{4}{ (2\pi)^3\rho_B}\int d^2k_\perp \int dk^+
\theta\left(k_F^2-k_\perp^2-(k^+-E^{*}(k))^2\right) \delta(y-
\frac{k^{+}+g_{v}V^+}{\bar{M}_{N}}) .\label{eq:preshift}
\end{eqnarray}
The use of the definition of the energy of a nucleon at the Fermi
surface,
\begin{equation}
E_F= E^*_F+g_{v}V^+=E^*_F+g_{v}V^0,
\end{equation}
allows one to achieve a simple expression for $f(y)$:
\begin{equation}
f(y)= \frac{3}{ 4} \frac{\bar{M}_{N}^3}{
k_F^3}\theta((E_F+k_F)/\bar{M}_{N}-y)
\theta(y-(E_F-k_F)/\bar{M}_{N}))\left[
\frac{k_F^2}{\bar{M}_{N}^2}-(\frac{E_F}{\bar{M}_{N}}-y)^2\right].\label{eq:shiftm1}
\end{equation}
The essence of the relativistic version of the Hugenholtz-van Hove
theorem \cite{HvH} follows simply from the statement that the
pressure of a stable nucleus vanishes, and hence $P^{+} = P^{-}$.
Therefore, in the nuclear rest frame
\begin{eqnarray}
E_{A}(k_{F}) & = & \frac{1}{2}\left( P^{+} + P^{-} \right) = P^{+}
= P^{-} = M_{A}\nonumber\\
\frac{E_{A}(k_{F})}{A} & = & \frac{M_{A}}{A}\nonumber\\
E(k_{F}) \equiv E_{F} & = & \bar{M}_{N}
\end{eqnarray}
The result Eq.~(\ref{eq:shiftm1}) can be further simplified to
\begin{equation}
f(y) = \frac{3}{4\Delta_{F}^{3}} \theta(1+\Delta_{F}-y)
\theta(y-1+\Delta_{F}) \left[
\Delta_{F}^{2}-(1-y)^2\right],\label{eq:pmdA}
\end{equation}
where $\Delta_{F} = k_{F}/\bar{M}_{N}$, correct for any
relativistic mean field theory of infinite nuclear matter.
Different theories with the same binding energy and Fermi momentum
may have very different scalar and vector potentials, but must
have the same $f(y)$.

A result very similar to Eq.~(\ref{eq:pmdA}) was previously
obtained by Birse \cite{Birse:1993hu}. The difference between his
formula and ours is the appearance of $\bar{M}_{N}$ in the
function $f(y)$, whereas he uses $M_{N}$. This difference is a
small effect numerically, and therefore our conclusions will be
the same as his.

The baryon sum rule and momentum sum rules are derived by taking
the first two moments of $f(y)$. This gives:
\begin{eqnarray}
\int dy f(y)&=&1\label{eq:norm1}\\
\int dy\; y\;f(y)&=&1.\label{eq:norm2}
\end{eqnarray}
The latter equation is remarkable; it states that in deep
inelastic scattering the nucleons act as if they carry all of the
$P^+$ of the nucleus even though the mesonic fields may be very
prominent.

This is clearer if we re-interpret these sum rules in terms of a
probability $f(k^+)$ that a nucleon has a plus momentum $k^+\equiv
y\bar{M}_{N}$, with $f(k^+)\equiv A f(y\bar{M}_{N})/\bar{M}_{N},$
so that
\begin{eqnarray}
&&\int dk^+\;f(k^+)=A,\label{eq:baryonsr}\\
&&\int dk^+ \;k^+ \;f(k^+)= A\bar{M}_{N}=M_A\label{eq:momsr}
\end{eqnarray}
The momentum sum rule Eq.~(\ref{eq:momsr}) shows the total plus
momentum carried by the nucleons (as seen in deep inelastic
scattering) is also the total momentum carried by the nucleus.

The main result of this is that the nuclear structure function is
given by \ref{eq:fn}) with the function $f(y)$ obtained in
\ref{eq:pmdA}). This tells us that, despite the fact that there is
considerable binding energy, there is no EMC binding effect.
Indeed, $F_{2}^{(A)}$ depends on the Fermi momentum but does not
depend on the effective mass $M_{N}^*$.

\section{Nuclear Structure Functions}
\label{sec:numbers}

\begin{figure}
\centering
\includegraphics[scale=0.70]{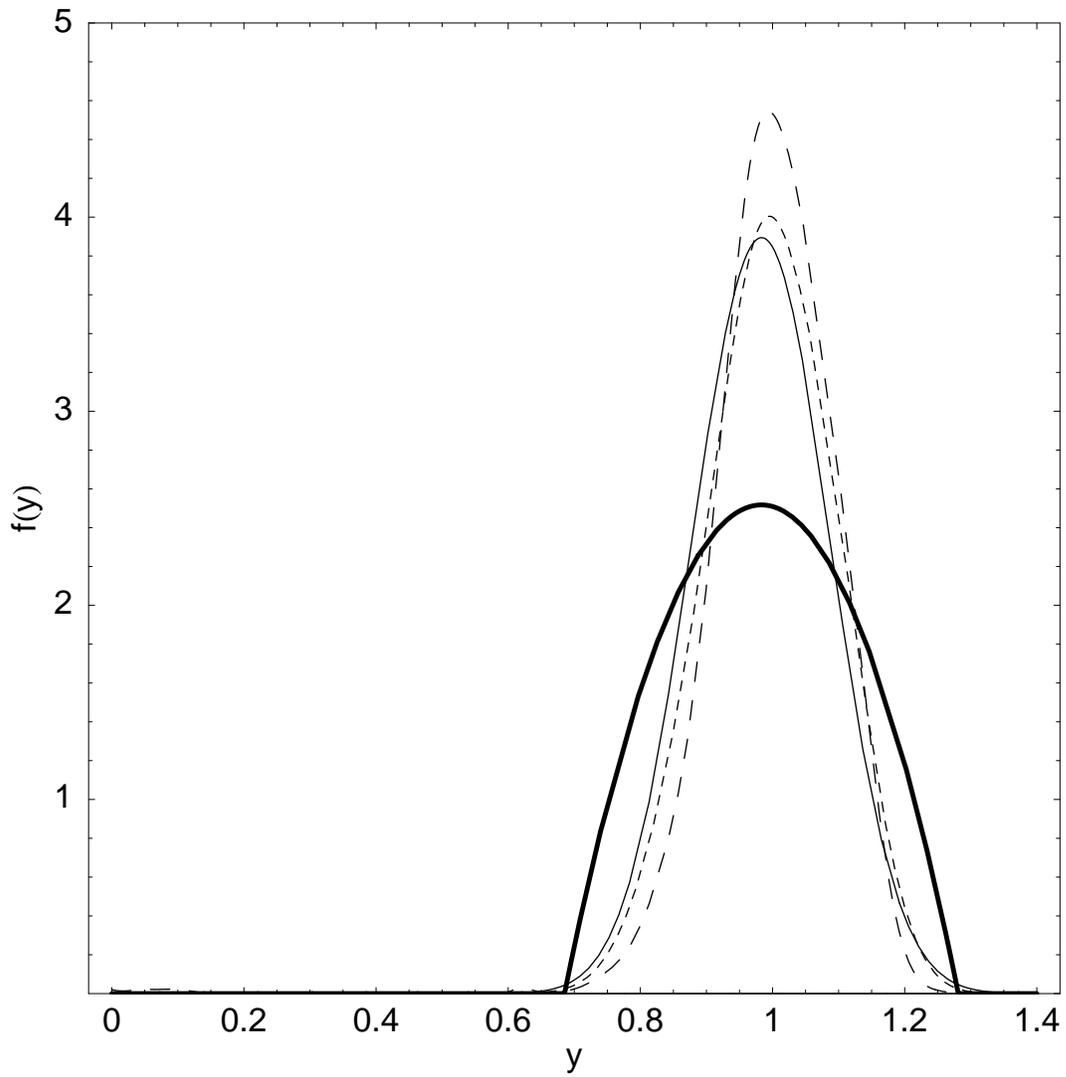}
\caption{Plus momentum distributions, $f(y)$, for $^{40}$Ca
(solid), $^{16}$O (short dashes), $^{208}$Pb (dot-dashes) and
nuclear matter (long dashes).} \label{fig:fn}
\end{figure}

We determine the wave functions appearing in
Eq.~(\ref{eq:fn-general}) numerically from a relativistic
self-consistent treatment following Horowitz and Serot
\cite{Horowitz:1981xw} using the same program \cite{cjht} which
includes electromagnetic effects. The plus momentum distribution
follows and is given in Fig.~\ref{fig:fn} for $^{16}$O, $^{40}$Ca,
$^{208}$Pb and in the nuclear matter limit. One can see that the
peaks appear near $y=1$ as required by the Hugenholtz-van Hove
theorem Eq.~(\ref{eq:HvH}).

It is worth noting that application of the Soper-Yan
transformation Eq.~(\ref{eq:soperyan}) to the $\Psi''$
wavefunctions obtained from the equal time wavefunctions
reproduces the plus momentum distributions, including the correct
asymmetry, of the light front calculations in
Ref.~\cite{Blunden:1999gq}, which did not use the approximation
Eq.~(\ref{eq:prime-doubleprime}).
\begin{figure}
\centering
\includegraphics[scale=0.70]{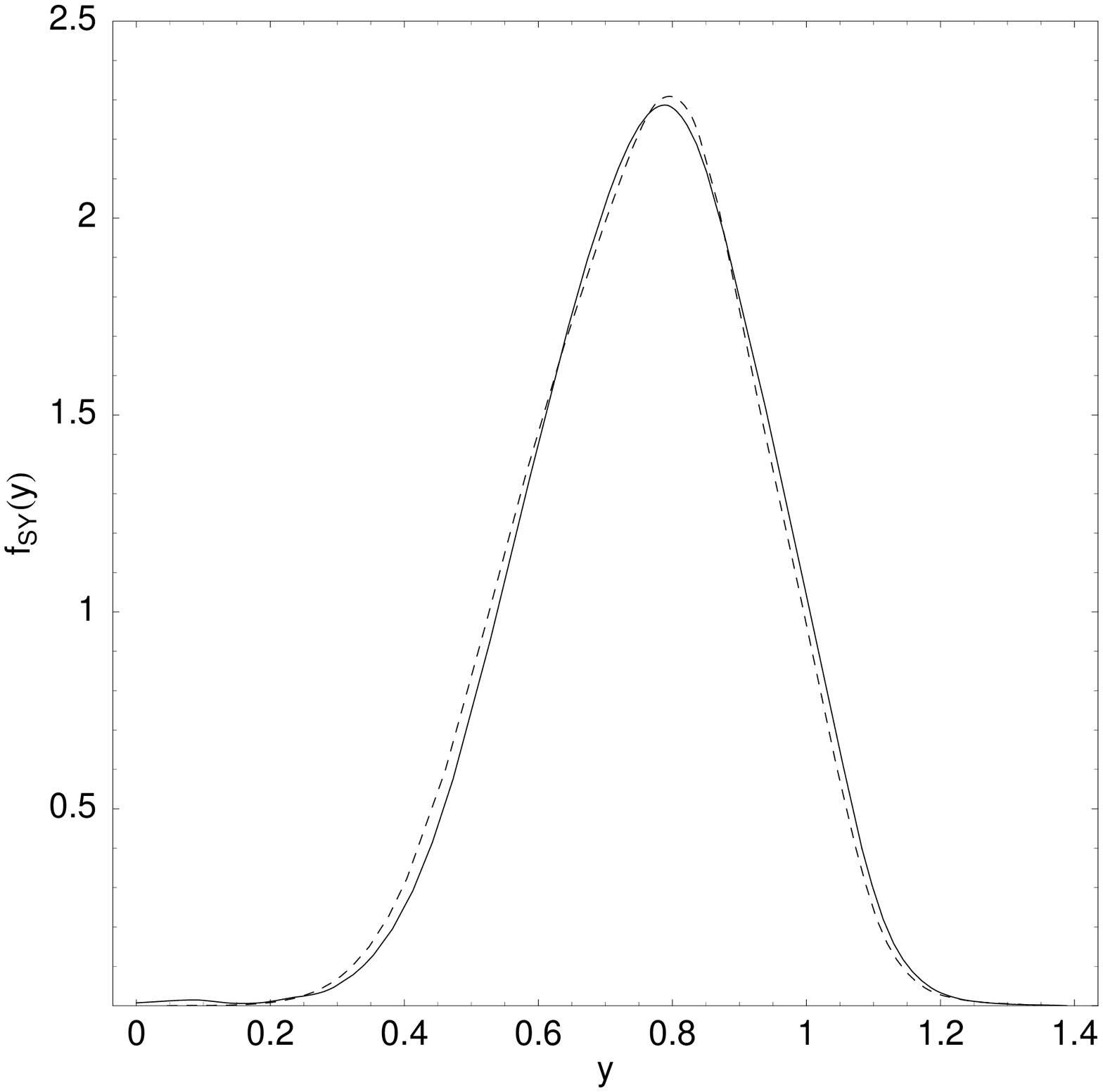}
\caption{$f(y)$ distribution for $^{16}$O (solid) after
application of the Soper-Yan transformation along with the
$^{16}$O distribution from Ref.~\cite{Blunden:1999gq} (short
dashes). Note that the peaks occur at $y<1$.} \label{fig:fblunden}
\end{figure}
We show the comparison of our Oxygen calculation to that of
Ref.~\cite{Blunden:1999gq} in Fig.~\ref{fig:fblunden}; the
agreement of these two curves demonstrates the excellence of the
approximation relating light front and equal time wavefunctions.
One can see that the effect in finite nuclei of the Soper-Yan
transformation is to shift and broaden the plus momentum
distribution, while in nuclear matter (also shown in
Fig.~\ref{fig:fblunden}) it is just a shift. If these
distributions were to be used in the nuclear structure function
Eq.~(\ref{eq:conv}) though, since $\langle y\rangle\simeq 0.8$ for
Oxygen, the ratio function (Eq.~(\ref{eq:ratio}) discussed later)
would fall precipitously to nearly zero at $x\simeq 0.6$ in stark
contradiction with experiment.

\begin{figure*}
\centering
\includegraphics[scale=0.7]{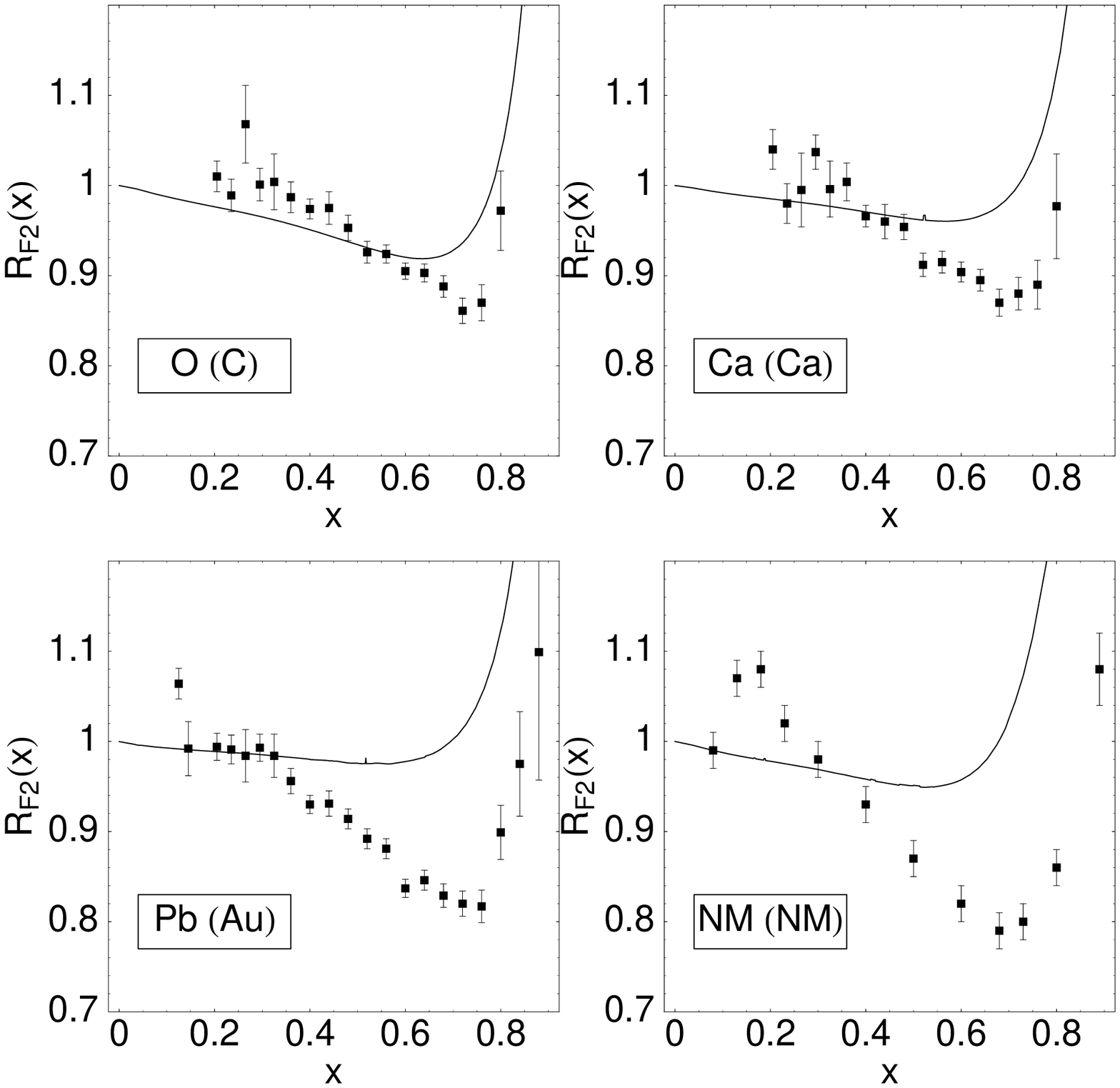}
\caption{Ratio functions for $^{16}$O, $^{40}$Ca and $^{208}$Pb
showing data for Carbon, Calcium and Gold, respectively, from
SLAC-E139 \cite{Gomez:1993ri}. The nuclear matter calculation
shows extrapolated data \cite{Sick:1992pw}.} \label{fig:ratio}
\end{figure*}

The structure function is given by the convolution
Eq.~(\ref{eq:conv}) with the generic function $q = F_{2}^{(N)}$.
The assumption that nuclear effects do not modify the structure of
the nucleon is embodied in Eq.~(\ref{eq:conv}) by the use of the
structure function of a free nucleon; we use the parameterization
\cite{deGroot:yb}
\begin{equation}
F_{2}^{(N)}(x)=0.58\sqrt{x}(1-x)^{2.8}+0.33\sqrt{x}(1-x)^{3.8}+0.49(1-x)^{8}.
\end{equation}
The experiments measure the ratio function Eq.~\ref{eq:ratio}. The
results of our calculations are plotted for $^{16}$O, $^{40}$Ca,
$^{208}$Pb and in the nuclear matter limit in Fig.~\ref{fig:ratio}
showing data for Carbon, Calcium and Gold from SLAC-E139
\cite{Gomez:1993ri} and an extrapolation \cite{Sick:1992pw} for
the nuclear matter calculation.
The most striking result is that these calculations fail to
reproduce the EMC effect; the curves consistently miss the minima
in the data, and the agreement gets worse with increasing $A$.
Another important result is that the ratio function does not fall
to zero as would be the case if the small effective mass ($\sim
0.56 M_{N}$ for nuclear matter in QHD-I) were the relevant
parameter describing the binding effect which would follow from
using Eq.~(\ref{eq:flf}) instead of Eq.~(\ref{eq:fn-general}). The
results also show a minimum near $x\simeq 0.6$ for Oxygen and
nuclear matter that is deeper than the Calcium and Lead
calculations. This is a curious feature that contradicts the trend
in experimental data, and is due to the effects of two parameters.

The first, and most important, is that of the location of the peak
of the plus momentum distribution given by Eq.~(\ref{eq:HvH}),
which gradually approaches $y=1$ as the nuclear matter limit is
reached. This is due to the fact that scalar mesons carry a small
amount of plus momentum \cite{Blunden:1999gq} that vanishes as
$A\rightarrow\infty$. The closer to $y=1$ the peak is in
Fig.~\ref{fig:fn}, the less pronounced the minimum in
Fig.~\ref{fig:ratio}, all else remaining constant. The second
effect is due to $\bar{M}_{N}$, which reaches a minimum at
$^{56}$Fe corresponding to a more pronounced minimum of the ratio
function than for $A<56$ or $A>56$, keeping the scalar meson
contribution constant.

\begin{figure}
\centering
\includegraphics[scale=0.7]{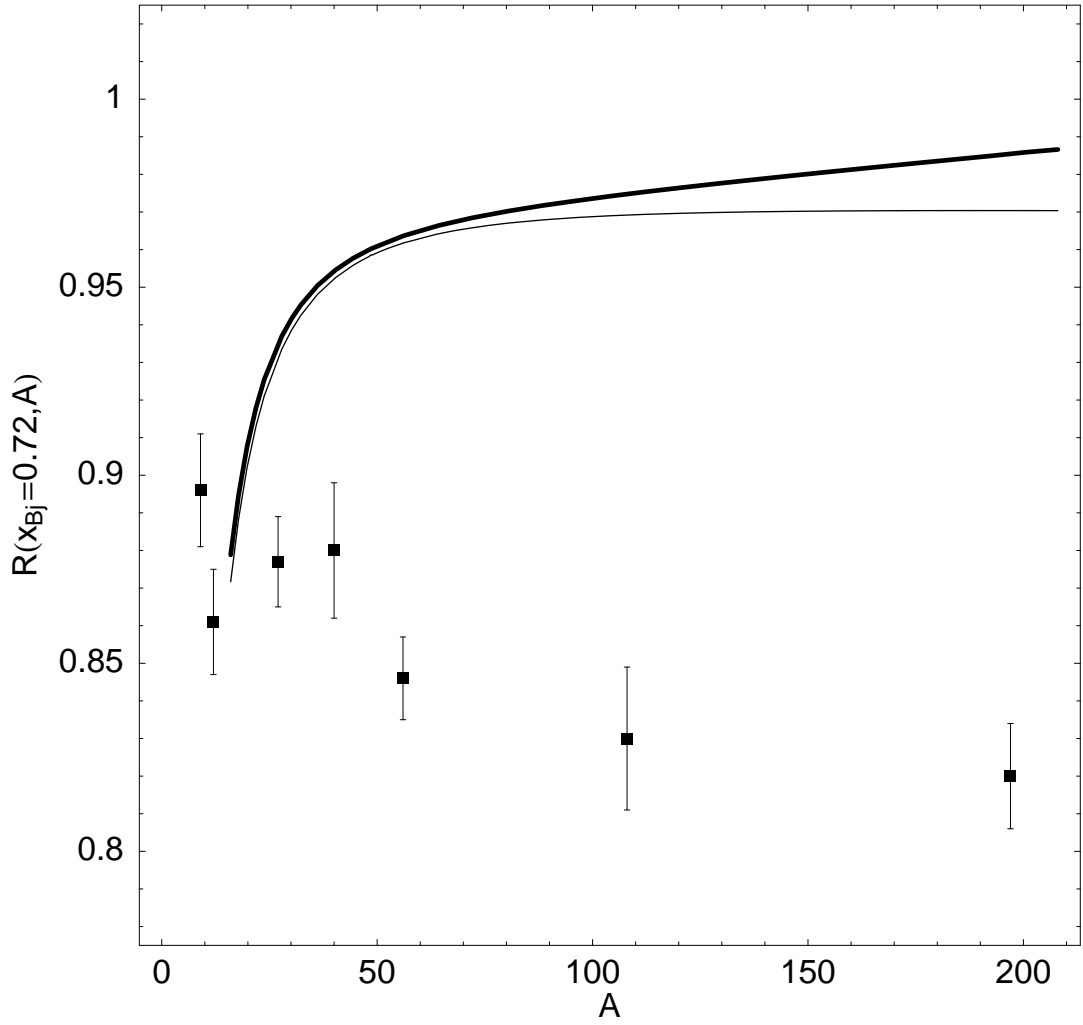}
\caption{$R(x=0.72)$ as a function of $A$ including scalar meson
and binding effects (solid line), and leaving binding energy per
nucleon constant at $-8.5\text{ MeV}$ (dashed line). The data are
from SLAC-E139 \cite{Gomez:1993ri}.} \label{fig:rminimum}
\end{figure}

Using a Gaussian parameterization of the plus momentum
distribution and the experimental binding energy per nucleon via
the semi-empirical mass formula, we have modeled the dependence of
the minimum of the ratio function, $R(x\simeq0.72)$, on the number
of nucleons in the nucleus in Fig.~\ref{fig:rminimum}.
The motivation for the use of Gaussian plus momentum distributions
is based on the expansion \cite{Frankfurt:nt}
\begin{eqnarray}
F_{2}^{(A)}(x_{A}) & = & F_{2}^{(N)}(x_{A})+\epsilon x_{A}
\frac{\partial F_{2}^{(N)}}{\partial x}\Big|_{x =
x_{A}}\nonumber\\* & & + \gamma \left[2 x_{A} \frac{\partial
F_{2}^{(N)}}{\partial x}\Big|_{x = x_{A}} + x_{A}^{2}
\frac{\partial^{2} F_{2}^{(N)}}{\partial x^{2}}\Big|_{x =
x_{A}}\right]\\* \text{where } \epsilon & \equiv & 1-\int dy y
f(y)\\* \gamma & \equiv & \int dy (y-1)^{2} f(y)
\end{eqnarray}
The Gaussian parameterization uses the peak location and width,
$\langle y\rangle$ and $(\langle y^{2}\rangle-\langle
y\rangle^{2})^{1/2}$ respectively, from the relativistic Hartree
calculations in Fig.~\ref{fig:fn}, and is normalized to unity.
This allows us to obtain a plus momentum distribution for any $A$
with minimal effort. We show the combined effect of scalar mesons
and binding energy per nucleon on the ratio function along with
the effect of scalar mesons alone using a constant binding energy
per nucleon of $-8.5\text{ MeV}$ independent of $A$. It can be
seen that a changing $\bar{M}_{N}$ with $A$ has the most effect
for nuclei much larger than Iron, but does not change the general
trend that the minimum of the ratio function becomes less
pronounced as $A$ increases due to the vanishing scalar meson
contribution and the peak of the plus momentum distribution
approaching unity. This dependence of the binding effect on $A$ is
quite different, both in magnitude and shape, than the trend in
experimental data summarized in Ref.~\cite{Sick:1992pw} which
satisfies $R(x\simeq 0.72)\sim A^{-1/3}$, so that the minimum
becomes more pronounced as $A$ increases. This fully demonstrates
the inadequacy of conventional nucleon-meson dynamics to explain
the EMC effect.

\section{Scalar Mesons and More General Considerations}
\label{sec:scalar}

The average value of $y$, given by Eq.~(\ref{eq:HvH}), yields the
nucleon contribution to the plus momentum, and is less than one
which can be seen in Fig.~\ref{fig:fn}. We now address the
remaining plus momentum in finite nuclei. Previous results
\cite{Blunden:1999gq} show that a small fraction ($\delta y\sim
0.005$) of the plus momentum is carried by the scalar mesons which
vanishes as the nuclear matter limit is approached. This is due to
the fact that scalar mesons couple to gradients in the scalar
density (arising mainly from  the surface of finite nuclei) which
vanish as $A\rightarrow\infty$. The question is: why are scalar
mesons allowed to carry plus momentum and not vector mesons?

The simplest answer lies in the Dirac structure of
Eq.~(\ref{eq:fn}); the $\gamma^{+}$ in the trace picks out terms
in the full interacting Green's function with an odd number of
gamma matrices which includes all Lorentz vector interactions and
excludes Lorentz scalar interactions. The Dirac structure of
$f(y)$ is directly related to the Dirac structure of the energy
momentum tensor, so the answer also lies there and illuminates a
problem with conventional nucleon-meson dynamics. The component of
the energy momentum tensor relevant to the plus momentum, from a
chiral Lagrangian containing isoscalar vector mesons, scalar
mesons and pions, is given by \cite{Miller:1997cr,Miller:1999ap}
\begin{eqnarray}
T^{++} & = & V^{+\mu}V_{\mu}^{\:\: +} + m_{v}^{2}V^{+}V^{+} +
\overline{\Psi}\gamma^{+}i\partial^{+}\Psi
 + \partial^{+}\phi\partial^{+}\phi \nonumber\\*
& & + \partial^{+}\bm{\pi} \cdot
\partial^{+}\bm{\pi} + \bm{\pi}\cdot\partial^{+}\bm{\pi}
\frac{\bm{\pi}\cdot\partial^{+}\bm{\pi}}{\pi^{2}}
(1-\frac{f^{2}}{\pi^{2}}\sin^{2}\frac{\pi}{f}).\label{eq:t++}
\end{eqnarray}
Since each of the terms in Eq.~(\ref{eq:t++}) involves one of the
fields, it is natural to associate each term with a particular
contribution to the plus momentum. This decomposition, though, is
not well defined; field equations relate various components. We
see the first three terms of Eq.~(\ref{eq:t++}) appear in $\langle
y \rangle$, which defines the nucleon contribution to the total
nuclear plus momentum, in the derivation of the Hugenholtz-van
Hove theorem Eq.~(\ref{eq:HvH}); we are not allowed to have the
vector mesons contribute a well defined fraction of plus momentum.
This means that one could trade certain mesonic degrees of freedom
for nucleons by replacing mesonic vertices with nucleon point
couplings, for example, in line with the general concept of
effective field theory. In our case the first three terms are
related by the vector meson field equation, but the fourth is left
out since the scalar mesons couple to the scalar density
$\overline{\Psi}\Psi$ which is not present in Eq.~(\ref{eq:t++}).
Therefore the scalar mesons (and pions) contribute a well defined
fraction of plus momentum. These explicit meson contributions
create an EMC binding effect, but the pionic contributions  are
also limited by nuclear Drell-Yan experiments \cite{Alde:im} to
carrying about 2\% of the plus momentum which is insufficient to
account for the entire EMC effect which corresponds to about 5\%
of the plus momentum for Iron.

\vita{ Jason Smith was born in Mattoon, Illinois on December 20,
1976, although he spent the vast majority of his childhood in The
Woodlands, Texas, a suburb of Houston. He has been slowly educated
since 1982, earning along the way a Bachelor of Science in Physics
and a Bachelor of Science in Mathematics from the University of
Texas at Austin in May of 1999, and a Doctor of Philosophy in
physics from the University of Washington, Seattle in August of
2005. }

\end{document}